\documentclass[journal]{IEEEtran}

\usepackage[prependcaption,colorinlistoftodos]{todonotes}

\usepackage{xcolor}
\usepackage{cite}
\usepackage{amssymb,amsfonts}
\usepackage{graphicx}
\usepackage{textcomp}
\usepackage{booktabs}
\usepackage{multirow}
\usepackage{tabularx}
\usepackage{siunitx}
\DeclareSIUnit \voltampere { VA }
\usepackage{lipsum}
\usepackage{bbm}
\usepackage{comment}
\usepackage{makecell}
\usepackage{bm}
\usepackage{adjustbox}

\usepackage{enumitem}

\usepackage{amsthm}
\usepackage[cmex10]{amsmath}

\newtheorem{remark}{Remark}

\usepackage{tikz}
\usetikzlibrary{positioning}
\usetikzlibrary{arrows}
\usetikzlibrary{circuits.ee.IEC}

\usepackage{algorithm}
\usepackage{algpseudocode}

\definecolor{backgroundblue}{rgb}{0, 0.4470, 0.7410}
\definecolor{hydro}{rgb}{0.101,0.5,1}
\definecolor{backgroundgreen}{rgb}{0.4660, 0.6740, 0.1880}
\definecolor{backgroundred}{rgb}{0.6196, 0.1294,0.0902}
\definecolor{steam}{rgb}{0.9290, 0.6940, 0.1250}
\definecolor{PV}{rgb}{1, 0.752, 0}
\definecolor{battery}{rgb}{1, 0.4, 1}
\definecolor{wind}{rgb}{0.4, 0.8, 1}
\definecolor{SC}{rgb}{0.572, 0.816, 0.313}
\definecolor{STATCOM}{rgb}{0, 0.8, 0.6}
\definecolor{wind_form}{rgb}{0.4, 0.8, 1}
\definecolor{PV_form}{rgb}{1, 0.752, 0}
\definecolor{battery_form}{rgb}{1, 0.5686, 1}
\definecolor{wind_foll}{rgb}{0, 0.6352, 1}
\definecolor{PV_foll}{rgb}{1, 0.549, 0}
\definecolor{battery_foll}{rgb}{0.9176, 0.1725, 0.9333}
\definecolor{redred}{rgb}{0.8,0.192,0.1}

\usepackage[capitalize,nameinlink,compress]{cleveref}

\crefname{figure}{Fig.}{Figures} % Use Figure instead of Fig.
\crefname{line}{line}{lines} % Make sure line is not capitalized
\crefname{claim}{Claim}{Claims} % Make sure line is not capitalized
\crefname{equation}{}{} % No Eq. for equations
\crefname{problem}{Problem}{Problems}
\crefname{assumption}{Assumption}{Assumptions}

\usepackage{cuted}
\usepackage{flushend}
\usepackage{stfloats}

\usepackage[font=footnotesize]{caption}
\usepackage[font=footnotesize]{subcaption}

\usepackage[bottom]{footmisc}

\usepackage{accents}

\usepackage{soul}
\usepackage{framed}
\colorlet{shadecolor}{yellow}
\soulregister\cite7
\soulregister\cref7
\soulregister\ref7
\soulregister\pageref7
\soulregister\eqref7

\renewcommand{\baselinestretch}{0.98}

\begin{document}

\title{Grid-Forming and Spatially Distributed Control Design of Dynamic Virtual Power Plants}

\author{Verena~Häberle,~\IEEEmembership{Student Member,~IEEE,}
        Ali~Tayyebi,
        Xiuqiang~He,~\IEEEmembership{Member,~IEEE,}\\
        Eduardo~Prieto-Araujo,~\IEEEmembership{Senior Member,~IEEE,}
        and~Florian~Dörfler,~\IEEEmembership{Senior Member,~IEEE} \\ $\vspace{-4.5mm}$ % <-this % stops a space
\thanks{This work was supported by the European Union's Horizon 2020 research and innovation program (Grant Agreement Number 883985). Verena Häberle, Ali Tayyebi, Xiuqiang He and Florian Dörfler are with the Automatic Control Laboratory, ETH Zurich, 8092 Zurich, Switzerland. Email:\{verenhae,xiuqhe,dorfler\}@ethz.ch. Ali Tayyebi is additionally with Hitachi Energy Research (HER), 72226 Väster\aa s, Sweden. Email: ali.tayyebi@hitachienergy.com. Eduardo Prieto-Araujo is a Serra Húnter Lecturer with CITCEA, Universitat Politècnica de Catalunya, 08028 Barcelona, Spain. Email: eduardo.prieto-araujo@upc.edu}% <-this % stops a space
}

\maketitle

\begin{abstract}
We present a novel \textit{grid-forming} control design approach for dynamic virtual power plants (DVPP). We consider a group of heterogeneous grid-forming distributed energy resources (DER) which collectively provide desired dynamic ancillary services, such as fast frequency and voltage control. To achieve that, we study the nontrivial aggregation of grid-forming DERs to establish the DVPP, and employ an adaptive divide-and-conquer strategy that disaggregates the desired control specifications of the aggregate DVPP via adaptive dynamic participation factors to obtain local desired behaviors of each DER. We then design local controllers at the DER level to realize these local desired behaviors. In the process, physical and engineered limits of each DER are taken into account. We extend the proposed approach to make it also compatible with grid-following DER controls, thereby establishing the concept of so-called hybrid DVPPs. Furthermore, we generalize the DVPP design to spatially dispersed DER locations in power grids with different voltage levels {and $R/X$ ratios}. Finally, the DVPP control performance is verified via numerical case studies in the IEEE nine-bus transmission grid with an interconnected medium voltage distribution grid. 
\end{abstract}

\begin{IEEEkeywords}
Dynamic virtual power plant, dynamic ancillary services, grid-forming control
\end{IEEEkeywords}

\IEEEpeerreviewmaketitle

\section{Nomenclature}
\noindent\begin{tabular}{ll}
         AC& Alternating current\\
         ADPF&Adaptive dynamic participation factor\\
         BESS& Battery energy storage system\\
         BPF& Band-pass filter\\
         COI& Center of inertia\\
         DC & Direct current\\
         DER& Distributed energy resource\\
         DVPP& Dynamic virtual power plant  \\
         EMT& Electromagnetic transients\\
         HPF& High-pass filter\\
         HV, MV, LV& High voltage, medium voltage, low voltage\\
         LPF& Low-pass filter\\
         LPV& Linear parameter-varying\\
         PCC& Point of common coupling\\
         PI & Proportional integral\\
         PLL& Phase-locked loop\\
         POC& Point of coupling\\
         PV & Photovoltaic\\
         RoCoF & Rate-of-change-of-frequency\\
         SG& Synchronous generator
\end{tabular}

\section{Introduction}
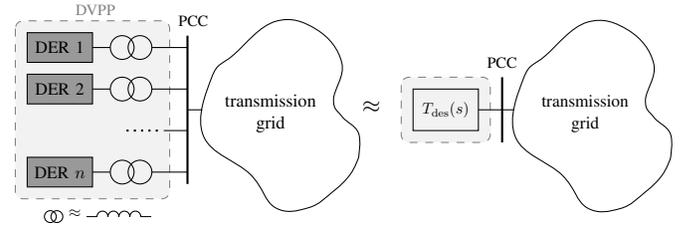
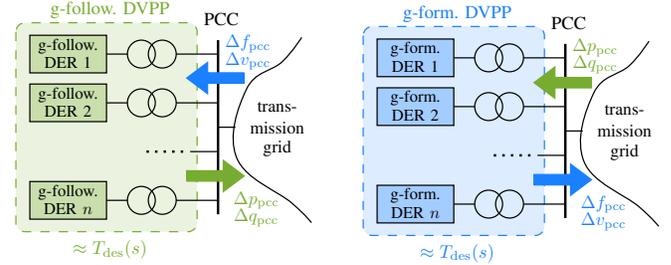
\begin{figure}[t!]
\centering
\vspace{-2mm}
\begin{subfigure}{0.5\textwidth}
    \centering
    \resizebox {0.98\columnwidth} {!} {
\begin{tikzpicture}[circuit ee IEC,scale=0.8, every node/.style={scale=0.6}]
\draw  [dashed, rounded corners=3,color=black!50, fill=black!5](-10.4,3.85) rectangle (-7.8,0.85);

\draw  plot[rotate=-90,smooth, tension=.7,scale=1.4] coordinates {(-1.1,-3.3) (-1.4,-3.3) (-1.6,-3.5) (-1.8,-3.6)(-2.2428,-3.5) (-2.6428,-3.8) (-2.7428,-4.3) (-2.3428,-4.8) (-2.0667,-5.0667) (-1.3667,-5.1667) (-0.9,-4.6) (-0.5,-4.3) (-0.5,-3.9) (-0.7,-3.6) (-0.9,-3.4) (-1.1,-3.3)};
\node [scale = 1.15]at (-6.1,2.5) {transmission};
\node [scale = 1.15]at (-6.1,2.1) {grid};
\draw (-7.5,2.35) -- (-7.25,2.35);

\draw [thick](-7.5,3.6) -- (-7.5,1.1); 
\draw (-7.5,2.7)  -- (-8.1,2.7) node (v2) {};

\draw (-7.5,1.3) -- (-8.1,1.3) node (v3) {}; 
\draw (-7.5,3.4) -- (-8.1,3.4);

\draw  (-8.6,3.4) ellipse (0.2 and 0.2);
\draw (-8.3,3.4) ellipse (0.2 and 0.2); 

\draw  (-9.81,0.55) ellipse (0.1 and 0.1);
\draw (-9.7,0.55) ellipse (0.1 and 0.1); 

\draw  (-8.6,1.3) ellipse (0.2 and 0.2);
\draw (-8.3,1.3) ellipse (0.2 and 0.2); 

\draw  (-8.6,2.7) ellipse (0.2 and 0.2);
\draw (-8.3,2.7) ellipse (0.2 and 0.2); 

\draw (-8.8,3.4) -- (-9.1,3.4); 
\draw (-8.8,2.7)  -- (-9.1,2.7); 
\draw (-8.8,1.3)  -- (-9.1,1.3); 
\draw [fill=black!40] (-10.2,3.65) rectangle (-9.1,3.15); 
\draw  [fill=black!40](-10.2,2.95) rectangle (-9.1,2.45); 
\draw [fill=black!40] (-10.2,1.55) rectangle (-9.1,1.05);
\node [scale = 1.1]at (-9.65,3.4) {DER 1};
\node [scale = 1.1]at (-9.65,2.7) {DER 2};
\node [scale = 1.1]at (-9.65,1.3) {DER $n$};
\draw (-7.5,2) -- (-7.9,2);
\draw[dotted,thick] (-8,2) -- (-8.6,2);

\node [color=black!50] at (-9.05,4.03) {DVPP};

\node at (-7.4,3.85) {PCC};
\draw  plot[rotate=-90,smooth, tension=.7,scale=1.4] coordinates {(-1.0905,0.4525) (-1.3905,0.4525) (-1.5905,0.2525) (-1.7905,0.1525)(-2.2333,0.2525) (-2.6333,-0.0475) (-2.7333,-0.5475) (-2.3333,-1.0475) (-2.0572,-1.3142) (-1.3572,-1.4142) (-0.8905,-0.8475) (-0.4977,-0.5475) (-0.4905,-0.1475) (-0.6905,0.1525) (-0.8905,0.3525) (-1.0905,0.4525)};
\node [scale = 1.1]at (-0.8,2.5) {transmission};
\node [scale = 1.1]at (-0.8,2.1) {grid};

\draw  [rounded corners=3,dashed, color=black!50, fill=black!5]  (-3.9,2.9) rectangle (-2.4,1.8);
\draw(-3.7,2.7) rectangle (-2.6,2);
\node at (-3.15,2.35) {$T_\mathrm{des}(s)$};
\draw (-2,2.35) -- (-2.6,2.35) node (v1) {};

\node [scale=1.5] at (-4.4,2.4) {$\approx$};
\draw [thick](-2.2,2.9) -- (-2.2,1.8);
\node at (-2.2,3.15) {PCC};
%\node at (-9.4,0.6) {$\stackrel{\wedge}{=}$};
\node at (-9.4,0.6) {$\approx$};
\draw(-9.2,0.55) to [inductor={}] (-8.11,0.55);
\end{tikzpicture}}
           \vspace{-9.5mm}
    \caption{{Sketch of a DVPP composed of heterogeneous DERs to match a desired dynamic behavior $T_\mathrm{des}(s)$. The latter may encode a fast frequency and voltage response for dynamic ancillary services provision, e.g., by relating active and reactive power with frequency and voltage magnitude, accordingly.}}
    \label{fig:DVPP_at_PCC_schematic}
        \vspace{1.2mm}
\end{subfigure}
\begin{subfigure}{0.23\textwidth}
    \centering
    \resizebox {0.98\columnwidth} {!} {
\begin{tikzpicture}[circuit ee IEC,scale=0.8, every node/.style={scale=0.6}]
\draw  [dashed, rounded corners=3,color=backgroundgreen, fill=backgroundgreen!15](-10.4,3.85) rectangle (-7.8,0.85);

\draw  plot[rotate=-90,smooth, tension=.7,scale=1.4] coordinates { (-2.6,-4.5) (-2.35,-4.7) (-2.0667,-5.0667) (-1.3667,-5.1667) (-0.9,-4.6) (-0.7,-4.4)};
\node [scale = 0.95] at (-6.65,2.65) {trans-};
\node [scale = 0.95] at (-6.65,2.35) {mission};
\node [scale = 0.95] at (-6.65,2.05) {grid};
\draw (-7.5,2.35) -- (-7.25,2.35);

\draw [thick](-7.5,3.6) -- (-7.5,1.1); 
\draw (-7.5,2.7)  -- (-8.1,2.7) node (v2) {};

\draw (-7.5,1.3) -- (-8.1,1.3) node (v3) {}; 
\draw (-7.5,3.4) -- (-8.1,3.4);

\draw  (-8.6,3.4) ellipse (0.2 and 0.2);
\draw (-8.3,3.4) ellipse (0.2 and 0.2);

\draw  (-8.6,1.3) ellipse (0.2 and 0.2);
\draw (-8.3,1.3) ellipse (0.2 and 0.2); 

\draw  (-8.6,2.7) ellipse (0.2 and 0.2);
\draw (-8.3,2.7) ellipse (0.2 and 0.2); 

\draw (-8.8,3.4) -- (-9.1,3.4); 
\draw (-8.8,2.7)  -- (-9.1,2.7); 
\draw (-8.8,1.3)  -- (-9.1,1.3); 
\draw [fill=backgroundgreen!40] (-10.2,3.67) rectangle (-9.1,3.13); 
\draw  [fill=backgroundgreen!40](-10.2,2.97) rectangle (-9.1,2.43); 
\draw [fill=backgroundgreen!40] (-10.2,1.57) rectangle (-9.1,1.03);
\node [scale = 0.9]at (-9.65,3.52) {g-follow.};
\node [scale = 0.9]at (-9.65,3.28) {DER 1};
\node [scale = 0.9]at (-9.65,2.82) {g-follow.};
\node [scale = 0.9]at (-9.65,2.58) {DER 2};
\node [scale = 0.9]at (-9.65,1.42) {g-follow.};
\node [scale = 0.9]at (-9.65,1.18) {DER $n$};
\draw (-7.5,2) -- (-7.9,2);
\draw[dotted,thick] (-8,2) -- (-8.6,2);

\node [color=backgroundgreen] at (-9.05,4.03) {g-follow. DVPP};
\node [color=backgroundgreen] at (-9.05,0.6) {$\approx T_\mathrm{des}(s)$};

\node at (-7.45,3.9) {PCC};
\draw [scale=1.5, fill = hydro, color = hydro](-4.75,2.0874) -- (-5.1,2.0874) -- (-5.1,2.1874) -- (-5.3,2.0374) -- (-5.1,1.8874) -- (-5.1,1.9874) -- (-4.75,1.9874)--(-4.75,2.0874);
\draw [scale=1.5, fill = backgroundgreen, color = backgroundgreen, rotate=180](5.3,-1.05) -- (4.95,-1.05) -- (4.95,-0.95) -- (4.75,-1.1) -- (4.95,-1.25) -- (4.95,-1.15) -- (5.3,-1.15)--(5.3,-1.05);
\node [scale = 0.97, color = hydro]at (-7.07,3.55) {$\Delta f_\mathrm{pcc}$};
\node [scale = 0.97, color = hydro] at (-7.07,3.3) {$\Delta v_\mathrm{pcc}$};
\node [scale = 0.97, color = backgroundgreen] at (-6.95,1.3) {$\Delta p_\mathrm{pcc}$};
\node [scale = 0.97, color = backgroundgreen] at (-6.95,1.05) {$\Delta q_\mathrm{pcc}$};
%\draw [-stealth, hydro,thick](-8.85,2.85) -- (-9.2,2.85);
%\draw [-stealth, hydro,thick](-8.85,1.45) -- (-9.2,1.45);
%\draw [-stealth, hydro,thick](-8.85,3.55) -- (-9.2,3.55);
%\draw [-stealth, backgroundgreen,thick] (-9.2,1.15)--(-8.85,1.15);
%\draw [-stealth, backgroundgreen,thick] (-9.2,3.25)--(-8.85,3.25);
%\draw [-stealth, backgroundgreen,thick] (-9.2,2.55)--(-8.85,2.55);
%\node [backgroundgreen] at (-9.325,0.1) {$\begin{bmatrix}\Delta p_\mathrm{pcc}(s)\\ \Delta q_\mathrm{pcc}(s)\end{bmatrix}$};
%\node at (-8.025,0.1) {$\approx \,T_\mathrm{des}(s)$};
%\node [hydro] at (-6.825,0.1) {$\begin{bmatrix}\Delta f_\mathrm{pcc}(s)\\ \Delta v_\mathrm{pcc}(s)\end{bmatrix}$};
\end{tikzpicture}}
           \vspace{-10mm}
    \caption{{\textit{Grid-following} (g-follow.) DVPP composed of heterogeneous grid-following DERs. The desired dynamic behavior $T_\mathrm{des}(s)$ relies on a grid-following signal causality, with input being the frequency and voltage magnitude measurements $\Delta f_\mathrm{pcc}$ and $\Delta v_\mathrm{pcc}$ at the PCC, and output being the active and reactive power deviations $\Delta p_\mathrm{pcc}$ and $\Delta q_\mathrm{pcc}$.}}
    \label{fig:gfoll_DVPP_PCC}
    \vspace{0.25mm}
\end{subfigure}
\hspace{2mm}
\begin{subfigure}{0.23\textwidth}
    \centering
    \resizebox {0.98\columnwidth} {!} {
\begin{tikzpicture}[circuit ee IEC,scale=0.8, every node/.style={scale=0.6}]
\draw  [dashed, rounded corners=3,color=hydro, fill=hydro!15](-10.4,3.85) rectangle (-7.8,0.85);

\draw  plot[rotate=-90,smooth, tension=.7,scale=1.4] coordinates { (-2.6,-4.5) (-2.35,-4.7) (-2.0667,-5.0667) (-1.3667,-5.1667) (-0.9,-4.6) (-0.7,-4.4)};
\node [scale = 0.95] at (-6.65,2.65) {trans-};
\node [scale = 0.95] at (-6.65,2.35) {mission};
\node [scale = 0.95] at (-6.65,2.05) {grid};
\draw (-7.5,2.35) -- (-7.25,2.35);

\draw [thick](-7.5,3.6) -- (-7.5,1.1); 
\draw (-7.5,2.7)  -- (-8.1,2.7) node (v2) {};

\draw (-7.5,1.3) -- (-8.1,1.3) node (v3) {}; 
\draw (-7.5,3.4) -- (-8.1,3.4);

\draw  (-8.6,3.4) ellipse (0.2 and 0.2);
\draw (-8.3,3.4) ellipse (0.2 and 0.2);

\draw  (-8.6,1.3) ellipse (0.2 and 0.2);
\draw (-8.3,1.3) ellipse (0.2 and 0.2); 

\draw  (-8.6,2.7) ellipse (0.2 and 0.2);
\draw (-8.3,2.7) ellipse (0.2 and 0.2); 

\draw (-8.8,3.4) -- (-9.1,3.4); 
\draw (-8.8,2.7)  -- (-9.1,2.7); 
\draw (-8.8,1.3)  -- (-9.1,1.3); 
\draw [fill=hydro!40] (-10.2,3.67) rectangle (-9.1,3.13); 
\draw  [fill=hydro!40](-10.2,2.97) rectangle (-9.1,2.43); 
\draw [fill=hydro!40] (-10.2,1.57) rectangle (-9.1,1.03);
\node [scale = 0.9]at (-9.65,3.52) {g-form.};
\node [scale = 0.9]at (-9.65,3.28) {DER 1};
\node [scale = 0.9]at (-9.65,2.82) {g-form.};
\node [scale = 0.9]at (-9.65,2.58) {DER 2};
\node [scale = 0.9]at (-9.65,1.42) {g-form.};
\node [scale = 0.9]at (-9.65,1.18) {DER $n$};
\draw (-7.5,2) -- (-7.9,2);
\draw[dotted,thick] (-8,2) -- (-8.6,2);

\node [color=hydro] at (-9.05,4.03) {g-form. DVPP};
\node [color=hydro] at (-9.05,0.6) {$\approx T_\mathrm{des}(s)$};

\node at (-7.45,3.9) {PCC};
\draw [scale=1.5, fill = backgroundgreen, color = backgroundgreen](-4.75,2.0794) -- (-5.1,2.0794) -- (-5.1,2.1794) -- (-5.3,2.0294) -- (-5.1,1.8794) -- (-5.1,1.9794) -- (-4.75,1.9794)--(-4.75,2.0794);
\draw [scale=1.5, fill = hydro, color = hydro, rotate=180](5.3,-1.05) -- (4.95,-1.05) -- (4.95,-0.95) -- (4.75,-1.1) -- (4.95,-1.25) -- (4.95,-1.15) -- (5.3,-1.15)--(5.3,-1.05);
\node [scale = 0.97, color = backgroundgreen]at (-7.07,3.55) {$\Delta p_\mathrm{pcc}$};
\node [scale = 0.97, color = backgroundgreen] at (-7.07,3.3) {$\Delta q_\mathrm{pcc}$};
\node [scale = 0.97, color = hydro] at (-6.95,1.3) {$\Delta f_\mathrm{pcc}$};
\node [scale = 0.97, color = hydro] at (-6.95,1.05) {$\Delta v_\mathrm{pcc}$};
%\draw [-stealth, backgroundgreen,thick](-8.85,2.85) -- (-9.2,2.85);
%\draw [-stealth, backgroundgreen,thick](-8.85,1.45) -- (-9.2,1.45);
%\draw [-stealth, backgroundgreen,thick](-8.85,3.55) -- (-9.2,3.55);
%\draw [-stealth, hydro,thick] (-9.2,1.15)--(-8.85,1.15);
%\draw [-stealth, hydro,thick] (-9.2,3.25)--(-8.85,3.25);
%\draw [-stealth, hydro,thick] (-9.2,2.55)--(-8.85,2.55);
%\node [hydro] at (-9.325,0.1) {$\begin{bmatrix}\Delta f_\mathrm{pcc}(s)\\ \Delta v_\mathrm{pcc}(s)\end{bmatrix}$};
%\node at (-8.025,0.1) {$\approx \,T_\mathrm{des}(s)$};
%\node [backgroundgreen] at (-6.825,0.1) {$\begin{bmatrix}\Delta p_\mathrm{pcc}(s)\\ \Delta q_\mathrm{pcc}(s)\end{bmatrix}$};
\end{tikzpicture}}
           \vspace{-10.5mm}
    \caption{{\textit{Grid-forming} (g-form.) DVPP composed of heterogeneous grid-forming DERs. The desired dynamic behavior $T_\mathrm{des}(s)$ relies on a grid-forming signal causality, with input being the active and reactive power measurements $\Delta p_\mathrm{pcc}$ and $\Delta q_\mathrm{pcc}$ at the PCC, and output being the imposed frequency and voltage magnitude $\Delta f_\mathrm{pcc}$ and $\Delta v_\mathrm{pcc}$.}}
    \label{fig:gform_DVPP_PCC}
    \vspace{0.25mm}
\end{subfigure}
\caption{DVPP configurations with heterogeneous DERs connected at one single bus in the transmission grid. {DVPPs with spatially dispersed DER locations are considered later in our novel results in \cref{sec:spatially_distributed}}.}
\label{fig:DVPP_at_PCC}
\vspace{-5mm}
\end{figure}

\IEEEPARstart{I}{n} future power systems, dynamic ancillary services have to be shouldered by non-synchronous, distributed energy resources (DER), implying great challenges to cope with the fluctuating nature of renewable energy sources, as well as their device-specific limitations \cite{milano2018foundations}. Recently, the concept of \textit{dynamic virtual power plants (DVPP)} has been proposed to pave the way for future ancillary services provision by DERs \cite{marinescu2021dynamic,haberle2021control,joak,zhong2021coordinated,zhong2021impact}. DVPPs are ensembles of heterogeneous DERs (all with individual constraints), aggregated to collectively provide desired \textit{dynamic} ancillary services such as fast frequency and voltage control {(\cref{fig:DVPP_at_PCC_schematic})}. In particular, while none of the DERs in isolation can provide these services consistently across all power and energy levels or all time scales, a sufficiently heterogeneous group of DERs is able to do so.

All of the few existing control design methods for DVPPs found in literature rely on \textit{grid-following} DER aggregations (\cref{fig:gfoll_DVPP_PCC})\cite{haberle2021control,joak,zhong2021coordinated,zhong2021impact}. In this case, frequency and voltage magnitude are measured at the point of common coupling (PCC) (via an explicit synchronization scheme, e.g., a phase-locked loop (PLL)) and treated as inputs to the DVPP units. Based on these inputs, the units' active and reactive power outputs are modified, and sum up to the aggregated dynamic power output of the DVPP, which is intended to match the desired dynamic behavior $T_\mathrm{des}(s)$. However, due to their dependency on the measured frequency, grid-following DVPPs require a stiff grid to operate (in terms of a stable frequency and voltage) \cite{kenyon2020stability,kenyon2020grid}. Namely, their responsiveness and PLL tracking performance can deteriorate or even result in instabilities when the DVPPs are integrated into grids with a high share of grid-following DERs, lacking in external assets which form a stable ac grid frequency. Another limitation of grid-following DER controls, and therefore of grid-following DVPPs, is their inability to operate in an isolated mode or black start a network \cite{rocabert2012control}. 

In the course of future power systems to more and more incorporate grid-forming DERs \cite{milano2018foundations,paolone2020fundamentals,lasseter2019grid}, we resolve these issues by envisioning \textit{grid-forming} DERs to construct more reliable DVPP configurations, which establish an independent frequency and voltage magnitude while providing dynamic grid support {(\cref{fig:gform_DVPP_PCC})}\cite{kenyon2020stability,kenyon2020grid}. In particular, given the voltage source behavior of grid-forming DERs \cite{osmoseh2020}, corresponding DVPPs {composed of grid-forming DERs} do not need a stiff grid to operate, and typically exhibit a grid-friendly response behavior, superior stability characteristics under weak grid conditions, as well as black start capabilities \cite{rocabert2012control}. Beyond that, the practical relevance of grid-forming DVPPs is also supported by future grid-code specifications which are envisioned to require enough grid-forming assets that impose an independent frequency and voltage magnitude while providing dynamic grid-support\cite{osmoseh2020,entso2021grid}. Nevertheless, considering the imminent transition to future power systems, they are expected to contain a mixture of grid-following and grid-forming DERs \cite{kenyon2020stability}, thus also raising the need for DVPPs that combine both types of DER controls, which we refer to as \textit{hybrid DVPPs}. 

In this paper, we present a novel control design approach for grid-forming DVPPs that enables desired frequency and volt- age control services on fast timescales in an aggregated fash- ion. In particular, we consider an aggregated dynamic ancillary services provision of the DVPP that relies on a grid-forming signal causality, i.e., the DVPP frequency and voltage magni- tude are imposed while being controlled as a function of power measurements. As the main difference to our grid-following DVPP proposal in\cite{haberle2021control}, in this work, we study the aggregation of \textit{grid-forming} DERs, which requires a special treatment and results in nontrivial aggregation conditions of the local DER dynamics. Once the aggregation is obtained, similar to\cite{haberle2021control}, we resort to an adaptive divide-and-conquer strategy, which disag- gregates the overall DVPP specification via adaptive dynamic participation factors to obtain local desired behaviors of each DVPP unit, while taking (possibly time-varying) DER limita- tions into account. Moreover, as a second contribution of this paper, we extend the proposed grid-forming control setup to also include grid-following DER controls, thereby establishing the {novel} concept of hybrid DVPPs. As a starting point, we consider all DVPP units to be connected at one single bus in the transmission system (\cref{fig:DVPP_at_PCC}), where we specify a \textit{decoupled} desired behavior for frequency and voltage regulation through measured active and reactive power injections changes. 

Several work on the aggregation of grid-forming DERs to provide dynamic ancillary services has been proposed in literature, e.g.,\cite{elkhatib2018evaluation,su2020adaptive}. However, they typically consider aggregations of identical DERs which are all equipped with the same pre-specified local control law. Hence, no aggregated control design of the individual DERs is explicitly pursued, such that a desired aggregated behavior for any kind of dynamic ancillary services provision cannot be achieved. This is in vast contrast to our approach, where we consider an aggregation of heterogeneous DERs, whose controllers are designed with the goal to match any desired aggregated behavior that has been specified by the system operator for dynamic ancillary services provision. Moreover, our controllers of the DVPP units are designed based on adaptive dynamic participation factors to take time-varying DER limitations (e.g., capacity, response time, weather-dependency, etc.) into account. In contrast, being restricted to aggregations of identical DERs only, the existing work cannot account for individual DER limitations by mutual compensation of different DERs. 

As the third contribution of this paper, we generalize the DVPP controls to \textit{spatially distributed} DER locations in the power grid, ranging from high-voltage transmission systems to low-voltage distribution grids {with different $R/X$ ratios}. In particular, we consider a DVPP configuration, where the DERs are spatially distributed within an area of the power system that is connected to the remaining power grid via one or multiple point of couplings. The control of spatially distributed DVPP units has already been proposed in the work in\cite{zhong2021impact}, where different DVPP topologies in the transmission and distribution system are investigated for grid-following primary frequency control. However, the method in \cite{zhong2021impact} relies on the decoupled power flow assumption, and thus does not account for non-inductive line impedances between the spatially distributed DERs and associated power losses, which are especially relevant in medium-voltage (MV) or low-voltage (LV) distribution grids {with high $R/X$ ratios\cite{de2007voltage,yao2010design,bevrani2013intelligent}. In particular, high $R/X$ ratios may cause a nodal active power injection difference between the local node of a DER and the DVPP's point(s) of coupling.} In this regard, the method in \cite{zhong2021impact} cannot provide a desired aggregate behavior at the coupling point(s) of the DVPP area, and affects the overall frequency response {of the power system} only in an imprecise way. This is, however, contradicting the literal definition of a (D)VPP, being a DER aggregation with the same technological functionality as a single conventional power plant \cite{marinescu2021dynamic,saboori2011virtual}. 

In contrast, in our proposed method for spatially distributed DVPP control, we resort to \textit{coupled} desired specifications for frequency and voltage regulation through both active and reactive power. In particular, {the coupled control of active and reactive power has been reported to be an effective solution for frequency and voltage regulation especially in MV or LV grids with high $R/X$ ratios\cite{rocabert2012control,de2007voltage,yao2010design,bevrani2013intelligent}. Since such a coupled control allows to account for non-inductive grid impedances between the DVPP units, we can propose a more versatile control strategy, ensuring an accurate matching of a desired response behavior at the coupling point(s) of the DVPP area.}

This paper is organized as follows: In \cref{sec:grid_forming_and_hybrid_DVPP}, we introduce the grid-forming DVPP control setup for fast frequency and voltage control, as well as its extension to a hybrid DVPP configuration. We provide a simplified setup using the formalism of linearized systems, which makes it convenient to develop our control design. \cref{sec:spatially_distributed} presents the generalization of the DVPP control to spatially distributed DER locations. In \cref{sec:testcase}, we demonstrate the performance of our DVPP control design via numerical case studies in the IEEE nine-bus system with an interconnected medium voltage grid, using detailed and nonlinear system and device models. Finally, \cref{sec:conclusion} summarizes the main results and discusses open questions. 

\section{Grid-Forming {\& Hybrid} DVPP Control}\label{sec:grid_forming_and_hybrid_DVPP}

\subsection{Grid-Forming DVPP Control Setup}\label{sec:grid_forming_DVPP_control}
We consider a grid-forming DVPP as a collection $\mathcal{F}_\mathrm{orm}$ of heterogeneous grid-forming DERs. To start with, we assume that all DVPP units are connected in parallel at the same bus of a transmission grid (Fig. \ref{fig:gform_DVPP_PCC}), which we refer to as point of common coupling (PCC). While being connected at one bus, it is essential to ensure a sufficiently large electrical distance (e.g., via transformers or cables) between the grid-forming units, since they are operated as parallel voltage sources and cannot simultaneously impose different voltages at the PCC.

To compensate for ancillary services conventionally provided by synchronous generators in transmission networks during normal operating conditions, a \textit{decoupled} frequency and voltage control behavior is specified for the aggregate DVPP as a desired diagonal transfer function matrix
    \begin{align}\label{eq:aggregate_specification}
        \begin{bmatrix}
            \Delta f_\mathrm{pcc}(s)\\ \Delta v_\mathrm{pcc}(s)
        \end{bmatrix}
        =\underset{=T_\mathrm{des}(s)}{\underbrace{\begin{bmatrix}
            T_\mathrm{des}^\mathrm{pf}(s)&0\\0&T_\mathrm{des}^\mathrm{qv}(s)
        \end{bmatrix}}}
        \begin{bmatrix}
           \Delta p_\mathrm{pcc}(s) \\ \Delta q_\mathrm{pcc}(s)
        \end{bmatrix},
    \end{align}
where $\Delta p_\mathrm{pcc}$ and $\Delta q_\mathrm{pcc}$ are the measured active and reactive power injection changes at the PCC (deviating from the respective power setpoint), and $\Delta f_\mathrm{pcc}$ and $\Delta v_\mathrm{pcc}$ denote the imposed frequency and voltage magnitude deviation at the PCC. In this work, we primarily consider small-signal changes during normal operating conditions as usual for dynamic ancillary services provision. Moreover, notice that our formalism directly extends to \textit{coupled} control specifications for $T_\mathrm{des}(s)$, which are potentially relevant in other types of networks (see \cref{sec:spatially_distributed} for details).
\begin{remark}
    The aggregate DVPP specification in \eqref{eq:aggregate_specification} relies on a grid-forming signal causality, where frequency and voltage magnitude at the PCC are imposed while being controlled as a function of active and reactive power measurements. In contrast, for the grid-following DVPP developed in our work in\cite{haberle2021control}, we consider an inverted, i.e., grid-following, signal causality of the aggregate DVPP specification, where the ag- gregated power injection at the PCC is controlled as a function of frequency and voltage magnitude measurements, i.e.,
\begin{align}\label{eq:aggregate_specification_gfoll}
        \begin{bmatrix}
            \Delta p_\mathrm{pcc}(s)\\ \Delta q_\mathrm{pcc}(s)
        \end{bmatrix}
        =\underset{=T_\mathrm{des}(s)}{\underbrace{\begin{bmatrix}
            T_\mathrm{des}^\mathrm{fp}(s)&0\\0&T_\mathrm{des}^\mathrm{vq}(s)
        \end{bmatrix}}}
        \begin{bmatrix}
           \Delta f_\mathrm{pcc}(s) \\ \Delta v_\mathrm{pcc}(s)
        \end{bmatrix}.
    \end{align}
    An illustrative comparison of both DVPP concepts can be found in \cref{fig:comparison_DVPP}.
\end{remark}

The desired behavior $T_\mathrm{des}(s)$ in \cref{eq:aggregate_specification} is intended to be provided by the aggregated dynamics of the grid-forming DVPP units, which can be achieved by controlling their local closed-loop behaviors appropriately. We therefore assume that all DVPP units are controllable\footnote{The grid-forming DVPP control setup can be extended to also include \textit{non-controllable} (i.e., known pre-installed/fixed) local DER dynamics, e.g., synchronous generator swing dynamics with turbine and governor controls, or automatic voltage regulators, etc. as in\cite{haberle2021control}.} converter-based DERs, i.e., their local dynamic behaviors can be altered to collectively match the desired specification in \cref{eq:aggregate_specification}. In doing so, it is important to ensure that practical limitations of the DERs are not exceeded during normal operation, and device-level stability can be guaranteed.

Of course, to achieve this, the power park comprising the DVPP has to be sufficiently diverse covering all time scales and energy/power levels. Furthermore, it is assumed that $T_\mathrm{des}(s)$ is provided by the power system operator, who is encoding grid-code requirements in the form of desired transfer functions (e.g., virtual inertia and droop control; see \eqref{eq:T_des_casestudy1}). In this regard, $T_\mathrm{des}(s)$ is supposed to be reasonably specified so that its collective realization is feasible for the DVPP units during normal operating conditions, while rendering the closed-loop power system stable, and being robust to model uncertainties and parameter variations in the rest of the grid. In other words, system-level stability is ensured by properly specifying the desired ancillary service transfer function $T_\mathrm{des}(s)$, which is also part of our ongoing research.

In what follows, we present detailed control setups of the aggregated grid-forming DVPP dynamics for both frequency and voltage regulation to match the desired behavior in \cref{eq:aggregate_specification}. Notice that the frequency and voltage regulation of the grid-forming DVPP have to be addressed separately (unlike in the grid-following DVPP setup\cite{haberle2021control}; see \cref{fig:DVPP_ctrl_setup_gfoll}).

\renewcommand{\arraystretch}{1.2}
\begin{table}[t!]\scriptsize
    \centering
    \setlength{\tabcolsep}{1mm}
     \caption{List of notation for the grid-forming DVPP frequency control setup.}
    \vspace{-1mm}
    \begin{tabular}{c||c}
     \toprule
         Description & Symbol  \\ \hline
         Set of grid-forming DVPP units &$\mathcal{F}_\mathrm{orm}$\\
         Laplacian matrix of DVPP interconnection network & $L_\mathrm{dvpp}$\\\hline
         Active power injection change at the PCC & $\Delta p_\mathrm{pcc}$\\
         Imposed frequency deviation at the PCC & $\Delta f_\mathrm{pcc}$\\ \hline
         Local active power injection disturbance of unit $i$ & $\Delta p_{\mathrm{d},i}$\\
         Local active power output deviation of unit $i$ & $\Delta p_i$\\
         Local active power flow fluctuation of unit $i$ & $\Delta p_{\mathrm{e},i}$\\
         Local frequency deviation of unit $i$ & $\Delta f_i$\\\hline 
         Vector of local active power injection disturbances & $\Delta p_{\mathrm{d}}= [\Delta p_{\mathrm{d},1}\,...\, \Delta p_{\mathrm{d},n}]^\top$\\
         Vector of local active power output deviations & $\Delta p = [\Delta p_1 \,...\, \Delta p_n]^\top$\\
         Vector of local active power flow fluctuations & $\Delta p_{\mathrm{e}}=[\Delta p_{\mathrm{e},1}\,...\, \Delta p_{\mathrm{e},n}]^\top$\\
         Vector of local frequency deviations & $\Delta f = [\Delta f_1\,...\, \Delta f_n]^\top$\\ \hline
         Local $\mathrm{p}$-$\mathrm{f}$ closed-loop transfer function of unit $i$ & $T_i^\mathrm{pf}(s)$\\
         Desired DVPP transfer function for $\mathrm{p}$-$\mathrm{f}$ control & $T_\mathrm{des}^\mathrm{pf}(s)$\\
    \bottomrule
    \end{tabular}
\label{tab:gform_freq_symbols}
\vspace{-3mm}
\end{table}
\renewcommand{\arraystretch}{1} \normalsize

\renewcommand{\arraystretch}{2}
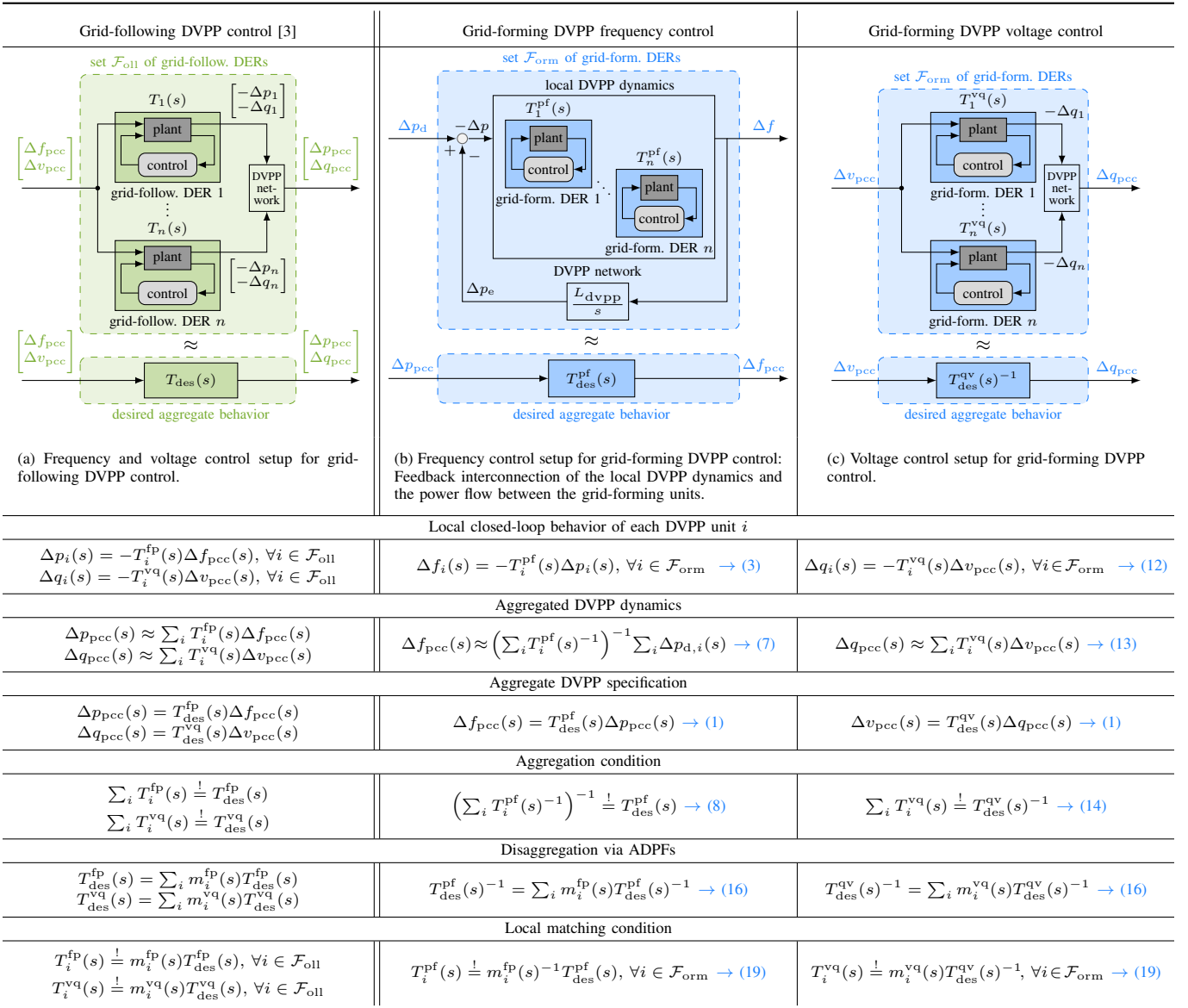
\begin{figure*}[t!]
\setlength{\tabcolsep}{1mm}
       \centering \scriptsize
    \begin{tabular}{c||c|c}
     \toprule
{Grid-following DVPP control\cite{haberle2021control}}&{Grid-forming DVPP frequency control} & {Grid-forming DVPP voltage control} \\ \hline

\begin{tikzpicture}[scale=0.9,every node/.style={scale=0.9}]
\tikzstyle{roundnode}=[circle,draw=black!60,fill=black!5,scale=0.75]
\draw [dashed,rounded corners = 3,backgroundgreen, fill=backgroundgreen!15] (0.1,0.15) rectangle (3.8,-4.3);
\draw [dashed, rounded corners = 3,backgroundgreen, fill=backgroundgreen!15] (0.1,-4.7) rectangle (3.8,-5.5);
\draw [fill=backgroundgreen!40] (0.7,-0.5) rectangle (2.5,-1.7);
\draw  [fill=backgroundgreen!40](0.7,-2.7) rectangle (2.5,-3.9);
\draw [fill=black!40] (1.2,-0.6) rectangle (2,-1);
\node at (1.6,-0.8) {plant};
\draw [rounded corners = 3,fill=black!15] (1.1,-1.2) rectangle (2.1,-1.6);
\node at (1.6,-1.4) {control};

\draw [-latex](2,-0.9) -- (2.4,-0.9) -- (2.4,-1.4) -- (2.1,-1.4); 
\draw[-latex] (1.1,-1.4) -- (0.8,-1.4) -- (0.8,-0.9) -- (1.2,-0.9); 

\node at (1.6,-0.3) {$T_1(s)$};
\node at (1.6,-1.9) {grid-follow. DER 1};

\draw [fill=black!40] (1.2,-2.8) rectangle (2,-3.2);
\node at (1.6,-3) {plant};
\draw [rounded corners = 3,fill=black!15] (1.1,-3.4) rectangle (2.1,-3.8);
\node at (1.6,-3.6) {control};

\draw [-latex](2,-3.1) -- (2.4,-3.1) -- (2.4,-3.6) -- (2.1,-3.6); 
\draw[-latex] (1.1,-3.6) -- (0.8,-3.6) -- (0.8,-3.1) -- (1.2,-3.1); 

\node at (1.6,-2.5) {$T_n(s)$};
\node at (1.6,-4.1) {grid-follow. DER $n$};

\node [scale = 0.7] at (1.6,-2.1) {$\bm{\vdots}$};
\draw [-latex](-0.9,-1.8) -- (0.4,-1.8)  -- (0.4,-0.7) -- (1.2,-0.7); 
\draw  [-latex](0.4,-1.8)-- (0.4,-2.9) -- (1.2,-2.9);
\fill [black] (0.4,-1.8) circle(0.4mm);

\draw [-latex](2,-0.7) -- (3.3,-0.7) -- (3.3,-1.4); 
\draw [-latex](2,-2.9) -- (3.3,-2.9) -- (3.3,-2.2);

\draw [-latex](3.6,-1.8) -- (4.9,-1.8);

\node[backgroundgreen] at (-0.5,-1.3) {$\begin{bmatrix}\vspace{-7mm}\\\Delta f_\mathrm{pcc}\vspace{-2.5mm}\\ \Delta v_\mathrm{pcc}\end{bmatrix}$};
\node[backgroundgreen] at (4.4,-1.3) {$\begin{bmatrix}\vspace{-7mm}\\\Delta p_\mathrm{pcc}\vspace{-2.5mm}\\\Delta q_\mathrm{pcc}\end{bmatrix}$};
\node at (3.15,-0.3) {$\begin{bmatrix}\vspace{-7mm}\\-\Delta p_1\vspace{-3mm}\\-\Delta q_1 \vspace{-0.5mm}\end{bmatrix}$};
\node at (3.15,-3.3) {$\begin{bmatrix}\vspace{-7mm}\\-\Delta p_n\vspace{-3mm}\\ -\Delta q_n\vspace{-0.5mm}\end{bmatrix}$};

\node [backgroundgreen] at (1.8,0.35) {set $\mathcal{F}_\mathrm{oll}$ of grid-follow. DERs};
\node [scale=1.2] at (2,-4.5) {$\approx$};

\draw  [fill=backgroundgreen!40](1.2,-4.8) rectangle (2.8,-5.4);
\node at (2,-5.1) {$T_\mathrm{des}(s)$};
\draw[-latex] (-0.9,-5.1) -- (1.2,-5.1); 
\draw[-latex] (2.8,-5.1) -- (4.9,-5.1);
\node[backgroundgreen] at (4.4,-4.6) {$\begin{bmatrix}\vspace{-7mm}\\\Delta p_\mathrm{pcc}\vspace{-2.5mm}\\\Delta q_\mathrm{pcc} \end{bmatrix}$};
\node[backgroundgreen] at (-0.5,-4.6) {$\begin{bmatrix}\vspace{-7mm}\\\Delta f_\mathrm{pcc}\vspace{-2.5mm}\\ \Delta v_\mathrm{pcc}\end{bmatrix}$};
\node[backgroundgreen] at (2,-5.7) {desired aggregate behavior};
\draw  (3,-1.4) rectangle (3.6,-2.2);
\node [scale=0.8]at (3.3,-1.6) {DVPP};
\node [scale=0.9]at (3.3,-1.8) {net-};
\node [scale=0.9]at (3.3,-2) {work};
\end{tikzpicture}

&

\begin{tikzpicture}[scale=0.95,every node/.style={scale=0.9}]
\tikzstyle{roundnode}=[circle,draw=black!60,fill=black!5,scale=0.75]
\draw [dashed,rounded corners = 3,hydro, fill=hydro!15] (-1,1.2) rectangle (3.9,-3);
\draw [dashed, rounded corners = 3,hydro, fill=hydro!15] (-1,-3.4) rectangle (3.9,-4.2);

\draw [fill=hydro!40] (1.9,-0.4) rectangle (3.3,-1.5);
\draw [fill=black!40] (2.3,-0.5) rectangle (2.9,-0.9);
\node at (2.6,-0.7) {plant};
\draw [rounded corners = 3,fill=black!15] (2.2,-1) rectangle (3,-1.4);
\node at (2.6,-1.2) {control};
\draw [-latex](2.9,-0.7) -- (3.2,-0.7) -- (3.2,-1.2) -- (3,-1.2); 
\draw[-latex] (2.2,-1.2) -- (2,-1.2) -- (2,-0.7) -- (2.3,-0.7); 
\node at (2.6,-0.2) {$T_n^\mathrm{pf}(s)$};
\node at (2.6,-1.7) {grid-form. DER $n$};

\draw [fill=hydro!40] (0.1,0.4) rectangle (1.5,-0.7);
\draw [fill=black!40] (0.5,0.3) rectangle (1.1,-0.1);
\node at (0.8,0.1) {plant};
\draw [rounded corners = 3,fill=black!15] (0.4,-0.2) rectangle (1.2,-0.6);
\node at (0.8,-0.4) {control};
\draw [-latex](1.1,0.1) -- (1.4,0.1) -- (1.4,-0.4) -- (1.2,-0.4); 
\draw[-latex] (0.4,-0.4) -- (0.2,-0.4) -- (0.2,0.1) -- (0.5,0.1); 
\node at (0.8,0.6) {$T_1^\mathrm{pf}(s)$};
\node at (0.8,-0.9) {grid-form. DER 1};

\node [scale=1.2] at (1.5,-3.2) {$\approx$};

\draw  [fill=hydro!40](0.8,-3.5) rectangle (2.2,-4.1);
\node at (1.5,-3.8) {$T_\mathrm{des}^\mathrm{pf}(s)$};
\draw[-latex] (-1.8,-3.8) -- (0.8,-3.8); 
\draw[-latex] (2.2,-3.8) -- (4.7,-3.8);
\node[hydro] at (4.3,-3.6) {$\Delta f_\mathrm{pcc}$};
\node[hydro] at (-1.4,-3.6) {$\Delta p_\mathrm{pcc}$};
\node[hydro] at (1.5,-4.4) {desired aggregate behavior};

\node at (1.7,-0.6) {$\ddots$};
\draw  (-0.1,0.8) rectangle (3.5,-1.85);
\draw  (1.1,-2.25) rectangle (2.1,-2.85);
\node [scale=1.2] at (1.6,-2.55) {$\frac{L_\mathrm{dvpp}}{s}$};
\node at (1.6,-2.05) {DVPP network};
\node at (1.8,0.95) {local DVPP dynamics};
\draw[-latex] (-0.5,0.1) -- (-0.1,0.1);
\node [roundnode] at (-0.6,0.1) {}; 
\draw[-latex] (1.1,-2.55) -- (-0.6,-2.55) -- (-0.6,0); 
\draw [-latex](-1.8,0.1) -- (-0.7,0.1);

\node at (-0.3,-2.35) {$\Delta p_\mathrm{e}$};
\node at (-0.45,0.3) {$-\Delta p $};
\node [hydro] at (4.3,0.3) {$\Delta f$};
\node [hydro] at (-1.4,0.3) {$\Delta p_\mathrm{d}$};

\node at (-0.8,-0.1) {$+$};
\node at (-0.4,-0.2) {$-$};
\draw [-latex](3.5,0.1) -- (3.7,0.1) node (v1) {} -- (3.7,-2.55) -- (2.1,-2.55);
\draw [-latex] (3.7,0.1)  -- (4.7,0.1);
\node [hydro] at (1.5,1.4) {set $\mathcal{F}_\mathrm{orm}$ of grid-form. DERs};
\end{tikzpicture}

&

\begin{tikzpicture}[scale=0.9,every node/.style={scale=0.9}]
\tikzstyle{roundnode}=[circle,draw=black!60,fill=black!5,scale=0.75]
\draw [dashed,rounded corners = 3,hydro, fill=hydro!15] (0,-0.2) rectangle (3.6,-4.4);
\draw [dashed, rounded corners = 3,hydro, fill=hydro!15] (0,-4.8) rectangle (3.6,-5.6);
\draw [fill=hydro!40] (0.9,-0.6) rectangle (2.7,-1.8);
\draw  [fill=hydro!40](0.9,-2.8) rectangle (2.7,-4);
\draw [fill=black!40] (1.4,-0.7) rectangle (2.2,-1.1);
\node at (1.8,-0.9) {plant};
\draw [rounded corners = 3,fill=black!15] (1.3,-1.3) rectangle (2.3,-1.7);
\node at (1.8,-1.5) {control};

\draw [-latex](2.2,-1) -- (2.6,-1) -- (2.6,-1.5) -- (2.3,-1.5); 
\draw[-latex] (1.3,-1.5) -- (1,-1.5) -- (1,-1) -- (1.4,-1); 

\node at (1.8,-0.4) {$T_1^\mathrm{vq}(s)$};
\node at (1.8,-2) {grid-form. DER 1};

\draw [fill=black!40] (1.4,-2.9) rectangle (2.2,-3.3);
\node at (1.8,-3.1) {plant};
\draw [rounded corners = 3,fill=black!15] (1.3,-3.5) rectangle (2.3,-3.9);
\node at (1.8,-3.7) {control};

\draw [-latex](2.2,-3.2) -- (2.6,-3.2) -- (2.6,-3.7) -- (2.3,-3.7); 
\draw[-latex] (1.3,-3.7) -- (1,-3.7) -- (1,-3.2) -- (1.4,-3.2); 

\node at (1.8,-2.6) {$T_n^\mathrm{vq}(s)$};
\node at (1.8,-4.2) {grid-form. DER $n$};

\node [scale = 0.7] at (1.8,-2.2) {$\bm{\vdots}$};
\draw [-latex](-0.8,-1.9) -- (0.4,-1.9)  -- (0.4,-0.8) -- (1.4,-0.8); 
\draw  [-latex](0.4,-1.9)-- (0.4,-3) -- (1.4,-3);
\fill [black] (0.4,-1.9) circle(0.4mm);

\draw [-latex](2.2,-0.8) -- (3.15,-0.8) -- (3.15,-1.5); 
\draw [-latex](2.2,-3) -- (3.15,-3) -- (3.15,-2.3);

\draw [-latex](3.45,-1.9) -- (4.5,-1.9);

\draw  (2.85,-1.5) rectangle (3.45,-2.3);

\node[hydro] at (-0.4,-1.7) {$\Delta v_\mathrm{pcc}$};
\node[hydro] at (4.1,-1.7) {$\Delta q_\mathrm{pcc}$};
\node at (3.2,-0.6) {$-\Delta q_1$};
\node at (3.2,-3.2) {$-\Delta q_n$};

\node [hydro] at (1.8,0) {set $\mathcal{F}_\mathrm{orm}$ of grid-form. DERs};
\node [scale=1.2] at (1.8,-4.6) {$\approx$};

\draw  [fill=hydro!40](1,-4.9) rectangle (2.6,-5.5);
\node at (1.8,-5.2) {$T_\mathrm{des}^\mathrm{qv}(s)^{-1}$};
\draw[-latex] (-0.8,-5.2) -- (1,-5.2); 
\draw[-latex] (2.6,-5.2) -- (4.5,-5.2);
\node[hydro] at (4.1,-5) {$\Delta q_\mathrm{pcc}$};
\node[hydro] at (-0.4,-5) {$\Delta v_\mathrm{pcc}$};
\node[hydro] at (1.8,-5.8) {desired aggregate behavior};
\node[scale=0.8] at (3.15,-1.7) {DVPP};
\node[scale=0.9] at (3.15,-1.9) {net-};
\node[scale=0.9] at (3.15,-2.1) {work};
\end{tikzpicture}

\\
\begin{subfigure}{0.29\textwidth}\captionsetup{font=scriptsize}
    \caption{{Frequency and voltage control setup for grid-following DVPP control.\\}}
     \label{fig:DVPP_ctrl_setup_gfoll}
\end{subfigure} 
& 
\begin{subfigure}{0.33\textwidth}\captionsetup{font=scriptsize}
    \caption{{Frequency control setup for grid-forming DVPP control: Feedback interconnection of the local DVPP dynamics and the power flow between the grid-forming units.}}
     \label{fig:DVPP_freq_ctrl_setup}
\end{subfigure}
&
\begin{subfigure}{0.27\textwidth}\captionsetup{font=scriptsize}
    \caption{{Voltage control setup for grid-forming DVPP control.}\\}
     \label{fig:DVPP_volt_ctrl_setup}
\end{subfigure}

\\ \hline 
\multicolumn{3}{c}{}\\[-0.7cm]
\multicolumn{3}{c}{Local closed-loop behavior of each DVPP unit $i$}\\[-0.5mm] \hline
\makecell{
\\[-2mm]$\Delta p_i(s) = -T_i^\mathrm{fp}(s)\Delta f_\mathrm{pcc}(s),\,\forall i \in \mathcal{F}_\mathrm{oll}$\\[0.5mm]
$\Delta q_i(s) = -T_i^\mathrm{vq}(s)\Delta v_\mathrm{pcc}(s),\,\forall i \in \mathcal{F}_\mathrm{oll}$\\[1mm]
}
&
$\Delta f_i(s) = -T_i^\mathrm{pf}(s)\Delta p_i(s),\,\forall i \in \mathcal{F}_\mathrm{orm}$ \color{hydro}$\,\,\rightarrow$ \cref{eq:local_closed_loop_tf_pf}\color{black}
&
$\Delta q_i(s) = -T_i^\mathrm{vq}(s)\Delta v_\mathrm{pcc}(s),\,\forall i \hspace{-0.5mm}\in \hspace{-0.5mm}\mathcal{F}_\mathrm{orm}$ \color{hydro}$\,\,\rightarrow$ \cref{eq:local_closed_loop_tf_vq}\color{black}
\\ \hline
\multicolumn{3}{c}{}\\[-0.7cm]
\multicolumn{3}{c}{Aggregated DVPP dynamics}\\ [-0.5mm]\hline
\makecell{
\\[-2mm]$\Delta p_\mathrm{pcc}(s) \approx \sum_{i} T_i^\mathrm{fp}(s)\Delta f_\mathrm{pcc}(s) $\\[0.5mm]
$\Delta q_\mathrm{pcc}(s) \approx \sum_{i} T_i^\mathrm{vq}(s)\Delta v_\mathrm{pcc}(s)$\\[1mm]
}
&
$\Delta f_\mathrm{pcc}(s) \hspace{-0.5mm}\approx\hspace{-0.5mm} \left(\sum_{i} \hspace{-0.5mm}T_i^\mathrm{pf}(s)^{-1} \right)^{-1} \hspace{-0.5mm}\sum_{i}\hspace{-0.5mm}\Delta p_{\mathrm{d},i}(s)$\color{hydro}$\,\,\rightarrow$ \cref{eq:coherent_dynamics_PCC}\color{black}
&
$\Delta q_\mathrm{pcc}(s) \approx \sum_{i}\hspace{-0.5mm} T_i^\mathrm{vq}(s)\Delta v_\mathrm{pcc}(s) $\color{hydro}$\,\,\rightarrow$ \cref{eq:aggregate_reactive_power_control}\color{black}
\\ \hline
\multicolumn{3}{c}{}\\[-0.7cm]
\multicolumn{3}{c}{Aggregate DVPP specification}\\ [-0.5mm]\hline
\makecell{
\\[-2mm]
$\Delta p_\mathrm{pcc}(s)=T_\mathrm{des}^\mathrm{fp}(s)\Delta f_\mathrm{pcc}(s)$\\[0.5mm]
$\Delta q_\mathrm{pcc}(s)=T_\mathrm{des}^\mathrm{vq}(s)\Delta v_\mathrm{pcc}(s)$\\[1mm]
}
&
$\Delta f_\mathrm{pcc}(s)=T_\mathrm{des}^\mathrm{pf}(s)\Delta p_\mathrm{pcc}(s)$\color{hydro}$\,\,\rightarrow$ \cref{eq:aggregate_specification}\color{black}
&
$\Delta v_\mathrm{pcc}(s)=T_\mathrm{des}^\mathrm{qv}(s)\Delta q_\mathrm{pcc}(s)$\color{hydro}$\,\,\rightarrow$ \cref{eq:aggregate_specification}\color{black}
\\ \hline
\multicolumn{3}{c}{}\\[-0.7cm]
\multicolumn{3}{c}{Aggregation condition}\\[-0.5mm] \hline
\makecell{
\\[-2mm]
$\sum_{i} T_i^\mathrm{fp}(s)\stackrel{!}{=}T_\mathrm{des}^\mathrm{fp}(s)$\\[0.5mm]
$\sum_{i} T_i^\mathrm{vq}(s)\stackrel{!}{=}T_\mathrm{des}^\mathrm{vq}(s)$\\[1mm]
}
&
$ \left(\sum_{i} T_i^\mathrm{pf}(s)^{-1} \right)^{-1} \stackrel{!}{=} T_\mathrm{des}^\mathrm{pf}(s)$\color{hydro}$\,\,\rightarrow$ \cref{eq:aggregation_condition_frequency}\color{black}
&
$ \sum_{i} T_i^\mathrm{vq}(s) \stackrel{!}{=} T_\mathrm{des}^\mathrm{qv}(s)^{-1}$\color{hydro}$\,\,\rightarrow$ \cref{eq:aggregation_condition_voltage}\color{black}
\\ \hline
\multicolumn{3}{c}{}\\[-0.7cm]
\multicolumn{3}{c}{Disaggregation via ADPFs}\\[-0.5mm] \hline
\makecell{
\\[-2mm]
$T_\mathrm{des}^\mathrm{fp}(s)=\sum_{i} m_i^\mathrm{fp}(s)T_\mathrm{des}^\mathrm{fp}(s)$\\[0.5mm]
$T_\mathrm{des}^\mathrm{vq}(s)=\sum_{i} m_i^\mathrm{vq}(s)T_\mathrm{des}^\mathrm{vq}(s)$\\[1mm]
}
&
$T_\mathrm{des}^\mathrm{pf}(s)^{-1}= \sum_{i}m_i^\mathrm{fp}(s)T_\mathrm{des}^\mathrm{pf}(s)^{-1}$\color{hydro}$\,\,\rightarrow$ \cref{eq:disaggregation}\color{black}
&
$T_\mathrm{des}^\mathrm{qv}(s)^{-1}= \sum_{i}m_i^\mathrm{vq}(s)T_\mathrm{des}^\mathrm{qv}(s)^{-1}$\color{hydro}$\,\,\rightarrow$ \cref{eq:disaggregation}\color{black}
\\ \hline
\multicolumn{3}{c}{}\\[-0.7cm]
\multicolumn{3}{c}{Local matching condition}\\ [-0.5mm]\hline
\makecell{
\\[-2mm]
$ T_i^\mathrm{fp}(s)\stackrel{!}{=}m_i^\mathrm{fp}(s)T_\mathrm{des}^\mathrm{fp}(s),\,\forall i \in \mathcal{F}_\mathrm{oll}$\\[0.5mm]
$ T_i^\mathrm{vq}(s)\stackrel{!}{=}m_i^\mathrm{vq}(s)T_\mathrm{des}^\mathrm{vq}(s),\,\forall i \in \mathcal{F}_\mathrm{oll}$\\[1mm]
}
&
$ T_i^\mathrm{pf}(s)\stackrel{!}{=}m_i^\mathrm{fp}(s)^{-1}T_\mathrm{des}^\mathrm{pf}(s),\,\forall i \in \mathcal{F}_\mathrm{orm}$\color{hydro}$\,\,\rightarrow$ \cref{eq:matching_conditions}\color{black}
&
$ T_i^\mathrm{vq}(s)\stackrel{!}{=}m_i^\mathrm{vq}(s)T_\mathrm{des}^\mathrm{qv}(s)^{-1}\hspace{-0.5mm},\,\forall i \hspace{-0.5mm}\in \hspace{-0.5mm}\mathcal{F}_\mathrm{orm}$\color{hydro}$\,\,\rightarrow$ \cref{eq:matching_conditions}\color{black}
\\
         \bottomrule
    \end{tabular}
     \captionsetup{justification=centering}
            \caption{Comparison of grid-following and grid-forming DVPP controls.}
    \label{fig:comparison_DVPP}
    \vspace{-3mm}
\end{figure*}
\renewcommand{\arraystretch}{1} \normalsize

\subsubsection*{Frequency Control} 
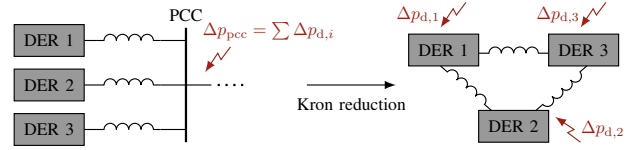
\begin{figure}[t!]
    \centering
    \usetikzlibrary{positioning}
\usetikzlibrary{arrows}
\usetikzlibrary{circuits.ee.IEC}

\resizebox {0.955\columnwidth} {!} {
\begin{tikzpicture}[circuit ee IEC,scale=0.8, every node/.style={scale=0.63}]
\draw (-7.5,2.7) -- (-7.1,2.7) node (v1) {};

\draw [thick](-7.5,3.6) -- (-7.5,1.8); 
\draw (-7.5,2.7)  -- (-7.9,2.7) node (v2) {};

\draw (-8.9,3.4) to [inductor={}](-7.9,3.4)  node (v4) {}; 
\draw (-7.5,2) -- (-7.9,2) node (v3) {}; 
\draw (-7.5,3.4) -- (-7.9,3.4);
\draw (-8.9,2.7)to [inductor={}]  (-7.9,2.7) node (v5) {};

\draw  (-8.9,2) to [inductor={}] (-7.9,2) node (v6) {};

\draw (-8.9,3.4) -- (-9.1,3.4); 
\draw (-8.9,2.7)  -- (-9.1,2.7); 
\draw (-8.9,2)  -- (-9.1,2); 
\draw [fill=black!40](-4,3.5) rectangle (-2.9,3); 
\draw  [fill=black!40](-2.9,2.3) rectangle (-1.8,1.8); 
\draw [fill=black!40] (-1.8,3.5) rectangle (-0.7,3);
\node at (-3.45,3.25) {DER 1};
\node at (-2.35,2.05) {DER 2};
\node at (-1.25,3.25) {DER 3};

\draw [fill=black!40] (-10.2,3.65) rectangle (-9.1,3.15); 
\draw  [fill=black!40](-10.2,2.95) rectangle (-9.1,2.45); 
\draw [fill=black!40] (-10.2,2.25) rectangle (-9.1,1.75);
\node at (-9.65,3.4) {DER 1};
\node at (-9.65,2.7) {DER 2};
\node at (-9.65,2) {DER 3};

\node at (-7.5,3.8) {PCC};
\draw [dotted, thick] (v1) -- (-6.6,2.7);
\draw[-latex, rotate=180, color = backgroundred] (6.8,-3.3) -- (6.95,-3.1) -- (7,-3.2) -- (7.2,-2.9);
\draw[-latex, rotate=90, color = backgroundred] (1.8,1.3) -- (1.95,1.5) -- (2,1.4) -- (2.2,1.7);
\draw[-latex, rotate=180, color = backgroundred] (3.1,-4) -- (3.25,-3.8) -- (3.3,-3.9) -- (3.5,-3.6);
\draw[-latex, rotate=180, color = backgroundred] (0.9,-4) -- (1.05,-3.8) -- (1.1,-3.9) -- (1.3,-3.6);
\node [color=backgroundred] at (-6.2,3.55) {$\Delta p_\mathrm{pcc}=\textstyle\sum\Delta p_{\mathrm{d},i}$};
\draw [-latex](-5.6,2.7) -- (-4.2,2.7);
\node at (-4.9,2.4) {Kron reduction};
\draw (-3.5,3) to [inductor={}] (-2.7,2.3);
\draw(-1.2,3) to [inductor={}] (-2,2.3);
\draw (-2.9,3.2) to [inductor={}] (-1.8,3.2);
\node [color=backgroundred] at (-3.85,3.8) {$\Delta p_{\mathrm{d},1}$}; 
\node  [color=backgroundred] at (-1.65,3.8) {$\Delta p_{\mathrm{d},3}$};
\node  [color=backgroundred] at (-0.95,2) {$\Delta p_{\mathrm{d},2}$};
\end{tikzpicture}

}
    \vspace{-8.5mm}
    \caption{Exemplary Kron-reduction of a parallel DER interconnection.}
    \label{fig:DVPP_kron_reduction}
\end{figure}

The DVPP control setup for frequency regulation is modelled as a feedback interconnection of the local DVPP dynamics and the power flow of the inductive lines (transformers or cables) between the DVPP units (Fig.~\ref{fig:DVPP_freq_ctrl_setup} and \cref{tab:gform_freq_symbols}). Namely, we consider the Kron-reduced setup \cite{dorfler2012kron} of the parallel DER connection in Fig.~\ref{fig:gform_DVPP_PCC} to eliminate the algebraic constraints of the PCC (Fig.~\ref{fig:DVPP_kron_reduction}). In this regard, the input signal $\Delta p_\mathrm{d}=[\Delta p_{\mathrm{d},1}\,...\,\Delta p_{\mathrm{d},n}]^\top$ in Fig.~\ref{fig:DVPP_freq_ctrl_setup} represents the vector of local active power injection disturbances $\Delta p_{\mathrm{d},i}$ at each DVPP unit $i$. The output signal $\Delta f = [\Delta f_1 \,...\, \Delta f_n ]^\top$ represents the vector of local frequency deviations $\Delta f_i$ of the DVPP units from its nominal value. The dynamics of the DVPP units that map the vector $\Delta p = [\Delta p_1\,...\,\Delta p_n]^\top$ of local active power deviation outputs $\Delta p_i$ to the vector $\Delta f$ of local frequency deviations $\Delta f_i$ are described by the local closed-loop transfer functions $T_i^\mathrm{pf}(s)$, i.e.,
\begin{align}\label{eq:local_closed_loop_tf_pf}
    \Delta f_i(s) = - T_i^\mathrm{pf}(s) \Delta p_i(s),\quad \forall i \in \mathcal{F}_\mathrm{orm}.
\end{align}
Notice that the local closed-loop transfer functions $T_i^\mathrm{pf}(s)$ capture all dynamics underlying the decoupled ``active power loop'' of a grid-forming DER power electronics architecture which maps from an active power measurement to the imposed frequency, i.e., the power converter dynamics, the filter dynamics, the grid-side converter control loops, the dc-side dynamics, and the dynamics of the primary source technology, e.g., a wind turbine or a PV system.

For constant voltage magnitudes (when neglecting fast filter dynamics) and inductive DVPP interconnection lines, the vector of active power fluctuations $\Delta p_\mathrm{e}=[\Delta p_{\mathrm{e},1}\,...\,\Delta p_{\mathrm{e},n}]^\top$ of the interconnection lines are approximated by the linearized power flow equations \cite{purchala2005usefulness}
\begin{align}\label{eq:laplacian}
    \Delta p_\mathrm{e}(s)=\tfrac{L_\mathrm{dvpp}}{s}\Delta f(s),
\end{align}
where $L_\mathrm{dvpp}$ is an undirected weighted Laplacian (also termed admittance) matrix of the Kron-reduced DVPP interconnection network (Fig. \ref{fig:DVPP_kron_reduction}, right) with nonnegative and real-valued eigenvalues $0=\lambda_1\leq \lambda_2\leq ... \lambda_n$.

Generally, a group of grid-forming DERs as in Fig. \ref{fig:DVPP_freq_ctrl_setup} is considered \textit{approximately coherent}, if all DERs have similar frequency responses $\Delta f_i$ under active power injection disturbances $\Delta p_{\mathrm{d},i}$ of any shape. We are interested in characterizing the dynamic response of the approximately coherent DVPP units, which we term \textit{coherent DVPP dynamics}. For the sake of providing a historic example, if $T_i^\mathrm{pf}(s)$ were to model generator swing dynamics, then we are in the classical slow coherency problem setup extensively studied in power system literature\cite{chow2013power}. In respect thereof, a pragmatical approach to derive the coherent DVPP dynamics is provided in the following. We hereby refer to the results in\cite{paganini2019global,min2019dynamics,min2020accurate,jiang2021grid,min2021coherence}, which rigorously formalize the classical slow coherency in the power system stability literature.

In particular, from Fig.~\ref{fig:DVPP_freq_ctrl_setup}, the overall frequency dynamics of the system can be stated as
\begin{align}\label{eq:freq_dynamics_gform1}
\begin{split}
    &\Delta f (s)= \mathrm{diag}\{T_i^\mathrm{pf}(s)\} \hspace{-0.5mm}\left(\hspace{-0.5mm} \Delta p_\mathrm{d}(s)-\tfrac{L_\mathrm{dvpp}}{s} \Delta f(s)\hspace{-0.5mm}\right)\hspace{-0.5mm}\\
   \Leftrightarrow\quad&\mathrm{diag}\{T_i^\mathrm{pf}(s)^{-1}\} \Delta f(s) = \hspace{-0.5mm}\left( \hspace{-0.5mm}\Delta p_\mathrm{d}(s)-\tfrac{L_\mathrm{dvpp}}{s}\Delta f(s)\hspace{-0.5mm}\right)\hspace{-0.5mm}.\hspace{-0.5mm}
\end{split}
\end{align}
Next, by left-multiplying both sides of \cref{eq:freq_dynamics_gform1} with the transposed vector of all ones $\mathbf{1}_n^\top$, we obtain
\begin{align}
        &\mathbf{1}_n^\top\mathrm{diag}\{T_i^\mathrm{pf}(s)^{-1}\} \Delta f(s) = \mathbf{1}_n^\top\hspace{-0.5mm}\left( \hspace{-0.5mm}\Delta p_\mathrm{d}(s)-\tfrac{L_\mathrm{dvpp}}{s}\Delta f(s)\hspace{-0.5mm}\right)\hspace{-0.5mm}\nonumber\\\label{eq:freq_dynamics_gform2}
        \Leftrightarrow\quad&\textstyle\sum_{i\in\mathcal{F}_\mathrm{orm}}T_i^\mathrm{pf}(s)^{-1}\Delta f_i (s) = \textstyle\sum_{i\in\mathcal{F}_\mathrm{orm}}\Delta p_{\mathrm{d},i}(s),
\end{align}
where we have used the fact that $\mathbf{1}_n$ corresponds to the eigenvector of $L_\mathrm{dvpp}$ that is associated with the dominant zero eigenvalue $\lambda_1=0$, and it therefore holds that  $\mathbf{1}_n^\top\tfrac{L_\mathrm{dvpp}}{s}\Delta f(s) = 0$. Referring back to classical slow coherency theory, the latter equation \eqref{eq:freq_dynamics_gform2} (i.e., projecting the full interconnected dynamics on the vector of all ones $\mathbf{1}_n$) results in the system center-of-inertia (COI) dynamics which describe the slow aggregate system dynamics. 

By assuming approximate coherency of the frequency outputs at the PCC, i.e., $\Delta f_i \approx \Delta f_\mathrm{pcc}$, we can derive the approx- imate coherent DVPP dynamics at the PCC from \eqref{eq:freq_dynamics_gform2} as
\begin{align}\label{eq:coherent_dynamics_PCC}
    \Delta f_\mathrm{pcc} (s) \approx \left( \textstyle\sum_{i\in\mathcal{F}_\mathrm{orm}}T_i^\mathrm{pf}(s)^{-1}\right)^{-1}\textstyle\sum_{i\in\mathcal{F}_\mathrm{orm}} \Delta p_{\mathrm{d},i}(s), 
\end{align}
where $\textstyle\sum_{i\in \mathcal{F}_\mathrm{orm}}\Delta p_{\mathrm{d},i}\approx\Delta p_\mathrm{pcc}$, given the inductive DVPP interconnection lines. Namely, for a stable interconnection network as in Fig. \ref{fig:DVPP_freq_ctrl_setup}, the approximate coherency of the frequency outputs $\Delta f_i \approx \Delta f_\mathrm{pcc}$, i.e., the frequency synchronization, corresponds to the natural long-term and aggregate dynamics\cite{dorfler2013synchronization}. Stability of the interconnection network is typically guaranteed by means of passivity (or more general dissipativity or integral quadratic constraint) arguments\cite{khalil2015nonlinear}, which, for our envisioned design of the local closed-loop transfer functions $T_i^\mathrm{pf}(s)$ can usually (or at least approximately) be satisfied.

A rigorous derivation of \eqref{eq:coherent_dynamics_PCC} in the limit of large connectivity of the interconnection network and under specific stability assumptions is given in\cite{paganini2019global,min2019dynamics,min2020accurate,jiang2021grid,min2021coherence}. It is shown how the connectivity of the network trades off with the width of the frequency range in which coherency is achieved, which is typically sufficient for power system networks. In particular, in classical slow coherency, the approximation quality depends on the spectral gap in the Laplacian eigenvalues $\lambda_2-\lambda_1 = \lambda_2 \geq 0$.

Finally, by matching \cref{eq:aggregate_specification} and \cref{eq:coherent_dynamics_PCC}, the DVPP \textit{aggregation condition for frequency control} is obtained as\footnote{The symbol $``\stackrel{!}{=}"$ indicates that the terms on the left side of the equality must be designed in such a way that the equality is satisfied.}
\begin{align}\label{eq:aggregation_condition_frequency}
        \left(\textstyle\sum_{i\in \mathcal{F}_\mathrm{orm}} T_i^\mathrm{pf}(s)^{-1} \right)^{-1} \stackrel{!}{=} T_\mathrm{des}^\mathrm{pf}(s).
\end{align}

\renewcommand{\arraystretch}{1.2}
\begin{table}[t!]\scriptsize
    \centering
     \caption{List of notation for the grid-forming DVPP voltage control setup.}
    \vspace{-1mm}
    \begin{tabular}{c||c}
     \toprule
         Reactive power injection change at the PCC & $\Delta q_\mathrm{pcc}$\\
         Voltage magnitude deviation at the PCC & $\Delta v_\mathrm{pcc}$\\
         Aggregated reactive power deviation output of the DVPP & $\Delta q_\mathrm{agg}$\\
         \hline
         Local reactive power output deviation of unit $i$ & $\Delta q_i$\\
         \hline 
         Local $\mathrm{v}$-$\mathrm{q}$ closed-loop transfer function of unit $i$ & $T_i^\mathrm{vq}(s)$\\
         Desired DVPP transfer function for $\mathrm{q}$-$\mathrm{v}$ control & $T_\mathrm{des}^\mathrm{qv}(s)$\\
    \bottomrule
    \end{tabular}
    	  \vspace{-3mm}
    \label{tab:gform_volt_symbols}
\end{table}
\renewcommand{\arraystretch}{1} \normalsize

The aggregation condition in \eqref{eq:aggregation_condition_frequency} has been derived under the approximate coherency assumption \cref{eq:coherent_dynamics_PCC} and thus has to be satisfied only during synchronized conditions of the DERs' frequency responses, i.e., only in the frequency band during which coherency of the grid-foriming DERs is achieved. Hence, the instantaneous, non-synchronized transient power injection of the grid-forming units, which originates from their voltage source behavior\cite{osmoseh2020} and network electromagnetic transients, is not part of the aggregated DVPP control.
\begin{remark}
    The aggregated grid-forming DVPP dynamics for frequency control in \eqref{eq:coherent_dynamics_PCC} are given by the synchronized fre- quency outputs of the grid-forming DVPP units in response to the active power injection disturbance at the PCC. In contrast, for the grid-following DVPP developed in our work in\cite{haberle2021control}, the aggregated grid-following DVPP dynamics are given by the sum of all local active power injections in response to a common frequency deviation measurement at the PCC, i.e.,
    \begin{align}
        \Delta p_\mathrm{pcc}\approx-\textstyle\sum_{i\in\mathcal{F}_\mathrm{oll}}\Delta p_i = \textstyle\sum_{i\in\mathcal{F}_\mathrm{oll}} T_i^\mathrm{fp}(s) \Delta f_\mathrm{pcc}.
    \end{align}
    Hence, also the resulting aggregation condition for the grid-following DVPP differs to the grid-forming one in \eqref{eq:aggregation_condition_frequency}, i.e.,
    \begin{align}
\textstyle\sum_{i\in\mathcal{F}_\mathrm{oll}}T_i^\mathrm{fp}(s)\stackrel{!}{=}T_\mathrm{des}^\mathrm{fp}(s).
    \end{align}
    An illustrative comparison of both DVPP concepts can be found in \cref{fig:comparison_DVPP}.
\end{remark}

\subsubsection*{Voltage Control}
In contrast to the local frequencies of the grid-forming DERs, the local voltage magnitudes generally do not yield a coherent dynamic behavior. We thus consider a separate voltage control setup in \cref{fig:DVPP_volt_ctrl_setup} and \cref{tab:gform_volt_symbols}, where all units receive a common input measurement of the voltage magnitude deviation $\Delta v_\mathrm{pcc}$ at the PCC. The reactive power deviation output $\Delta q_i$ of each unit $i$ sums up to the aggregate reactive power deviation output $\Delta q_\mathrm{agg}$ of the DVPP, i.e.,
    \begin{align}\label{eq:reactive_power_sum}
        \Delta q_\mathrm{agg} = \textstyle\sum_{i\in \mathcal{F}_\mathrm{orm}} \Delta q_i.
    \end{align}
The dynamics of the DVPP units that map the measured voltage magnitude deviation $\Delta v_\mathrm{pcc}$ at the PCC to the local reactive power deviation output $\Delta q_i$ are described by the local closed-loop transfer functions $T_i^\mathrm{vq}(s)$, i.e.,
\begin{align}\label{eq:local_closed_loop_tf_vq}
    \Delta q_i(s) = -T_i^\mathrm{vq}(s)\Delta v_\mathrm{pcc}(s),\quad\forall i \in \mathcal{F}_\mathrm{orm}.
\end{align}
Considering the local closed-loop transfer functions $T_i^\mathrm{vq}(s)$, {which are realized via grid-forming converter controls and capture all dynamics underlying the decoupled ``reactive power loop'' of the DER power converter control architecture (see \noindent\cref{sec:converter_model}),} the aggregate DVPP behavior is given by
    \begin{align}\label{eq:aggregate_reactive_power_control}
        \Delta q_\mathrm{agg}(s) = -\textstyle\sum_{i\in \mathcal{F}_\mathrm{orm}} T_i^\mathrm{vq}(s) \Delta v_\mathrm{pcc}(s).
    \end{align}
By approximating\footnote{The accuracy of this approximation depends on the power measurement strategy of the DER units (see \cref{fig:converter_model_gform}).} $\Delta q_\mathrm{pcc}\approx -\Delta q_\mathrm{agg}$ (or compensating for reactive losses across the DVPP interconnection lines), we can derive the DVPP \textit{aggregation condition for voltage control} as
    \begin{align}\label{eq:aggregation_condition_voltage}
        \textstyle\sum_{i\in \mathcal{F}_\mathrm{orm}} T_i^\mathrm{vq}(s)\stackrel{!}{=}T_\mathrm{des}^\mathrm{qv}(s)^{-1}.
    \end{align}
Note that the condition in \eqref{eq:aggregation_condition_voltage} requires $T_\mathrm{des}^\mathrm{qv}(s)$ to be invertible. Alternatively (e.g., because of unstable zeros, etc.), one might resort to a causal and stable approximation of ${T_\mathrm{des}^\mathrm{qv}}(s)^{-1}$.

\begin{remark}
The aggregated grid-forming DVPP dynamics for voltage control in \eqref{eq:aggregate_reactive_power_control} are given by the sum of all local reactive power outputs of the DVPP units in response to a common voltage magnitude deviation measurement at the PCC. This is indeed equivalent with the signal causality of the aggregated dynamics for voltage control of the grid-following DVPP in our work in\cite{haberle2021control} (see \cref{fig:comparison_DVPP} for an illustrative comparison). However, while in case of the grid-forming DVPP, the aggregated local reactive power outputs of the \emph{grid-forming DERs} are controlled to inversely match the desired grid-forming DVPP specification $T_\mathrm{des}^\mathrm{qv}(s)$ as in \eqref{eq:aggregation_condition_voltage}, the \emph{grid-following DERs} in the grid-following DVPP are controlled to satisfy the aggregation condition
\begin{align}\label{eq:aggregation_condition_voltage_gfoll}
    \textstyle\sum_{i\in\mathcal{F}_\mathrm{oll}}T_i^\mathrm{vq}(s)\stackrel{!}{=}T_\mathrm{des}^\mathrm{vq}(s).
\end{align}
Obviously, if one specifies that $T_\mathrm{des}^\mathrm{vq}(s) \equiv T_\mathrm{des}^\mathrm{qv}(s)^{-1}$, the aggregation conditions in \eqref{eq:aggregation_condition_voltage} and \eqref{eq:aggregation_condition_voltage_gfoll} become equivalent.
\end{remark}

Finally, given $T_\mathrm{des}(s)$ in \cref{eq:aggregate_specification}, the overall grid-forming DVPP control design problem is to find local device-level controllers, such that the two aggregation conditions in \eqref{eq:aggregation_condition_frequency} and \eqref{eq:aggregation_condition_voltage} are satisfied. In doing so, it is important to ensure that practical limitations of the grid-forming DERs are not exceeded during normal operation, and device-level stability can be guaranteed.

\subsection{Adaptive Divide-and-Conquer Strategy}\label{sec:adaptive_divide_and_conquer}
{Following our grid-following DVPP proposal in\cite{haberle2021control}, the approach to solve the previous \textit{grid-forming} DVPP control design problem is based on an adaptive divide-and-conquer strategy, which is composed of two steps:}
\begin{enumerate}
    \item {Disaggregate the desired behavior $T_\mathrm{des}(s)$ among the DVPP units using adaptive dynamic participation factors (ADPFs) to obtain local desired behaviors.}
    \item {Design a local feedback control for each DVPP unit to optimally match the local desired behavior.}
\end{enumerate}
\begin{remark}
   The adaptive divide-and-conquer strategy presented in the following is conceptually equivalent to our work in\cite{haberle2021control}. However, given the inverted signal causality of a grid-forming DVPP and the associated different aggregation strategy presented in \cref{sec:grid_forming_DVPP_control}, the disaggregation of the 
    grid-forming DVPP specification $T_\mathrm{des}(s)$, as well as the local matching conditions and the associated matching control design for each grid-forming DVPP unit are not immediate, and thus have to be considered separetely from the grid-following DVPP setup. An illustrative comparison of the adaptive divide-and-conquer strategy for both DVPP concepts can be found in \cref{fig:comparison_DVPP}, and a list of notation is given in \cref{tab:gform_adaptive_divide_and_conquer_symbols}.
\end{remark}
\renewcommand{\arraystretch}{1.2}
\begin{table}[t!]\scriptsize
    \centering
     \caption{List of notation for the adaptive divide-and-conquer strategy.}
    \vspace{-1mm}
    \begin{tabular}{c||c}
     \toprule
         Description & Symbol  \\ \hline
         Control channel index & $k\in\{ \mathrm{fp,vq}\}$\\\hline
         ADPF of unit $i$ for channel $k$ & $m_i^k(s)$\\ 
         Time constant of unit $i$ for channel $k$ & $\tau_i^k$\\
         Time-varying dc gain of unit $i$ for channel $k$ & $\mu_i^k(t)$\\
    \bottomrule
    \end{tabular}
    	\vspace{-3.5mm}
    \label{tab:gform_adaptive_divide_and_conquer_symbols}
\end{table}
\renewcommand{\arraystretch}{1} \normalsize
\subsubsection*{Disaggregation of $T_\mathrm{des}(s)$}
Given the aggregation conditions for frequency and voltage control in \eqref{eq:aggregation_condition_frequency} and \eqref{eq:aggregation_condition_voltage}, respectively, we disaggregate the desired diagonal transfer function matrix $T_\mathrm{des}(s)$ to the individual DVPP units $i$ as
\begin{align}\label{eq:disaggregation}
\begin{split}
 \hspace{-1.8mm}T_\mathrm{des}^\mathrm{pf}(s)^{-1}\hspace{-0.5mm}&= \hspace{-0.5mm}\textstyle\sum_{i\in \mathcal{F}_\mathrm{orm}}\hspace{-0.8mm} m_i^\mathrm{fp}\hspace{-0.2mm}(s)T_\mathrm{des}^\mathrm{pf}(s)^{-1}\hspace{-0.5mm}\stackrel{!}{=} \hspace{-0.5mm}\textstyle\sum_{i\in \mathcal{F}_\mathrm{orm}}\hspace{-0.8mm} T_i^\mathrm{pf}\hspace{-0.2mm}(s)^{-1}\hspace{-0.5mm},\hspace{-1mm}\\
 \hspace{-1.8mm}T_\mathrm{des}^\mathrm{qv}(s)^{-1}\hspace{-0.5mm}&= \hspace{-0.5mm}\textstyle\sum_{i\in \mathcal{F}_\mathrm{orm}}\hspace{-0.8mm} m_i^\mathrm{vq}\hspace{-0.2mm}(s)T_\mathrm{des}^\mathrm{qv}(s)^{-1}\hspace{-0.5mm}\stackrel{!}{=} \hspace{-0.5mm}\textstyle\sum_{i\in \mathcal{F}_\mathrm{orm}}\hspace{-0.8mm} T_i^\mathrm{vq}\hspace{-0.2mm}(s),\hspace{-1mm}
    \end{split}
\end{align}
where the transfer functions $m_i^\mathrm{fp}(s)$ and $m_i^\mathrm{vq}(s)$ are \textit{adaptive dynamic participation factors (ADPFs)}, required to satisfy the \textit{participation conditions} \begin{align}\label{eq:participation_conditions}
    \textstyle\sum_{i\in\mathcal{F}_\mathrm{orm}} m_i^\mathrm{fp}(s)\stackrel{!}{=} 1 \quad \text{and} \quad \textstyle\sum_{i\in\mathcal{F}_\mathrm{orm}} m_i^\mathrm{vq}(s)\stackrel{!}{=}1.
\end{align}
\subsubsection*{ADPF Selection} The ADPFs of the DVPP units are selected such that the participation conditions in \eqref{eq:participation_conditions} are satisfied, while simultaneously respecting the heterogeneous time scales of the local DER dynamics along with their steady-state power capacity limitations during normal operating conditions (i.e., during small-signal changes). As in \cite{haberle2021control}, we therefore specify the ADPFs by two parameters: a \textit{time constant} $\tau_i^\mathrm{fp}$ (or $\tau_i^\mathrm{vq}$) for the roll-off frequency to account for different time scales of DER active (or reactive) power injection dynamics (akin to classical ramping rates), and a \textit{dc gain} {$m_i^\mathrm{fp}(s=0):=\mu_i^\mathrm{fp}$ (or $m_i^\mathrm{vq}(s=0):=\mu_i^\mathrm{vq}$)} to account for the available active (or reactive) power capacity limits of the DERs during steady state, similar to droop gains or static distribution factors\cite{su2020adaptive} in traditional power systems. In particular, given the latter, both active and reactive power sharing among the grid-forming DERs within the DVPP can be ensured. Based on the previous two parameters, we divide the ADPFs of the DVPP units $i$ into three categories, i.e., we envision (see case studies in \cref{sec:testcase} for examples)
\begin{itemize}
	\item a \textit{low-pass filter (LPF)} participation behavior for units that can provide regulation on slow time scales including steady-state contributions\footnote{{The symbol ``$\sim$'' indicates that the crucial characteristic of $m_i^k(s)$ is given by the expression on the right.}}, i.e., 
 \begin{align*}
    m_i^k(s) \sim \tfrac{\mu_i^k}{(\tau_i^ks+1)^d},\quad k\in\{\mathrm{fp},\mathrm{vq}\},\, d \in \mathbb{N},
 \end{align*}
	\item a \textit{high-pass filter (HPF)} participation behavior for units able to provide regulation on fast time scales, i.e., 
  \begin{align*}
     m_i^k(s) \sim \tfrac{\tau_i^ks}{(\tau_i^ks+1)^d},\quad k\in\{\mathrm{fp},\mathrm{vq}\},\, d \in \mathbb{N},
 \end{align*}
	\item a \textit{band-pass filter (BPF)} participation behavior for the re- maining units able to cover the intermediate regime, i.e.,
  \begin{align*}
     m_i^k(s) \sim \tfrac{(\tau_{i,1}^k-\tau_{i,2}^k)s}{(\tau_{i,1}^ks+1)^{d_1}(\tau_{i,2}^ks+1)^{d_2}},\, k\in\{\mathrm{fp},\mathrm{vq}\},\, d_1,d_2 \in \mathbb{N}.
 \end{align*}
\end{itemize}
 Note that the ADPFs with a BPF or HPF behavior will always have a zero dc gain by definition, {i.e., $m_i^\mathrm{fp}(s=0)=0$ (or $m_i^\mathrm{vq}(s=0)=0$)}. In contrast, for all units participating as a LPF, the dc gains $\mu_i^\mathrm{fp}$ (or $\mu_i^\mathrm{vq}$) have to satisfy
\begin{align}\label{eq:DC_gain_condition}
    \textstyle\sum_{i\in\mathcal{F}_\mathrm{orm}} m_i^k(s=0)\stackrel{!}{=}1, \quad k\in\{\mathrm{fp},\mathrm{vq}\}
\end{align}
to meet the participation conditions in \eqref{eq:participation_conditions}.

\subsubsection*{Online Adaptation of LPF DC Gains}
As in \cite{haberle2021control}, we specify the LPF dc gains $\mu_i^\mathrm{fp}$ (or $\mu_i^\mathrm{vq}$) such that they can be adapted online, in proportion to the power capacity limits of the DERs. The latter might be time-varying based on the resource availability (e.g., weather conditions). During power system operation, the dc gains are updated in a centralized fashion, where the DVPP operator continuously collects the capacity limits of the DERs and communicates back the appropriate dc gains. If a distributed implementation is more desired, one could alternatively use a consensus algorithm via peer-to-peer communication (for details see\cite{haberle2021control}). 

\subsubsection*{Local Matching Control} 
As a last step, we need to find local feedback controllers for the DVPP units to ensure their local closed-loop transfer functions $T_i^\mathrm{pf}(s)$ and $T_i^\mathrm{vq}(s)$ match their local desired behaviors, i.e., we impose the \textit{local matching conditions}
\begin{align}\label{eq:matching_conditions}
\begin{split}
    T_i^\mathrm{pf}(s)&\stackrel{!}{=}m_i^\mathrm{fp}(s)^{-1}T_\mathrm{des}^\mathrm{pf}(s),\\
    T_i^\mathrm{vq}(s)&\stackrel{!}{=}m_i^\mathrm{vq}(s)T_\mathrm{des}^\mathrm{qv}(s)^{-1}, \,\forall i\in\mathcal{F}_\mathrm{orm}.
\end{split}
\end{align}

\begin{figure}[t!]
    \centering
    \resizebox {1\columnwidth} {!} {

\begin{tikzpicture}[scale=1,every node/.style={scale=0.7}]
\tikzstyle{roundnode}=[circle,draw=black!60,fill=black!5,scale=0.75]
\draw [top color=hydro!30, bottom color=white, color=white] (-1.2,-1.8) rectangle (6.9,1.6);
\draw  [top color=white, bottom color = black!25,color=white] (-1.2,-1.8) rectangle (6.9,-5.6);
\draw  [rounded corners,fill=hydro!15](-1.1,1.5) rectangle (6.8,-0.05);
\draw [rounded corners,fill=black!15] (0.8,-3.5) rectangle (4.9,-4.7);
\draw [rounded corners,fill=black!15] (1,-5) rectangle (4.7,-5.5);
\draw [rounded corners,fill=black!15] (-0.35,-2) rectangle (6.05,-3.2);
\draw  [rounded corners,hydro,fill=hydro!15](-1,1.4) rectangle (2.1,0.05);
\draw [rounded corners,fill=hydro!15] (1,-0.35) rectangle (4.7,-1.6);

\node [hydro] at (0.55,0.9) {$\Delta f_\mathrm{pcc}(s) = T_\mathrm{des}^\mathrm{pf}(s)\Delta p_\mathrm{pcc}(s)$};
\node [hydro] at (0.55,0.5) {$\Delta v_\mathrm{pcc}(s)=T_\mathrm{des}^\mathrm{qv}(s)\Delta q_\mathrm{pcc}(s)
$};

\draw[black!50,fill=black!50] (4.4,-3.8) -- (4.6,-3.6) -- (4.5,-3.6) -- (4.5,-3.3) -- (4.3,-3.3) -- (4.3,-3.6) -- (4.2,-3.6) -- (4.4,-3.8) ;
\draw[black!50,fill=black!50] (4.4,-0.65) -- (4.6,-0.45) -- (4.5,-0.45) -- (4.5,-0.15) -- (4.3,-0.15) -- (4.3,-0.45) -- (4.2,-0.45) -- (4.4,-0.65) ;
\draw[black!50,fill=black!50] (4.4,-5.3) -- (4.6,-5.1) -- (4.5,-5.1) -- (4.5,-4.8) -- (4.3,-4.8) -- (4.3,-5.1) -- (4.2,-5.1) -- (4.4,-5.3) ;
\draw[black!50,fill=black!50] (4.4,-2.3) -- (4.6,-2.1) -- (4.5,-2.1) -- (4.5,-1.7) -- (4.3,-1.7) -- (4.3,-2.1) -- (4.2,-2.1) -- (4.4,-2.3) ;
\node at (2.9,-2.5) {$
T_\mathrm{des}^\mathrm{pf}(s)^{-1}
=
\sum_{i\in \mathcal{F}_\mathrm{orm}}m_i^\mathrm{fp}(s)T_\mathrm{des}^\mathrm{pf}(s)^{-1}
\stackrel{!}{=} 
\sum_{i\in \mathcal{F}_\mathrm{orm}} T_i^\mathrm{pf}(s)^{-1}$};
\node at (2.8,-2.9) {$
T_\mathrm{des}^\mathrm{qv}(s)^{-1}
=
\sum_{i\in \mathcal{F}_\mathrm{orm}}m_i^\mathrm{vq}(s)T_\mathrm{des}^\mathrm{qv}(s)^{-1}
\stackrel{!}{=}
\sum_{i\in \mathcal{F}_\mathrm{orm}} T_i^\mathrm{vq}(s)
$};

%\node at (5.4,-2.8) {$\rightarrow (16)$};

\node at (2.3,-4) {$T_i^\mathrm{pf}(s)\stackrel{!}{=}m_i^\mathrm{fp}(s)^{-1}T_\mathrm{des}^\mathrm{pf}(s)$};
\node at (2.3,-4.4) {$T_i^\mathrm{vq}(s)\stackrel{!}{=}m_i^\mathrm{vq}(s)T_\mathrm{des}^\mathrm{qv}(s)^{-1}$};
\node at (4.3,-4.2) {$\forall i\in\mathcal{F}_\mathrm{orm}$};

\node [scale=0.8,hydro] at (0.55,1.2) {aggregate DVPP specification \eqref{eq:aggregate_specification}};
\node[scale=0.8] at (2.8,-5.25) {local matching control [3]};

\node [scale=0.8] at (2.8,-2.2) {disaggregation via ADPFs \eqref{eq:disaggregation}};
\node [scale=0.8] at (2.8,-3.7) {local matching conditions \eqref{eq:matching_conditions}};

%\node at (1.5,0.8) {$\rightarrow (1)$};
\draw  [rounded corners,fill=hydro!15](2.2,1.4) rectangle (6.7,0.05);
\node [scale=0.8] at (4.45,1.2) {aggregated DVPP dynamics \eqref{eq:coherent_dynamics_PCC}, \eqref{eq:aggregate_reactive_power_control}};

\node at (4.45,0.8) {$\Delta f_\mathrm{pcc}(s)\approx \left(\textstyle\sum_{i\in\mathcal{F}_\mathrm{orm}}\hspace{-1mm}T_i^\mathrm{pf}(s)^{-1} \right)^{-1}\hspace{-3mm}\Delta p_\mathrm{pcc}(s)$};
\node at (4.15,0.37) {$\Delta q_\mathrm{pcc}(s) \approx \textstyle\sum_{i\in\mathcal{F}_\mathrm{orm}}\hspace{-1mm}T_i^\mathrm{vq}(s)\Delta v_\mathrm{pcc}(s)$};
%\node at (6.3,0.95) {$\rightarrow (7)$};
%\node at (6.1,0.55) {$\rightarrow (13)$};

\node [scale=0.8] at (2.8,-0.55) {aggregation conditions \eqref{eq:aggregation_condition_frequency}, \eqref{eq:aggregation_condition_voltage}};
\node at (2.74,-0.9) {$ \left(\textstyle\sum_{i\in\mathcal{F}_\mathrm{orm}}\hspace{-1mm}T_i^\mathrm{pf}(s)^{-1} \right)^{-1}\hspace{-1mm}\stackrel{!}{=}T_\mathrm{des}^\mathrm{pf}(s)$};

\node at (3.2,-1.32) {$ \textstyle\sum_{i\in\mathcal{F}_\mathrm{orm}}\hspace{-1mm}T_i^\mathrm{vq}(s)\stackrel{!}{=}T_\mathrm{des}^\mathrm{qv}(s)^{-1}$};

%\node at (4.1,-0.75) {$\rightarrow (8)$};
%\node at (4.4,-1.2) {$\rightarrow (14)$};

\draw[dashed] (-1.2,-1.8) node (v1) {} -- (6.9,-1.8);

%\draw[-latex] (7.2,1.3) -- (7.2,-1.7);
%\node [rotate=90,scale=0.9] at (7,-0.2) {aggregation of grid-form. DERs};
%\draw[-latex] (7.2,-1.9) -- (7.2,-5.4);
%\node[rotate=90,scale=0.9] at (7,-3.7) {adaptive divide \& conquer strategy [3]};
\node [hydro,scale=0.8] at (6.1,-1.4) {aggregation of};
\node  [hydro,scale=0.8] at (6.1,-1.6) {grid-form.DERs};

\node [black!80,scale=0.8] at (6.1,-5) {adaptive};
\node  [black!80,scale=0.8] at (6.1,-5.2) {divide \& conquer};
\node [black!80,scale=0.8] at (6.1,-5.4) {strategy, cf. [3]};
\node [hydro, scale = 0.65] at (0.55,0.2) {(provided by DVPP operator)};
\end{tikzpicture}

}
       \vspace{-8mm}
    \caption{Schematic of the proposed grid-forming DVPP control design from grid-level to device-level. The system operator provides the aggregate DVPP specification to encode grid code requirements in the form of a desired transfer function, while the DVPP operator takes care of all subsequent design steps.}
    \vspace{-3mm}
    \label{fig:flow_chart}
\end{figure}

Inspired by existing methods on multivariable\cite{huang2020h,kammer2018convex,chen2021generalized} and adaptive\cite{erfanmanesh2015performance},\cite{muhando11} $\mathcal{H}_\infty$ control of power converters, we address the local matching control with robust and optimal linear parameter-varying (LPV) $\mathcal{H}_\infty$ methods, which {ensure device-level stability} and are well-suited to deal with the ADPFs, being parameter-varying due to the time-varying dc gains $\mu_i^\mathrm{fp}(t)$ and $\mu_i^\mathrm{vq}(t)$. Details on how to incorporate such LPV $\mathcal{H}_\infty$ matching control implementations within the converter control architecture of the DVPP units are described in \cref{sec:converter_model} and illustrated in \cref{fig:converter_model_gform}. Beyond that, a rigorous LPV $\mathcal{H}_\infty$ control design procedure can be found in\cite{haberle2021control}.

Finally, a schematic overview of the grid-forming DVPP control design is illustrated in \cref{fig:flow_chart}.
\begin{figure}[b!]
    \centering
        \vspace{-3mm}
    \resizebox {\columnwidth} {!} {
\footnotesize
\tikzstyle{roundnode}=[circle,draw=black!60,fill=black!5,scale=0.75]
\begin{tikzpicture}[scale=0.95,every node/.style={scale=0.9}]
\draw  [rounded corners=3,dashed, color = black!50, fill=black!5](-3,0) rectangle (4.7,-4.65);

\draw  [rounded corners=3,dashed, hydro,fill=hydro!15](-1.775,-0.65) rectangle (0.8,-2.3);
\node at (-1.5,-1.2) {$\ddots$};
\node at (0.5,-1.9) {$\ddots$};
\draw [scale=0.75,fill=hydro!40] (-1.6333,-1.45) rectangle (0.2667,-2.9);
\draw [fill=black!40] (-0.8,-1.175) rectangle (-0.2,-1.575);
\node [scale = 0.9] at (-0.5,-1.375) {plant};
\draw [rounded corners = 3,fill=black!15] (-0.9,-1.675) rectangle (-0.1,-2.075);
\node [scale = 0.9]at (-0.5,-1.875) {control};
\draw [-latex](-0.2,-1.375) -- (0.1,-1.375) -- (0.1,-1.875) -- (-0.1,-1.875); 
\draw[-latex] (-0.9,-1.875) -- (-1.1,-1.875) -- (-1.1,-1.375) -- (-0.8,-1.375); 
\node at (-0.5,-0.85) {$T_i^\mathrm{pf}(s)$};

\draw  [rounded corners=3,dashed, backgroundgreen,fill=backgroundgreen!15](1.325,-1.55) rectangle (3.9,-3.2);
\node at (1.6,-2.1) {$\ddots$};
\node at (3.6,-2.8) {$\ddots$};
\draw [scale=0.75,fill=backgroundgreen!40] (2.5001,-2.65) rectangle (4.4001,-4.1);
\draw [fill=black!40] (2.3,-2.075) rectangle (2.9,-2.475);
\node [scale = 0.9]at (2.6,-2.275) {plant};
\draw [rounded corners = 3,fill=black!15] (2.2,-2.575) rectangle (3,-2.975);
\node [scale = 0.9]at (2.6,-2.775) {control};
\draw [-latex](2.9,-2.275) -- (3.2,-2.275) -- (3.2,-2.775) -- (3,-2.775); 
\draw[-latex] (2.2,-2.775) -- (2,-2.775) -- (2,-2.275) -- (2.3,-2.275); 
\node at (2.6,-1.75) {$1/T_i^\mathrm{fp}(s)$};

\draw  (-2,-0.2) rectangle (4.1,-3.4);
\draw  (0.5,-3.9) rectangle (1.5,-4.5);
\node [scale=1.2] at (1,-4.2) {$\frac{L_\mathrm{dvpp}}{s}$};

\draw [-latex](4.1,-1) -- (5.45,-1); 
\draw [-latex](4.4,-1) -- (4.4,-4.2) -- (1.5,-4.2);
\draw [-latex](0.5,-4.2) -- (-2.6,-4.2) -- (-2.6,-1.1);
\node [roundnode] at (-2.6,-1) {};
\draw [-latex](-2.47,-1) -- (-2,-1); 
\draw [-latex](-3.6,-1) -- (-2.7,-1);
\node  at (1,-3.7) {linearized power flow};
%\node  at (1.05,0) {local DVPP dynamics};
\fill[black](4.4,-1)circle (0.4mm);

\node at (-3.4,-0.8) {$\Delta p_\mathrm{d}$};
\node at (-2.4,-0.8) {$-\Delta p$};
\node at (-2.3,-3.95) {$\Delta p_\mathrm{e}\hspace{-0.5mm}$};
\node at (5.05,-0.8) {$\Delta f$};
\node at (-2.85,-1.15) {+};
\node at (-2.75,-1.3) {$-$};

\node [scale=1.5] at (1,-4.95) {$\approx$};
\draw [dashed, rounded corners=3, color=black!50, fill=black!5] (-3,-5.2) rectangle (4.7,-6);

\draw  (0.3,-5.3) rectangle (1.7,-5.9);
\node at (1,-5.6) {$T_\mathrm{des}^\mathrm{pf}(s)$};
\draw [-latex](-3.6,-5.6) -- (0.3,-5.6);
\draw [-latex](1.7,-5.6) -- (5.4,-5.6);
\node at (-3.5,-5.3) {$\Delta p_\mathrm{pcc}$};
\node at (5.2,-5.3) {$\Delta f_\mathrm{pcc}$};

\node [hydro] at (-0.5,-0.5) {grid-forming DERs $\mathcal{F}_\mathrm{orm}$};
\node [backgroundgreen] at (2.6,-1.4) {grid-following DERs $\mathcal{F}_\mathrm{oll}$};

\footnotesize

\end{tikzpicture}}
       \vspace{-7mm}
    \caption{Grid-forming frequency control architecture for hybrid DVPPs.}

    \label{fig:DVPP_freq_ctrl_setup_hybrid}
\end{figure}
\subsection{Hybrid DVPP Control}\label{sec:hybrid_DVPP}
We can extend the {previous} grid-forming control setup to also include grid-following DER controls, thereby obtaining a so-called \textit{hybrid DVPP} control configuration. 

In particular, as illustrated in \cref{fig:DVPP_freq_ctrl_setup_hybrid}, we include the grid-following units with an inverse signal causality into the grid-forming frequency control architecture in \cref{fig:DVPP_freq_ctrl_setup} as
\begin{align}\label{eq:freq_dep_load}
    \Delta f_i(s) = -T_i^\mathrm{fp}(s)^{-1}\Delta p_i(s),\quad i\in \mathcal{F}_\mathrm{oll},
\end{align}
where $T_i^\mathrm{fp}(s)$ are the local closed-loop transfer functions of the grid-following units that map the measured frequency deviation $\Delta f_i$ to the local active power deviation output $\Delta p_i$. For the setup in \cref{fig:DVPP_freq_ctrl_setup_hybrid}, we obtain the synchronized frequency response of the hybrid DVPP as a generalization of \cref{eq:coherent_dynamics_PCC}, and arrive at the \textit{aggregation condition} for frequency control
\begin{align}\label{eq:aggregation_condition_freq_hybrid}
\left( \textstyle\sum_{i\in\mathcal{F}_\mathrm{orm}}T_i^\mathrm{pf}(s)^{-1}+\sum_{i\in\mathcal{F}_\mathrm{oll}}T_i^\mathrm{fp}(s) \right)^{-1} \stackrel{!}{=} T_\mathrm{des}^\mathrm{pf}(s).
\end{align}

For the voltage control, we consider the local closed-loop transfer functions $T_i^\mathrm{vq}(s)$ of the grid-following units that map the measured voltage magnitude deviation $\Delta v_\mathrm{pcc}$ to the local reactive power deviation output $\Delta q_i$, i.e.,
\begin{align}\label{eq:local_voltage_dyn_gfoll}
    \Delta q_i(s) = -T_i^\mathrm{vq}(s) \Delta v_\mathrm{pcc}(s), \quad \forall i\in\mathcal{F}_\mathrm{oll}.
\end{align}
Having the same signal causality as the local closed-loop dynamics of the grid-forming units in \eqref{eq:local_closed_loop_tf_vq}, we can directly include the local closed-loop dynamics of the grid-following units into the voltage control architecture in \cref{fig:DVPP_volt_ctrl_setup}, and modify the \textit{aggregation condition} in \cref{eq:aggregation_condition_voltage} accordingly as
\begin{align}\label{eq:aggregation_condition_volt_hybrid}
    \textstyle\sum_{i\in\mathcal{F}_\mathrm{orm}\color{black}\cup\mathcal{F}_\mathrm{oll}\color{black}}T_i^\mathrm{vq}(s)\stackrel{!}{=}T_\mathrm{des}^\mathrm{qv}(s)^{-1}.
\end{align}

Finally, given the aggregation conditions in \cref{eq:aggregation_condition_freq_hybrid,eq:aggregation_condition_volt_hybrid}, the adaptive divide-and-conquer strategy can be similarly applied as described in \cref{sec:adaptive_divide_and_conquer}. In particular, we require the \textit{local matching conditions} of the grid-following units as
\begin{align}\label{eq:matching_conditions_hybrid}
\begin{split}
    T_i^\mathrm{fp}(s)&\stackrel{!}{=}m_i^\mathrm{fp}(s)T_\mathrm{des}^\mathrm{pf}(s)^{-1},\\
    T_i^\mathrm{vq}(s)&\stackrel{!}{=}m_i^\mathrm{vq}(s)T_\mathrm{des}^\mathrm{qv}(s)^{-1}, \,\forall i\in\mathcal{F}_\mathrm{oll},
\end{split}
\end{align}
where, similar to $T_\mathrm{des}^\mathrm{qv}(s)$ in \cref{eq:aggregation_condition_voltage}, now also $T_\mathrm{des}^\mathrm{pf}(s)$ has to be invertible or approximated in a causal and stable way.

\subsubsection*{Tolerating Mismatch} Since the measurement unit (e.g., a PLL) for the frequency and voltage magnitude measurement of the grid-following units is limited by some approximate bandwidth $1/\tau_\mathrm{pll}$, it suffices if the grid-following units satisfy the local matching conditions in \cref{eq:matching_conditions_hybrid} only for the frequency range below $1/\tau_\mathrm{pll}$ (see \cite{haberle2021control}). In this regard, we append the local reference models by a filter, i.e., for $i \in \mathcal{F}_\mathrm{oll}$ and some $d\in \mathbb{N}$, we require
\begin{align}
    T_i^{k}(s) \stackrel{!}{=} \tfrac{m_i^k(s)T_\mathrm{des}^{\bar{k}}(s)^{-1}}{(\tau_\mathrm{pll}s+1)^d},\,\,k\in\{\mathrm{fp,vq}\},\,\bar{k}\in\{\mathrm{pf,qv}\},
\end{align}
thereby allowing for a mismatch in the high frequency range of the Bode plot, thus relaxing the restrictions on the local matching control. As a further consequence, the reference models of the grid-following units are {ensured to be} causal.

\section{Spatially Distributed DVPPs}\label{sec:spatially_distributed}

\begin{figure}[b!]
    \centering
     \vspace{-4mm}
    \usetikzlibrary{circuits.ee.IEC}
\usetikzlibrary{arrows}
\resizebox {\columnwidth} {!} { % line included by Florian

\begin{tikzpicture}[circuit ee IEC,scale=0.8, every node/.style={scale=0.6}]

\draw  plot[rotate= -30,smooth, tension=.7,scale=1] coordinates {(-5.7134,-0.6) (-5.2134,-0.95) (-4.0134,-0.85)(-2.4134,0.25)(-0.1334,0.05)(0.4066,0.15)};
\draw  plot[smooth, tension=.7,scale=1.4] coordinates {(-1.5762,0.2667) (-1.7114,0.3) (-1.9114,0.32) (-2.2762,0.2667)(-2.719,0.0667) (-3.2,-0.25) (-3.35,-0.75) (-2.819,-1.2333) (-2.5429,-1.5) (-1.8429,-1.6) (-1.3762,-1.0333) (-0.9834,-0.7333) (-0.9762,-0.3333) (-1.1762,-0.0333) (-1.3762,0.1667) (-1.5762,0.2667)};
\draw  (-0.86,-0.22) ellipse (0.15 and 0.15);
\draw  (-0.66,-0.17) ellipse (0.15 and 0.15);
\draw  (-4.42,0.64) ellipse (0.15 and 0.15);
\draw  (-4.4,0.42) ellipse (0.15 and 0.15);
\draw (-4.52,1.42) -- (-4.44,0.78);
\draw (-4.36,0.26) -- (-4.3,-0.22) node (v1) {};
\draw (-1.4,-0.4) node (v2) {} -- (-1.01,-0.28);
\draw (-0.53,-0.12) -- (0,0.06);
\draw  [very thick](-4.2998,-0.48) -- (-4.2998,-0.92);
\draw (-1.4,-0.4)  -- (-1.7,-0.4);
\draw  [very thick](-1.7198,-0.2) -- (-1.7198,-0.58);
\draw [very thick] (-3.7798,-0.12) -- (-3.7798,-0.48);
\draw [very thick] (-3.2798,0.06) -- (-2.8398,0.06); 
\draw  [very thick](-3.6398,-1.34) -- (-3.6398,-1.76);
\draw  [very thick](-3.3198,-0.92) -- (-2.8198,-0.92); 
\draw [very thick] (-2.8798,-0.4) -- (-2.4598,-0.4);
\draw  [very thick](-2.9598,-1.72) -- (-2.9598,-2.16);
\draw [very thick](-2.5198,-1.14) -- (-2.5198,-1.54); 
\draw (-4.2998,-0.62) -- (-4.1998,-0.62) -- (-3.8798,-0.3) -- (-3.7798,-0.3) node (v6) {};
\draw (-3.7798,-0.22) node (v8) {} -- (-3.6398,-0.22) -- (-3.1598,-0.04) -- (-3.1598,0.06) -- (-2.9798,0.06) -- (-2.9798,-0.04) -- (-2.7798,-0.3) -- (-2.7798,-0.4);
\draw (-3.7798,-0.38) -- (-3.6398,-0.38) -- (-3.1798,-0.8) -- (-3.1798,-0.9) node (v3) {};
\draw (-4.2998,-0.78) -- (-4.1998,-0.78) -- (-3.6998,-1.54) -- (-3.6398,-1.54);
\draw (-3.6398,-1.46) -- (-3.5398,-1.46) -- (-3.1798,-1.02) --(-3.1798,-0.9) ; 
\draw (-2.5198,-1.32) -- (-2.4198,-1.32) -- (-1.7998,-0.46) -- (-1.7198,-0.46); 
\draw (-2.5798,-0.38) -- (-2.5798,-0.3) -- (-1.7198,-0.3);
\draw (-3.6398,-1.62) -- (-3.5398,-1.62) -- (-3.0198,-1.94) -- (-2.9598,-1.94) node (v4) {};
\draw (-2.9598,-0.92) -- (-2.9598,-1.02) -- (-2.6398,-1.24) -- (-2.5198,-1.24);
\draw [fill =SC] (-2.8198,-1.82) rectangle (-2.4798,-2.06);
\draw (-2.9598,-1.94) -- (-2.8198,-1.94);
\draw  [fill = battery](-4.0798,-0.58) rectangle (-3.7598,-0.82);
\draw [fill = wind] (-1.6398,-0.58) rectangle (-1.3998,-0.9);
\draw  [fill = STATCOM](-3.0398,-1.36) rectangle (-2.6798,-1.6);
\draw [fill = PV] (-2.8398,-0.56) rectangle (-2.4998,-0.8);
\draw  (-4.2998,-0.7) node (v9) {}  -- (-4.0798,-0.7);
\draw (-2.6798,-0.42) -- (-2.6798,-0.56);
\draw (-1.7198,-0.5) -- (-1.5398,-0.5) -- (-1.5398,-0.58);
\draw (-2.6798,-1.46) -- (-2.5198,-1.46);
\draw  [fill=SC](-2.7998,0.34) rectangle (-2.4798,0.12);
\draw (-3.0798,0.08) -- (-3.0798,0.24) -- (-2.7998,0.24);
\draw[-latex](-3.7798,-0.3)-- (-3.3798,-0.3);
\draw[-latex] (-2.5198,-1.4) -- (-2.3798,-1.4) -- (-2.3798,-1.76);
\draw [-latex](-2.8998,0.06) -- (-2.8998,-0.04) -- (-2.4598,-0.04);
\node at (-5,0.5) {POC 1};
\draw [very thick] (-1.23,-0.18) -- (-1.13,-0.48);
\node at (-0.6,-0.6) {POC $r$};
\node [scale = 1.2] at (-2.61,2.16) {remaining};
\node  [scale = 1.2] at (-1.96,1.86) {power system};
\draw [very thick](-4.62,0.92) -- (-4.3,0.96);
\node at (-1.36,-1.94) {DVPP area};
\draw [very thick](-4.48,0.1) -- (-4.18,0.14);
\draw[very thick] (-0.43,0.08) -- (-0.33,-0.22);

\draw  plot[smooth, tension=.7] coordinates {(-4.84,-1.8) (-4.7159,-1.524) (-4.46,-1.7) (-4.1759,-1.324)};
\node at (-4.62,-2.1) {$R/X\approx$ const.};

\draw (-3.7798,-0.22)-- (-4.04,-0.22) -- (-4.3,-0.22);
\draw[-latex] (-4.2998,-0.7) -- (-4.45,-0.7) -- (-4.45,-1.15);

\node [scale = 1.5] at (0.6,-0.8) {$\approx$};

\draw [dashed, rounded corners = 3, color = black!50, fill = black!5] (2.4,-0.4) rectangle (3.94,-1.49);
\draw  (2.64,-0.59) rectangle (3.7,-1.3);
\node[scale=0.8] at (1.91,-0.75) {$\displaystyle\sum_{j=1}^r\hspace{-1.2mm}\Delta p'_{\mathrm{poc},j}$};
\node[scale=0.8] at (4.32,-0.75) {$\Delta f_{\mathrm{poc},j}$};
\draw[-latex] (1.95,-0.95) -- (2.65,-0.95);
\node at (3.2,-0.95) {$T_\mathrm{des}^\mathrm{p'f}(s)$};
\node [color=black!50] at (3.2,-0.2) {DVPP};
\draw[-latex] (3.7,-0.95) -- (4.3,-0.95);

\draw  plot[rotate= -30,smooth, tension=.7,scale=1] coordinates {(-0.5172,2.4) (-0.0172,2.05) (1.1828,2.15)(2.7828,3.25)(5.0628,3.05)(5.6028,3.15)};
\draw  plot[smooth, tension=.7,scale=1.4] coordinates {(2.7096,0.2667) (2.5744,0.3) (2.3744,0.32) (2.0096,0.2667)(1.5668,0.0667) (1.05,-0.25) (0.95,-0.75) (1.4668,-1.2333) (1.7429,-1.5) (2.4429,-1.6) (2.9096,-1.0333) (3.3024,-0.7333) (3.3096,-0.3333) (3.1096,-0.0333) (2.9096,0.1667) (2.7096,0.2667)};
\draw  (5.14,-0.22) ellipse (0.15 and 0.15);
\draw  (5.34,-0.17) ellipse (0.15 and 0.15);
\draw  (1.58,0.64) ellipse (0.15 and 0.15);
\draw  (1.6,0.42) ellipse (0.15 and 0.15);
\draw (1.48,1.42) -- (1.56,0.78);
\draw (1.64,0.26) -- (1.7,-0.15) node (v1) {};
\draw (4.6,-0.4) node (v2) {} -- (4.99,-0.28);
\draw (5.47,-0.12) -- (6,0.06);

\node at (1,0.5) {POC 1};
\draw [very thick] (4.77,-0.18) -- (4.87,-0.48);
\node at (5.4,-0.6) {POC $r$};
\node [scale = 1.2] at (3.39,2.16) {remaining};
\node  [scale = 1.2] at (4.04,1.86) {power system};
\draw [very thick](1.38,0.92) -- (1.7,0.96);
\node at (4.64,-1.94) {DVPP area};
\draw [very thick](1.52,0.1) -- (1.82,0.14);
\draw[very thick] (5.57,0.08) -- (5.67,-0.22);

\draw (-2.3,1.48) -- (-2.35,1.05);
\draw[dotted] (-2.35,1.05) -- (-2.4,0.65);

\draw (3.7,1.48) -- (3.65,1.05);
\draw[dotted] (3.65,1.05) -- (3.6,0.65);
\end{tikzpicture}

}
    \vspace{-9mm}
    \caption{Spatially distributed DVPP within a DVPP area that is connected to the remaining power grid. The coloured boxes indicate different DVPP units.}
    \label{fig:spatially_distributed_DVPP}
    \vspace{-1mm}
\end{figure}
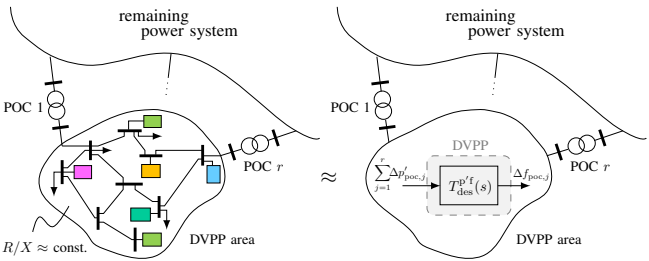
So far, the DVPP control setup has been restricted to DER aggregations at one single bus in the transmission system {(\cref{fig:DVPP_at_PCC})}. We now generalize our DVPP control to \textit{spatially distributed} DER locations in a general power grid (from HV transmission to LV distribution grids). Namely, we consider a generalized DVPP configuration, where the DERs are spatially distributed within an area of the power system that is connected to the remaining power grid via one or multiple point of couplings (POCs) $j\in\{1,...,r\}$ (\cref{fig:spatially_distributed_DVPP}, left and \cref{tab:spatially_distr_symbols}). A typical DVPP area could be given by a transmission or a distribution system area that is connected to the remaining grid (see example in \cref{sec:case_study_III}). Independent of the type of area, the overall goal is to design local controllers of the distributed DVPP units, such that we obtain a desired aggregated response behavior of the entire DVPP area at the POCs (\cref{fig:spatially_distributed_DVPP}, right). 

Assuming approximate coherency as in \cref{eq:coherent_dynamics_PCC}, we only design a DVPP for an aggregated frequency control at the POCs, while employing independent local voltage controls to the DVPP units, reason being that voltages are distinctly local quantities at different buses in the power grid \cite{kundur2007power}. Further, we consider all DERs in the DVPP area which provide dynamic ancillary services to be part of the DVPP. More specifically, in case of unknown, pre-installed dynamics within the DVPP area, the aggregated DVPP response will become deteriorated. However, major dynamic sources (i.e., larger DERs providing ancillary services, synchronous generators, etc.) can typically assumed to be known (or at least an approximation thereof), such that their dynamic behavior can be taken into account during the DVPP control design (see, e.g., case study I in our previous work in\cite{haberle2021control}). Beyond that, for all the remaining unknown dynamics, e.g., resulting from small dynamic loads, their impact on the power flow, and with this the aggregated DVPP response, is expected to be minor when compared to the known dominant dynamics of the major sources\cite{kundur2007power}. An investigation of such scenarios will be part of future work.

To regulate the frequency at the POCs, a classical approach is to employ a decoupled $\mathrm{p}$-$\mathrm{f}$ control of the DVPP units as in \cite{zhong2021impact}. With this, however, one does not account for non-inductive line impedances between the spatially distributed DERs and associated active power losses (which are especially relevant in MV or LV grids with high $R/X$ ratios), leading to a less effective frequency regulation at the POCs \cite{rocabert2012control,de2007voltage,yao2010design,bevrani2013intelligent}. Instead, as suggested by the latter references, we therefore consider rotational powers $p'$ and $q'$ via the rotation matrix $\mathcal{R}$
    \begin{align}\label{eq:rotation_matrix}
        \begin{bmatrix}
            p'\\q'
        \end{bmatrix}
        =\mathcal{R} \begin{bmatrix}
            p\\q
        \end{bmatrix} = \begin{bmatrix}
            \tfrac{X}{Z}&-\tfrac{R}{Z}\\ \tfrac{R}{Z}& \tfrac{X}{Z}
        \end{bmatrix}
        \begin{bmatrix}
            p\\q
        \end{bmatrix}\hspace{-1mm},\quad Z=\sqrt{R^2+X^2},
    \end{align}
where $Z$ is the line impedance, and $p$ and $q$ are the actual active and reactive powers. As a result, we obtain lossless power flow equations between two buses $l$ and $m$ of the form
\begin{align}\label{eq:decoupled_power_flow}
        \sin \delta &= \tfrac{Z p'}{v_lv_m},\quad\quad v_l-v_m \cos \delta = \tfrac{Z q'}{v_l},
\end{align}
where $v_l$ and $v_m$ are the voltage magnitudes, $\delta$ is the power angle, and $p'$ and $q'$ are the rotational active and reactive power injected at one of the two buses. As can be seen from \cref{eq:decoupled_power_flow}, for a sufficiently small power angle $\delta$ and voltage magnitude difference $v_l\hspace{-1mm}-\hspace{-1mm}v_m$, the definition of rotational powers $p'$ and $q'$ permits to independently influence the grid frequency (via $\delta$) and voltage magnitude, in analogy to the classical frequency and voltage regulation through respectively active and reactive power in HV grids, where $p'\hspace{-0.5mm}\approx\hspace{-0.5mm}p$ and $q'\hspace{-0.5mm}\approx\hspace{-0.5mm}q$, since $R/X\hspace{-0.5mm}\approx\hspace{-0.5mm}0$. 
\renewcommand{\arraystretch}{1.2}
\begin{table}[t!]\scriptsize
    \setlength{\tabcolsep}{1mm}
    \centering
     \caption{List of notation for the spatially distributed DVPP control.}
    \vspace{-1mm}
    \begin{tabular}{c||c}
     \toprule
         Description & Symbol  \\ \hline
         Line resistance, line inductance, line impedance & $R, X, Z$\\
         Rotation matrix to obtain rotational powers & $\mathcal{R}$\\ \hline
         Bus voltage magnitudes with bus indices $l$ and $m$ & $v_l$, $v_m$\\
         Power angle & $\delta$ \\
         Active and reactive power injected at one bus & $p, q$\\
         Rotational active and reactive power injected at one bus & $p', q'$\\\hline
         Imposed frequency deviation at POC $j$ & $\Delta f_{\mathrm{poc},j}$\\
         Active and reactive power injection change at POC $j$ & $\Delta p_{\mathrm{poc},j}, \Delta q_{\mathrm{poc},j}$ \\
         Rotational active power injection change at POC $j$ & $\Delta p_{\mathrm{poc},j}'$\\\hline
         Desired DVPP transfer function for $\mathrm{p'}$-$\mathrm{f}$ control& $T_\mathrm{des}^\mathrm{p'f}(s)$ \\ 
         ADPF of unit $i$ for $\mathrm{f}$-$\mathrm{p'}$ channel& $m_i^\mathrm{fp'}(s)$\\
         Local $\mathrm{p'}$-$\mathrm{f}$ closed-loop transfer function of unit $i$& $T_i^\mathrm{p'f}$\\
         Local $\mathrm{v}$-$\mathrm{q'}$ closed-loop transfer function of unit $i$& $T_i^\mathrm{vq'}(s)$\\   
    \bottomrule
    \end{tabular}
    	 \vspace{-4mm}
    \label{tab:spatially_distr_symbols}
\end{table}
\renewcommand{\arraystretch}{1} \normalsize

For the DVPP area, by applying the transformation matrix $\mathcal{R}$ at all controllable buses, and assuming an approximately homogeneous $R/X$ ratio for all transmission lines, we obtain a lossless $p'$ transmission. This allows to apply a modified version of the previous frequency control setup in \cref{fig:DVPP_freq_ctrl_setup,fig:DVPP_freq_ctrl_setup_hybrid}, where we replace all active power quantities by rotational active power quantities to obtain a decoupled $\mathrm{p}'$-$\mathrm{f}$ DVPP control (instead of the previous $\mathrm{p}$-$\mathrm{f}$ control). We hence aim to match a $\mathrm{p}$-$\mathrm{q}$-$\mathrm{f}$ \textit{coupled} desired specification for the DVPP control at the POCs $j$, given by the \textit{rotational power control}
\begin{align}\label{eq:coupled_specs}
%\begin{split}
    \Delta f_{\mathrm{poc},j}(s) &= T_\mathrm{des}^\mathrm{p'f}(s)\textstyle\sum_{j=1}^r \Delta p_{\mathrm{poc},j}'(s)\\
    &=\hspace{-0.5mm}\underset{=T_\mathrm{des}(s)}{\underbrace{\begin{bmatrix}
        \tfrac{X}{Z} T_\mathrm{des}^\mathrm{p'f}(s) & -\tfrac{R}{Z} T_\mathrm{des}^\mathrm{p'f}(s)
    \end{bmatrix}}} \hspace{-0.5mm}\begin{bmatrix}
        \textstyle\sum_{j=1}^r\Delta p_{\mathrm{poc},j}(s)\\ \nonumber\textstyle\sum_{j=1}^r\Delta q_{\mathrm{poc},j}(s)
    \end{bmatrix}\hspace{-1.25mm},
%    \end{split}
\end{align}
which differs from the decoupled specification in \eqref{eq:aggregate_specification}.
\begin{remark}
    If one was to install a DVPP where all DERs are connected \emph{at the same bus} in a MV or LV distribution grid with high $R/X$ ratio (i.e., no spatially distributed DVPP configuration), one should resort to a desired DVPP specification $T_\mathrm{des}(s)$ for \emph{both} frequency and voltage regulation in terms of a decoupled $\mathrm{p}'$-$\mathrm{f}$ and $\mathrm{q}'$-$\mathrm{v}$ control, which encodes a \emph{coupled} $\mathrm{p}$-$\mathrm{q}$-$\mathrm{f}$ and $\mathrm{p}$-$\mathrm{q}$-$\mathrm{v}$ desired specification in the same vein as \eqref{eq:coupled_specs}.
\end{remark}
Finally, the divide-and-conquer strategy can be similarly applied as in \cref{sec:adaptive_divide_and_conquer}, however, by selecting ADPFs $m_i^\mathrm{fp'}(s)$ to account for DER limitations on the rotational active power $\Delta p_i'$ (instead of $\Delta p_i$).

\subsubsection*{Practically Feasible Choice of ADPFs} Although we select the ADPFs $m_i^\mathrm{fp'}$ for the $\mathrm{p}'$-$\mathrm{f}$ control, we are {still} interested in the actual physical DER limitations in $\Delta p_i$ and $\Delta q_i$. For example, from the equation $\Delta p_i = (\tfrac{X}{Z}\Delta p_i'+\tfrac{R}{Z}\Delta q_i')$ which can be derived from \cref{eq:rotation_matrix}, we can see that $\Delta p_i$, which is limited in response time and capacity for each DER, depends on both $\Delta p_i'$ and $\Delta q_i'$, scaled by $X$ and $R$, respectively. However, while $\Delta p_i'$ is part of the DVPP control, and hence shaped by the ADPF $m_i^\mathrm{fp'}(s)$, the control of $\Delta q_i'$ is handled locally (i.e., individually) for each DVPP unit. Consequently, the local $\Delta q_i'$ injection change will be different for each unit, as determined by the local voltage controls $\Delta q_i'(s)=-T_i^\mathrm{vq'}(s)\Delta v_i(s)$. Thus, to ensure all limitations of $\Delta p_i$ (and $\Delta q_i$) are properly {addressed}, $m_i^\mathrm{fp'}(s)$ and $T_i^\mathrm{vq'}(s)$ need to be selected carefully and possibly conservatively in terms of their response time and dc gain. However, while the assumptions underlying the implementation of the rotational power control for DVPPs seem to be rather restrictive, we will observe in \cref{sec:testcase} that this method is quite effective in a practical setup. 

\section{Test Case}\label{sec:testcase}
To verify our DVPP controls, we use Simscape Electrical in MATLAB/Simulink to perform an electromagnetic transients (EMT) simulation based on the IEEE nine-bus system using nonlinear system and device models. In particular, the IEEE nine-bus system has been established as a typical benchmark system in the DVPP literature\cite{haberle2021control,zhong2021coordinated,zhong2021impact}, and therefore allows for comparable studies with prior work. To demonstrate the basic idea of our DVPP control strategies in a conceptual way, we use a deliberately simple test system assembled with DVPPs containing only a few DERs, where we perform plain and instructive simulation events. Our methods, however, can be easily extended to larger and more complex power systems comprising multiple DVPPs or DVPPs with a larger number of DERs, especially since the deployed local DER matching controls are independent of the size of the DVPP and the power system to which it is connected to. Moreover, the network size/topology does only affect the DVPP input signal, but not the internal DVPP behavior.

In a first case study, we start with a tutorial example of a grid-forming ``base DVPP'' connected at one bus to replace the fast frequency and voltage control of a thermal-based generator, while additionally including online-adaptation of the ADPFs to handle temporal variability of weather-dependent DVPP units. In a second case study, we extend the base DVPP to a hybrid setup by splitting each DER into a grid-forming and a grid-following unit. For different shares of grid-forming/-following units, we investigate the DVPP response during load variations and an outage of all synchronous generators (SG) in the grid. Finally, in a third case study, we demonstrate the performance of a spatially distributed arrangement of the base DVPP in a MV distribution grid during disturbances both in- and outside the DVPP area. 

\subsection{System Model}
The implementation of the initial IEEE nine-bus system (\cref{fig:system_model_case_study_I} without DVPP, \cref{tab:9bus_parameters}), containing three thermal-based SGs, is based on the system and device models in \cite{haberle2021control}, where the transmission lines are modelled via nominal $\pi$ sections, and the transformers via three-phase linear models. The loads are modelled as constant impedance loads. We adopt an 8th-order model for the synchronous machines equipped with a ST1A excitation system with built-in automatic voltage regulator and a power system stabilizer. The governors are modelled as a proportional speed-droop control with first-order delay, and the steam turbine parameters are from\cite{kundur2007power}. 

\subsection{Converter Interface of DERs}\label{sec:converter_model}
\begin{figure}[t!]
\centering
\begin{subfigure}{0.485\textwidth}
    \centering
    
    \resizebox {1\columnwidth} {!} {
\usetikzlibrary{circuits.ee.IEC}
\tikzstyle{roundnode} =[circle, draw=black!60, fill=hydro!5, scale = 0.5]
\begin{tikzpicture}[circuit ee IEC,scale=0.45, every node/.style={scale=0.64}]
\draw (-3.1,3.2) rectangle (-1.5,2);
\node at (-2.3,2.6) {$\frac{1}{\tau_{\mathrm{dc}}s+1}$};
\draw  (-0.4,2.6) ellipse (0.35 and 0.35);
\draw [-latex](-0.4,2.3) node (v2) {} -- (-0.4,2.9) node (v1) {};
\draw(-0.4,2.9)-- (-0.4,3.6) -- (1.35,3.6) node (v3) {} -- (1.35,3.1);
\draw  (1.15,3.1) rectangle (1.55,2.1);
\draw (1.35,2.1) -- (1.35,1.6) node (v4) {} -- (-0.4,1.6) -- (-0.4,2.3); 
\draw (1.35,3.6)-- (2.9,3.6) node (v5) {}--(2.9,2.65);
\draw (2.9,2.55) -- (2.9,1.6) node (v6) {} -- (1.35,1.6) ; 
\draw (2.65,2.65) -- (3.15,2.65);
\draw (2.65,2.55) -- (3.15,2.55); 
\draw (2.9,3.6) -- (4.2,3.6);
\draw (2.9,1.6) -- (4.2,1.6);
\draw  (4.2,3.8) node (v13) {} rectangle (5.8,1.4);
\draw (5.4,3.4) -- (5.4,3.1) -- (4.9,2.85) node (v3) {};
\draw[-latex](4.9,2.35) -- (5.4,2.1) ;
\draw (5.4,2.1) -- (5.4,1.6);
\draw (4.9,3.1) -- (4.9,2.1);
\draw (4.8,3.1)--(4.8,2.1);
\draw (4.4,2.6) -- (4.8,2.6);
\draw (7.3,2.6) node (v7) {} to [inductor={}] (9.5,2.6) node (v8) {};
\draw  (6.3,2.8) rectangle (7.3,2.4);
\draw (5.8,2.6) -- (6.3,2.6); 
\draw (7.3,2.6) -- (7.3,2.6); 
\draw (9.5,2.6) -- (11.15,2.6) node (v14) {};
\fill[black](9.8,2.6) node (v10) {}circle (0.5 mm); 
\draw [-latex](-1.5,2.6) -- (-0.75,2.6);  
\draw [dashed,-latex,rounded corners = 3, black!50] (1.7,4) rectangle (-3.6,1.3);

\node at (-1,3.1) {$i_{\mathrm{dc},i}$};
\node at (0.6,2.6) {$G_{\mathrm{dc}}$};
\node at (2.36,2.35) {$C_{\mathrm{dc}}$};
\node at (3.3,3.2) {$+$};
\node at (3.3,2) {$-$};
\node [black!50] at (3.65,2.6) {$v_{\mathrm{dc},i}$};
\node at (6.8,3.1) {$R_{\mathrm{f}}$};
\node at (8.4,3.1) {$L_{\mathrm{f}}$};
\fill[black](6.5,-2) circle (0.5 mm); 
\fill[black](2.9,1.6) node (v9) {} circle (0.5 mm); 
\fill[black](1.35,3.6) circle (0.5 mm); 
\fill[black](1.35,1.6) circle (0.5 mm); 
\draw [-latex](2.9,0.4) -- (1.7,0.4);
\draw [rounded corners = 3,-latex] (-2.6,0.8) rectangle (1.7,0);
\draw [-latex](-2.6,0.4) -- (-3.8,0.4) -- (-3.8,2.2)--(-3.8,2.6) -- (-3.1,2.6);
\node  at (-0.5,0.4) {dc-voltage control};
\node  at (2.3,0.75) {$v_{\mathrm{dc},i}$};
\node [black!50] at (-0.95,4.4) {dc energy source model};

\node at (4.9,4.4) {power};
\node at (4.9,4.1) {converter $i$};

\draw  (11.5,2.6) ellipse (0.35 and 0.35);
\draw (11.9,2.6) ellipse (0.35 and 0.35); 
\draw (12.2,2.6) -- (14,2.6);
\node at (11.6,3.25) {LV/MV};

\draw [thick](13.6,3.5) -- (13.6,1.7);
\node at (14.3,1.7) {PCC};
\draw[dotted] (14,2.6) -- (14.9,2.6);
\draw (9.55,1.7) -- (10.05,1.7);
\draw (9.55,1.8) -- (10.05,1.8);
\draw (9.8,2.6)  -- (9.8,1.8); 
\draw (9.8,1.7) -- (9.8,1);
\draw (9.7,1) -- (9.9,1);
\node at (10.4,1.4) {$C_{\mathrm{f}}$};

\fill[black!50](13.1,2.6) node (v12) {} circle (1 mm);

\draw[-latex, black!50] (13.1,2.6) -- (13.1,4.2);

\node [black!50]at (13.1,4.4) {$p_i, q_i$};
\draw [-latex, black!50] (9.3,2.3) -- (9.3,1.1);
\draw  [-latex, black!50](13.4,2.3) -- (13.4,1.1);
\node  [black!50] at (9,1.8) {$v_i$};
\node [black!50] at (12.9,1.8) {$v_\mathrm{pcc}$};

\draw [-latex](5,0.7) -- (5,1.4);
\draw  (3.6,0.7) rectangle (6.4,-0.1);
\node at (5,0.3) {modulation};
\draw  (4.5,-0.8) rectangle (5.5,-1.6);
\draw (4.5,-1.6) -- (5.5,-0.8);
\node [scale=0.7] at (4.82,-0.99) {abc};
\node  [scale=0.7] at (5.2,-1.4) {dq};
\draw[-latex] (5,-0.8) -- (5,-0.1);
\draw  (6.7,-0.7) rectangle (7.3,-1.7);
\node [scale =1.2]at (7,-1.2) {$\tfrac{1}{s}$};
\draw[-latex] (6.7,-1.2) -- (5.5,-1.2);

\draw [-latex](8.1,-1.2) -- (7.3,-1.2);
\node [roundnode] at (8.2,-1.2) {};
\draw [-latex](8.2,-0.3) -- (8.2,-1.1); 
\draw [-latex](9.2,-1.2) -- (8.3,-1.2);
\draw [-latex, fill = hydro!40] (10.8,-0.7) rectangle (14.3,-1.7);
\node [scale=0.95] at (12.6,-1.2) {$m_i^\mathrm{fp}(s)^{\hspace{-0.3mm}-\hspace{-0.3mm}1}T_\mathrm{des}^\mathrm{pf}(s)$}; 
\draw [-latex](15.5,-1.2) -- (14.3,-1.2); 
\draw [-latex,dashed, hydro](5.6,-3.6) -- (5.3,-3.6) node (v15) {} -- (5.3,-1.6);
\draw [fill=hydro!15, rounded corners = 3] (5.6,-2.7) rectangle (7.4,-4.5);
\node at (6.5,-3.1) {current};
\node at (6.5,-3.5) {control};
\node at (6.5,-4.05) {(PI)};
\draw [fill=hydro!15, rounded corners = 3] (8,-2.7) rectangle (9.8,-4.5);
\node at (8.9,-3.1) {voltage};
\node at (8.9,-3.5) {control};
\node at (8.9,-4.05) {(PI)};

\draw[dashed, hydro,-latex]  (5.6,-6.6)-- (4.7,-6.6)--(4.7,-1.6)  ;

\draw[-latex] (8,-3.6) -- (7.4,-3.6);
\draw [rounded corners = 3, fill = hydro!15] (10.3,-4.5) rectangle (15.5,-2.7);
\draw  (10.6,-3.3) rectangle (11.3,-3.9);
\node at (10.95,-3.6) {PI};
\draw [-latex](10.6,-3.6) -- (9.8,-3.6);

\draw [fill = hydro!40]  (11.9,-3.1) rectangle (15.3,-4.1);
\node [scale=0.95] at (13.65,-3.6) {$m_i^\mathrm{vq}\hspace{-0.3mm}(s)T_\mathrm{des}^\mathrm{qv}\hspace{-0.3mm}(s)^{\hspace{-0.3mm}-\hspace{-0.3mm}1}$};

\draw[-latex] (11.9,-3.6) -- (11.3,-3.6) node (v17) {};
\draw[-latex] (16.6,-3.6) -- (15.5,-3.6);

\draw  (9.2,-0.9) rectangle (9.9,-1.5);
\node at (9.55,-1.2) {$2\pi$};
\draw [-latex](10.8,-1.2) -- (9.9,-1.2);
\node at (6.2,-0.9) {$\theta_i$};
\node at (8.6,-0.4) {$\omega_0$};
\node at (8.9,-1) {$-$};
\node at (8.6,-0.8) {+};
\node at (15,-0.85) {$\Delta p_i$};
\draw [-latex](11,-5.2) -- (11,-3.9);

\node at (11.6,-5.1) {$\Delta q_i$};
\node at (16.35,-3.2) {$|\Delta v_\mathrm{pcc}|$};
\draw [-latex](6.2,-1.2) -- (6.2,-2) node (v16) {} -- (8.9,-2) -- (8.9,-2.7);
\fill[black](2.9,3.6) circle (0.5 mm); 
\draw [-latex](6.5,-2) -- (6.5,-2.7);
\draw [-latex](6.5,-5.2) -- (6.5,-4.5);
\draw [-latex] (8.9,-5.2) -- (8.9,-4.5);
\draw[fill = hydro!15, rounded corners = 3]  (5.6,-5.9) rectangle (9,-7.3);
\node at (7.3,-6.3) {$\mathcal{H}_\infty$ matching};
\node at (7.3,-6.8) {controller};
\draw [-latex](10.1,-6.6) -- (9,-6.6);

\node at (11.2,-6.6) {$i_{\mathrm{c},i},i_i, v_i$};
\draw[-latex, black!50] (7.6,2.4) -- (8.6,2.4);
\node [black!50]at (8,2.1) {$i_{\mathrm{c},i}$};
\draw[-latex,black!50] (10,2.4) -- (11,2.4);
\node  [black!50] at (10.4,2.1) {$i_i$};
\draw [-latex](10.2,-7.1) -- (9,-7.1);
\node at (11.1,-7.1) {$q_i, v_\mathrm{pcc}$};
\node at (7.3,-5.1) {$v_i, i_{\mathrm{c},i}$};
\node at (9.55,-5.1) {$i_{i},v_i$};
\draw [-latex](10.1,-6.1) -- (9,-6.1);
\node at (10.5,-6.1) {$\theta_i$};
\node [hydro,rotate=90,scale=0.75] at (4.5,-2.8) {option B};
\node  [hydro,rotate=90,scale=0.75] at (5.1,-2.8) {option A};
\end{tikzpicture}}
       \vspace{-9mm}
    \caption{One-line diagram of the three-phase power converter interface of each grid-forming DVPP unit $i$ (cf. \cref{tab:converter_parameters}).}
    \label{fig:converter_model_gform}
    \vspace{1.5mm}
\end{subfigure}

\begin{subfigure}{0.485\textwidth}
    \centering
   
    \resizebox {1\columnwidth} {!} {
\usetikzlibrary{circuits.ee.IEC}
\tikzstyle{roundnode} =[circle, draw=black!60, fill=backgroundgreen!5, scale = 0.5]
\begin{tikzpicture}[circuit ee IEC,scale=0.45, every node/.style={scale=0.64}]
\draw (-3.1,3.2) rectangle (-1.5,2);
\node at (-2.3,2.6) {$\frac{1}{\tau_{\mathrm{dc}}s+1}$};
\draw  (-0.4,2.6) ellipse (0.35 and 0.35);
\draw [-latex](-0.4,2.3) node (v2) {} -- (-0.4,2.9) node (v1) {};
\draw(-0.4,2.9)-- (-0.4,3.6) -- (1.35,3.6) node (v3) {} -- (1.35,3.1);
\draw  (1.15,3.1) rectangle (1.55,2.1);
\draw (1.35,2.1) -- (1.35,1.6) node (v4) {} -- (-0.4,1.6) -- (-0.4,2.3); 
\draw (1.35,3.6)-- (2.9,3.6) node (v5) {}--(2.9,2.65);
\draw (2.9,2.55) -- (2.9,1.6) node (v6) {} -- (1.35,1.6) ; 
\draw (2.65,2.65) -- (3.15,2.65);
\draw (2.65,2.55) -- (3.15,2.55); 
\draw (2.9,3.6) -- (4.2,3.6);
\draw (2.9,1.6) -- (4.2,1.6);
\draw  (4.2,3.8) node (v13) {} rectangle (5.8,1.4);
\draw (5.4,3.4) -- (5.4,3.1) -- (4.9,2.85) node (v3) {};
\draw[-latex](4.9,2.35) -- (5.4,2.1) ;
\draw (5.4,2.1) -- (5.4,1.6);
\draw (4.9,3.1) -- (4.9,2.1);
\draw (4.8,3.1)--(4.8,2.1);
\draw (4.4,2.6) -- (4.8,2.6);
\draw (7.3,2.6) node (v7) {} to [inductor={}] (9.5,2.6) node (v8) {};
\draw  (6.3,2.8) rectangle (7.3,2.4);
\draw (5.8,2.6) -- (6.3,2.6); 
\draw (7.3,2.6) -- (7.3,2.6); 
\draw (9.5,2.6) -- (10.4,2.6) node (v14) {};

\draw [-latex](-1.5,2.6) -- (-0.75,2.6);  
\draw [dashed,-latex,rounded corners = 3, black!50] (1.7,4) rectangle (-3.6,1.3);

\node at (-1,3.1) {$i_{\mathrm{dc},i}$};
\node at (0.6,2.6) {$G_{\mathrm{dc}}$};
\node at (2.36,2.35) {$C_{\mathrm{dc}}$};
\node at (3.3,3.2) {$+$};
\node at (3.3,2) {$-$};
\node [black!50] at (3.65,2.6) {$v_{\mathrm{dc},i}$};
\node at (6.8,3.1) {$R_{\mathrm{f}}$};
\node at (8.4,3.1) {$L_{\mathrm{f}}$};

\fill[black](2.9,1.6) node (v9) {} circle (0.5 mm); 
\fill[black](1.35,3.6) circle (0.5 mm); 
\fill[black](1.35,1.6) circle (0.5 mm); 
\draw [-latex](2.9,0.4) -- (1.7,0.4);
\draw [rounded corners = 3,-latex] (-2.6,0.8) rectangle (1.7,0);
\draw [-latex](-2.6,0.4) -- (-3.8,0.4) -- (-3.8,2.2)--(-3.8,2.6) -- (-3.1,2.6);
\node  at (-0.5,0.4) {dc-voltage control};
\node  at (2.3,0.75) {$v_{\mathrm{dc},i}$};
\node [black!50] at (-0.95,4.4) {dc energy source model};

\node at (4.9,4.4) {power};
\node at (4.9,4.1) {converter $i$};

\draw  (10.75,2.6) ellipse (0.35 and 0.35);
\draw (11.15,2.6) ellipse (0.35 and 0.35); 
\draw (11.45,2.6) -- (13.25,2.6);
\node at (10.85,3.25) {LV/MV};

\draw [thick](12.85,3.5) -- (12.85,1.7);
\node at (13.55,1.7) {PCC};
\draw[dotted] (13.25,2.6) -- (14.15,2.6);

\fill[black!50](12.35,2.6) node (v12) {} circle (1 mm);

\draw[-latex, black!50] (12.35,2.6) -- (12.35,4.2);

\node [black!50] at (12.35,4.4) {$p_i, q_i$};

\draw  [-latex, black!50](12.65,2.3) -- (12.65,1.1);

\node [black!50] at (12.15,1.8) {$v_\mathrm{pcc}$};

\draw [-latex](5,0.7) -- (5,1.4);
\draw  (3.6,0.7) rectangle (6.4,-0.1);
\node at (5,0.3) {modulation};
\draw  (4.5,-0.8) rectangle (5.5,-1.6);
\draw (4.5,-1.6) -- (5.5,-0.8);
\node [scale=0.7] at (4.82,-0.99) {abc};
\node  [scale=0.7] at (5.2,-1.4) {dq};
\draw[-latex] (5,-0.8) -- (5,-0.1);

\draw[-latex] (7.4,-1.2) -- (5.5,-1.2);

\draw [-latex,dashed, backgroundgreen](6,-3.6) -- (5.3,-3.6) node (v15) {} -- (5.3,-1.6);
\draw [fill=backgroundgreen!15, rounded corners = 3] (6,-2.7) rectangle (7.8,-4.6);
\node at (6.9,-3.2) {current};
\node at (6.9,-3.6) {control};
\node at (6.9,-4.15) {(PI)};

\draw[dashed, backgroundgreen,-latex]  (6,-6.6)-- (4.7,-6.6)--(4.7,-1.6)  ;

\draw [rounded corners = 3, fill = backgroundgreen!15] (8.4,-5) rectangle (14,-2.3);
\draw  (8.6,-3.9) rectangle (9.4,-4.5);
\node at (9,-4.2) {PI};

\draw [fill = backgroundgreen!40]  (10.2,-3.8) rectangle (13.5,-4.6);
\node [scale=0.95] at (11.9,-4.2) {$m_i^\mathrm{vq}\hspace{-0.3mm}(s)T_\mathrm{des}^\mathrm{qv}\hspace{-0.3mm}(s)^{\hspace{-0.3mm}-\hspace{-0.3mm}1}$};

\draw[-latex] (10.2,-4.2) -- (9.4,-4.2) node (v17) {};
\draw[-latex] (15.6,-4.2) -- (13.5,-4.2);

\node at (6.2,-0.9) {$\theta_{\mathrm{pll},i}$};

\draw [-latex](9,-5.2) -- (9,-4.5);

\node at (9.5,-5.3) {$\Delta q_i$};
\node at (14.9,-3.8) {$|\Delta v_\mathrm{pcc}|$};
\draw [-latex](6.9,-1.2) node (v11) {} -- (6.9,-2) -- (6.9,-2.7);
\fill[black](2.9,3.6) circle (0.5 mm); 

\draw [-latex](6.9,-5.2) -- (6.9,-4.6);
\draw[fill = backgroundgreen!15, rounded corners = 3]  (6,-5.9) rectangle (9.4,-7.3);
\node at (7.7,-6.3) {$\mathcal{H}_\infty$ matching};
\node at (7.7,-6.8) {controller};
\draw [-latex](10.5,-6.6) -- (9.4,-6.6);

\node at (10.9,-6.6) {$i_i$};

\draw[-latex,black!50] (9.2,2.4) -- (10.2,2.4);
\node  [black!50] at (9.6,2.1) {$i_i$};
\draw [-latex](10.6,-7.1) -- (9.4,-7.1);
\node at (12.5,-7.1) {$p_i,q_i,\omega_{\mathrm{pll},i}, v_\mathrm{pcc}$};
\node at (6.9,-5.4) {$v_\mathrm{pcc}, i_i$};

\draw [-latex](10.5,-6.1) -- (9.4,-6.1);
\node at (11.2,-6.1) {$\theta_{\mathrm{pll},i}$};
\node [backgroundgreen,rotate=90,scale=0.75] at (4.5,-2.8) {option B};
\node  [backgroundgreen,rotate=90,scale=0.75] at (5.1,-2.8) {option A};
\draw  (7.4,-0.7) rectangle (9,-1.7);
\node at (8.2,-1.2) {PLL};
\node at (14.9,-2.8) {$\Delta \omega_{\mathrm{pll},i}$};
\draw [-latex](9.9,-1.2) -- (9,-1.2);
\node at (9.7,-0.9) {$v_\mathrm{pcc}$};
\draw[-latex] (6.9,-1.2)-- (6.9,-0.2) -- (8.2,-0.2) -- (8.2,-0.7);
\fill[black] (6.9,-1.2) {} circle (0.5 mm); ;

\draw[fill=backgroundgreen!40]  (10.2,-2.7) rectangle (13.5,-3.5);
\node [scale=0.95] at (11.9,-3.1) {$m_i^\mathrm{fp}\hspace{-0.3mm}(s)T_\mathrm{des}^\mathrm{pf}\hspace{-0.3mm}(s)^{\hspace{-0.3mm}-\hspace{-0.3mm}1}$};
\draw  (8.6,-2.8) rectangle (9.4,-3.4);
\node at (9,-3.1) {PI};
\draw[-latex] (10.2,-3.1) -- (9.4,-3.1);

\draw [-latex](9,-2.1) -- (9,-2.8);
\draw [-latex](15.6,-3.1) -- (13.5,-3.1); 
\draw[-latex] (8.6,-3.1) -- (7.8,-3.1);
\draw[-latex] (8.6,-4.2) -- (7.8,-4.2);

\node at (9.5,-2) {$\Delta p_i$};
\end{tikzpicture}}
       \vspace{-9mm}
    \caption{One-line diagram of the three-phase power converter interface of each grid-following DVPP unit $i$ (cf. \cref{tab:converter_parameters}).}
    \label{fig:converter_model_gfoll}
    \vspace{-1mm}
\end{subfigure}
    \caption{Converter interfaces of grid-forming and grid-following DVPP units. Depending on the application, there are different alternatives for measuring the local active and reactive powers $p_i$ and $q_i$ on either the converter- or the grid-side of the LV/MV transformer. In case of the former, one would have to add an additional affine control term to the matching control in order to compensate for the reactive losses associated with the dominantly inductive transformer impedance. The active power losses are typically negligible.}
    \label{fig:converter_model}
     \vspace{-4mm}
\end{figure}
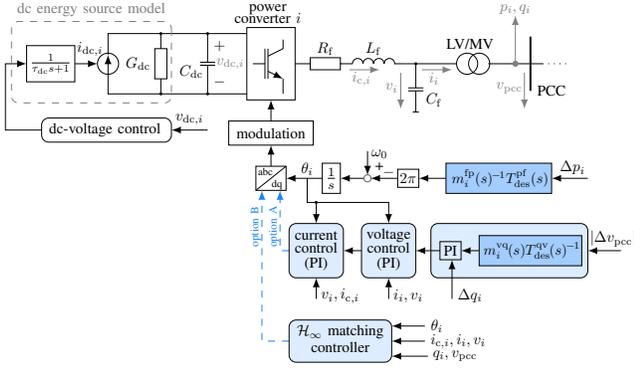
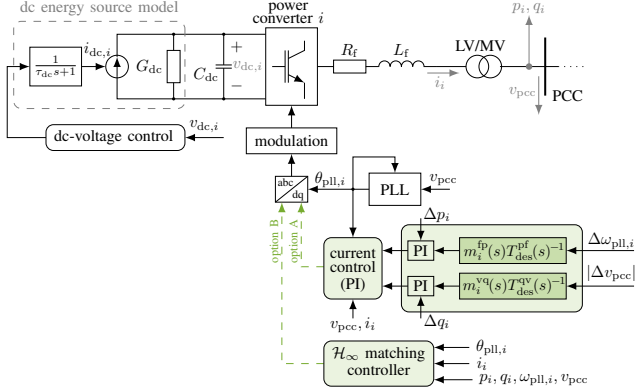

For each grid-forming and grid-following DER within the DVPP, we consider a uniform three-phase power converter interface as illustrated in \cref{fig:converter_model}. In both implementations, the grid-side converter model used for dynamic simulation represents an aggregation of multiple commercial converter modules\cite{purba2017reduced}, and is based on a state-of-the-art converter control scheme into which we have incorporated the matching control implementation for the desired frequency and voltage regulation of a grid-forming or grid-following DVPP unit, respectively (details follow subsequently).
\renewcommand{\arraystretch}{1.2}
\begin{table}[t!]\scriptsize
    \centering
           \caption{System model and DER parameters}
           \vspace{-1mm}
           \begin{subtable}{1\columnwidth}
               \caption{IEEE nine-bus system parameters \cite{anderson}}
               \vspace{-2mm}
               \begin{tabular}{c||c}
     \toprule
         Parameter & Value  \\ \hline
         System base power &  $100\,\text{MVA}$ \\
         System base voltage (ph-ph, rms)  & $230\,\text{kV}$\\
         System base frequency & $50\,\text{Hz}$ \\ \hline
         Power rating of SG1, SG2, SG3 & $250\,\text{MVA}$, $96\,\text{MVA}$, $64\,\text{MVA}$\\\hline
         Voltage rating (ph-ph, rms) of SG1, SG2, SG3 & $16.5\,\text{kV}$, $18\,\text{kV}$, $13.8\,\text{kV}$\\
         \bottomrule
    \end{tabular}
    \label{tab:9bus_parameters}
    \end{subtable}
        \begin{subtable}{1\columnwidth}
    \vspace{1.8mm}
    \caption{Converter parameters rated at $S_\mathrm{r}, v_\mathrm{r}\hspace{-0.5mm}=\hspace{-0.5mm}\sqrt{2/3}\,\text{kV}$ (ph-n, peak), $v_\mathrm{dc}^\star=3v_\mathrm{r}$}
         \vspace{-2mm}
    \begin{tabular}{c||c|c}
    \toprule
         Parameter & Symbol & Value  \\ \hline
         DC-link components & $C_\mathrm{dc}, G_\mathrm{dc}$ & $0.096\,\text{pu}, 0.05\,\text{pu}$ \\\hline
         \hspace{-3mm} $RLC$ filter components& $R_\mathrm{f}, L_\mathrm{f}, C_\mathrm{f}$ & $0.01\,\text{pu}, 0.11\,\text{pu}, 0.0942\,\text{pu}$ \\\hline
         LV/MV transformer components & $R_\mathrm{t}, L_\mathrm{t}$& $0.01\,\text{pu},\,0.1\,\text{pu}$ \\\hline
         PLL control gains & $k_\mathrm{p}^\mathrm{pll}, k_\mathrm{i}^\mathrm{pll}$  & $60, 240$ \\ \hline
         DC-voltage control gain & $k_\mathrm{dc}$ & $100$\\ \bottomrule
    \end{tabular}
     \label{tab:converter_parameters}
    \end{subtable}
    \begin{subtable}{1\columnwidth}
    \vspace{1.8mm}
    \caption{Base DVPP parameters}
         \vspace{-2mm}
    \begin{tabular}{c||c|c}
    \toprule
         Parameter & Symbol & Value  \\ \hline
         Power rating, DVPP & $S_\mathrm{dvpp,r}$ & $96\,\text{MVA}$ \\
         \hspace{-3mm} Power rating wind, $\hspace{-0.5mm}$PV, $\hspace{-0.5mm}$BESS $\hspace{-2mm}$& $\hspace{-1mm}S_\mathrm{wind,r}, S_\mathrm{pv,r}, S_\mathrm{bess,r}\hspace{-1.5mm}$ & $\hspace{-1.5mm}46\,\text{MVA}, 73\,\text{MVA}, 60\,\text{MVA}\hspace{-2mm}$ \\\hline
         DER time constants & $\tau_\mathrm{wind},\,\tau_\mathrm{pv},\,\tau_\mathrm{bess}$& $1.5\,\text{s},\,0.6\,\text{s},\,0.2\,\text{s}$ \\\hline
           $T_\mathrm{des}(s)$ parameters & $D_\mathrm{p},\,H_\mathrm{p},\,D_\mathrm{q}$ & $33.33,5.55\,\text{s}, 0.01$\\
    \bottomrule
    \end{tabular}
     \label{tab:DVPP_parameters}
    \end{subtable}
    \vspace{-3mm}
     \label{tab:system_parameters}
\end{table}
\renewcommand{\arraystretch}{1} \normalsize

Similar to\cite{tayyebi2020frequency}, for both the grid-forming and grid-following DER implementation, we assume that the dc current $i_{\mathrm{dc},i}$ of each unit $i$ is supplied by a controllable dc-current source, e.g., representing the primary-side converter of a wind power plant, a PV system, or an energy storage system. More specifically, we consider a coarse-grain model of the underlying primary source and model its response time by a first-order delay with time constant $\tau_\mathrm{dc}$, e.g., representing the resource associated dynamics, communication delays, and/or actuation delays.

The grid-side converter control is separated into two control loops for the dc and the ac side. For the sake of consistency, we employ the same dc-side control strategy for the grid-forming and grid-following DVPP units, i.e., we regulate the dc voltage through the controllable dc-current source of each DVPP unit $i$, accordingly. On the other hand, the ac-side control is implemented in a $\mathrm{dq}$-coordinate frame and includes the matching control implementation to provide the desired frequency and voltage regulation, therefore differing in case of a grid-forming or a grid-following DVPP unit, respectively. Namely, the $\mathrm{dq}$-coordinate frame of a grid-following DER is oriented via a PLL which tracks the system frequency after the LV/MV transformer, while keeping the converter synchronized with the grid voltage\cite{yazdani2010voltage}. In contrast, for a grid-forming DER, the angle reference of the $\mathrm{dq}$-coordinate frame is computed from a direct implementation of the desired local frequency control specification $m_i^\mathrm{fp}(s)^{-1}T_\mathrm{des}^\mathrm{pf}(s)$ in \eqref{eq:matching_conditions} (see \cref{fig:converter_model_gform}).

As indicated in \cref{fig:converter_model}, there are different options for the matching control implementation to provide the desired frequency and/or voltage regulation of the grid-forming and grid-following DERs as specified in \eqref{eq:matching_conditions} and \eqref{eq:matching_conditions_hybrid}, respectively. Notice that by design choice, for the grid-forming DER implementation, the matching controller only captures the voltage regulation to match the local desired behavior $m_i^\mathrm{vq}(s) T_\mathrm{des}^\mathrm{qv}(s)^{-1}$ in \eqref{eq:matching_conditions}, since the desired frequency behavior is directly incorporated when computing the reference angle of the $\mathrm{dq}$-coordinate frame. To implement the matching control, a conventional solution for both types of DERs is a direct and proportional-integral (PI)-based control implementation using a standard hierarchical converter control architecture (option A in \cref{fig:converter_model_gform} and \cref{fig:converter_model_gfoll}). This option, however, generally requires an involved tuning to ensure a smooth interaction of the cascaded PI loops\cite{wen2015analysis,pogaku2007modeling,suul2009design}, and, in case of ill-tuned control gains, often sacrifices performance or even causes instabilities\cite{suul2016impedance,huang2019grid}. Inspired by existing methods on multivariable\cite{huang2020h,kammer2018convex,chen2021generalized} and adaptive\cite{erfanmanesh2015performance,muhando11} $\mathcal{H}_\infty$ control of power converters, we thus resort to a more robust and optimal control implementation, by replacing the cascaded PI-loops with a multivariable LPV $\mathcal{H}_\infty$ state-feedback controller (option B in \cref{fig:converter_model_gform} and \cref{fig:converter_model_gfoll}). The latter is designed based on a linearized representation of the grid-forming or grid-following converter model, respectively, and then applied to the full nonlinear model of each converter type during simulations (see\cite{haberle2021control} for detailed implementation aspects). Note that for both the grid-forming and the grid-following converter interface, the multivariable $\mathcal{H}_\infty$ control design can be arbitrarily adapted from replacing only some of the inner control loops of the conventional converter control right through to capturing the entire converter control by one single controller. Further, the $\mathcal{H}_\infty$ control can also be adapted to other types of converter topologies, e.g., with more detailed primary source models. A comparison of both matching control implementations (options A and B) is given in the appendix.

\subsection{Case Study I: Synchronous Generator Replacement by Adaptive Grid-Forming DVPP Control}
\begin{figure}[t!]
    \centering

    \usetikzlibrary{arrows}
\resizebox {0.96\columnwidth} {!} {
\begin{tikzpicture}[scale=0.39, every node/.style={scale=0.62}]
	
	\draw(-16.45,30.3) -- (-8.45,30.3);
	\draw [ultra thick](-12.4,31) -- (-12.4,29.5);
	\draw [ultra thick](-18.5,31) -- (-18.5,29.5);
	
	\draw [ultra thick](-15.4,31) -- (-15.4,29.5);
	
	\draw [ultra thick](-6.3,31) -- (-6.3,29.5);
	
	\draw [ultra thick](-9.4,31) -- (-9.4,29.5);
	\draw[ultra thick] (-14.7,28.8) -- (-13.1,28.8);
	
	\draw [ultra thick](-11.7,28.8) -- (-10.1,28.8);
	\draw [ultra thick](-13.2,27) -- (-11.6,27);

	\fill[black] (-18.5,30.3)circle (0.7 mm); 
	\fill[black]  (-15.4,30.3)circle (0.7 mm); 
	\fill[black] (-12.4,30.3)circle (0.7 mm); 
	\fill[black]  (-9.4,30.3)circle (0.7 mm); 
	\fill[black] (-6.3,30.3) circle (0.7 mm);

	\fill[black] (-10.4,28.8)circle (0.7 mm); 
	\fill[black] (-11.4,28.8)circle (0.7 mm); 
	\fill[black] (-13.4,28.8)circle (0.7 mm); 
	\fill[black] (-14.4,28.8)circle (0.7 mm); 
	\fill[black] (-12.9,27)circle (0.7 mm); 
	\fill[black] (-11.9,27)circle (0.7 mm); 
	\fill[black] (-12.4,27)circle (0.7 mm);

	\node at (-18.5,31.5) {2};
	\node at (-15.4,31.5) {7};
	\node at (-12.4,31.5) {8};
	\node at (-9.4,31.5) {9};
	\node at (-6.3,31.5) {3};
	\node at (-15.2,28.8) {5};
	\node at (-9.6,28.8) {6};
	\node at (-11.1,27) {4};
	\node at (-15,26.6) {1};

	\node (v2) at (-15.4,29.8) {};
	\node at (-6.3,29.8) {};
	\draw (-15.4,29.8) -- (-14.4,29.8) -- (-14.4,28.8);
	\fill[black] (-15.4,29.8) circle (0.7 mm); 
	\draw (-13.4,28.8) -- (-13.4,28.1) -- (-12.9,27.7) -- (-12.9,27);
	\draw (-11.4,28.8) -- (-11.4,28.1) -- (-11.9,27.7) -- (-11.9,27);
	
	\draw (-12.4,27) -- (-12.4,25.4) node (v1) {}-- (-13.35,25.4);
	\draw (-9.4,29.8) -- (-10.4,29.8) -- (-10.4,28.8);
	\fill[black] (-9.4,29.8) circle (0.7 mm); 
	\draw [-latex](-12.4,29.8) -- (-11.9,29.8) -- (-11.9,29);
	\fill[black] (-12.4,29.8) circle (0.7 mm); 
	\draw [-latex](-13.9,28.8) -- (-13.9,27.6);
	\fill[black](-13.9,28.8)circle (0.7 mm); 
	\draw [-latex](-10.9,28.8) -- (-10.9,27.6);
	\fill[black](-10.9,28.8)circle (0.7 mm);

	%\draw [-latex] (-4.5,4) rectangle (-2.5,2.5);
	\draw[fill=black!20](-20,30.3) node (v3) {} circle (7 mm); 
	\draw(-17.2,30.3)  circle (3.5 mm); 
	\draw(-16.8,30.3)  circle (3.5 mm); 
	
	\draw  plot[smooth, tension=.7] coordinates {(-20.5,30.3) (-20.287,30.8) (v3) (-19.76,29.8) (-19.5,30.3)};
	
	%\node at (-3.5,3.25) {DVPP};
	\draw [fill=black!20](-5.1,30.3) node (v8) {} circle (7 mm); 
	\draw(-13.7,25.4)  circle (3.5 mm); 
	\draw(-14.1,25.4)  circle (3.5 mm); 
	\draw  plot[smooth, tension=.7] coordinates {(-5.69,30.3) (-5.4,30.8) (v8) (-4.8,29.8) (-4.53,30.3)};
	\draw(-7.7,30.3)  circle (3.5 mm); 
	\draw(-8.1,30.3)  circle (3.5 mm); 
	\draw (-19.3,30.3) -- (-17.55,30.3);
	\draw (-5.8,30.3) -- (-7.35,30.3);
	\draw (-14.45,25.4) -- (-15.7,25.4);

	\draw[fill=black!20](-16.4,25.5) node {} circle (7 mm); 
	%(-12.7,16.5)
	\draw  plot[smooth, tension=.7] coordinates {(-16.9,25.5) (-16.687,26)(-16.4,25.5)(-16.14,25) (-15.9,25.5)};
	\node at (-20,29.2) {SG 2};

	\node at (-5.2,29.2) {SG 3};

	\node at (-16.4,24.3) {SG 1};

	\draw [-latex](-9.4,30.7) -- (-10,30.7) -- (-10,31.5);

\draw  [rounded corners =3,dashed,black!50,fill=black!5](-24.6,28.5) rectangle (-20.4,24.8);
\draw [fill=wind_form!60] (-24.3,28.2) rectangle (-22.5,27.3);
\draw [fill=PV_form] (-24.3,27.1) rectangle (-22.5,26.2);
\draw [fill=battery_form!60] (-24.3,26) rectangle (-22.5,25.1);
\node at (-23.4,27.75) {wind};
\node  [scale=0.9] at (-23.4,26.65) {PV};
\node  [scale=0.9] at (-23.4,25.55) {BESS};
\draw (-20.7,26.7) node (v7) {} -- (-20.7,27.8) -- (-21.1,27.8);
\draw (-20.7,26.7) -- (-21.1,26.7) --(-20.7,26.7) node (v4) {}  -- (-20.7,25.5) -- (-21.1,25.5);
\node [color=black!50] at (-22.4,24.3) {grid-forming DVPP};
\draw[color=black!50] (-21,29.2) -- (-19.2,31.3);
\draw (-20.7,26.7) -- (-20.1,26.7) node (v5) {};
\draw[-latex] (-18.4,29.9) -- (-18,29.9) -- (-18,28.8);
\draw [dotted]  (-20.1,26.7) -- (-19.1,26.7) -- (-19.1,30) -- (-18.5,30);
\draw (-21.8,25.5)  circle (3 mm); 
\draw (-21.4,25.5)  circle (3 mm); 
\draw (-21.8,27.8)  circle (3 mm);
\draw (-21.4,27.8)  circle (3 mm);
\draw (-21.8,26.7)  circle (3 mm); 
\draw (-21.4,26.7)  circle (3 mm); 

\draw (-22.1,27.8) -- (-22.5,27.8);

\draw (-22.1,26.7) -- (-22.5,26.7);

\draw (-22.1,25.5) -- (-22.5,25.5);

\draw[ultra thick] (-15,26.2) -- (-15,24.7);

\end{tikzpicture}

}
       \vspace{-9.25mm}
    \caption{\textit{Case study I:} IEEE nine-bus system with a grid-forming DVPP replacing the synchronous generator (SG) at bus 2.}
    \label{fig:system_model_case_study_I}
    \vspace{-3mm}
\end{figure}
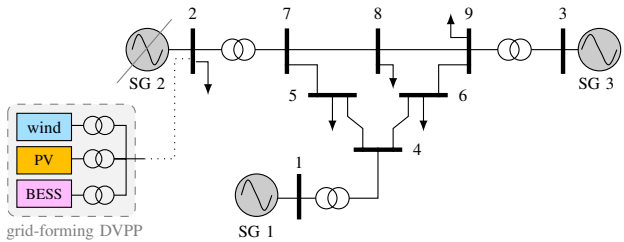

We investigate a grid-forming DVPP consisting of a wind power plant, a photovoltaic (PV) system and a battery energy storage system (BESS) that replaces the fast frequency and voltage control of SG 2 in the original nine-bus system (\cref{fig:system_model_case_study_I}, \cref{tab:DVPP_parameters}). Further, we exploit the complementary nature of wind and solar energy \cite{nehrir2011review} to compensate their fluctuations via an online adaptation of the ADPFs, while not affecting the overall DVPP response.

We want to substitute the services of SG 2, and specify a decoupled $\mathrm{p}$-$\mathrm{f}$ and $\mathrm{v}$-$\mathrm{q}$ control as
\begin{align}\label{eq:T_des_casestudy1}
    \hspace{-1mm}\begin{bmatrix}
        \Delta f_\mathrm{pcc}(s)\\ \Delta v_\mathrm{pcc}(s)
    \end{bmatrix}\hspace{-1mm}=T_\mathrm{des}(s)\hspace{-1mm}
    \begin{bmatrix}
        \Delta p_\mathrm{pcc}(s)\\ \Delta q_\mathrm{pcc}(s)\end{bmatrix}\hspace{-1mm}, \, T_\mathrm{des}(s)\hspace{-0.4mm}:=\hspace{-1mm}\begin{bmatrix}
            \tfrac{1}{H_\mathrm{p}s+D_\mathrm{p}}&0\\ 0 & D_\mathrm{q}
        \end{bmatrix}\hspace{-1mm},
\end{align}
where $H_\mathrm{p}$ and $D_\mathrm{p}$ are the virtual inertia and droop coefficients for the $\mathrm{p}$-$\mathrm{f}$ control, and $D_\mathrm{q}$ is a droop gain for the $\mathrm{q}$-$\mathrm{v}$ control. In particular, the droop gains in \eqref{eq:T_des_casestudy1} ensure both active and reactive power sharing of the DVPP with the rest of the grid.

\begin{figure}[t!]
    \vspace{-1mm}
    \centering
    \scalebox{0.53}{\includegraphics[]{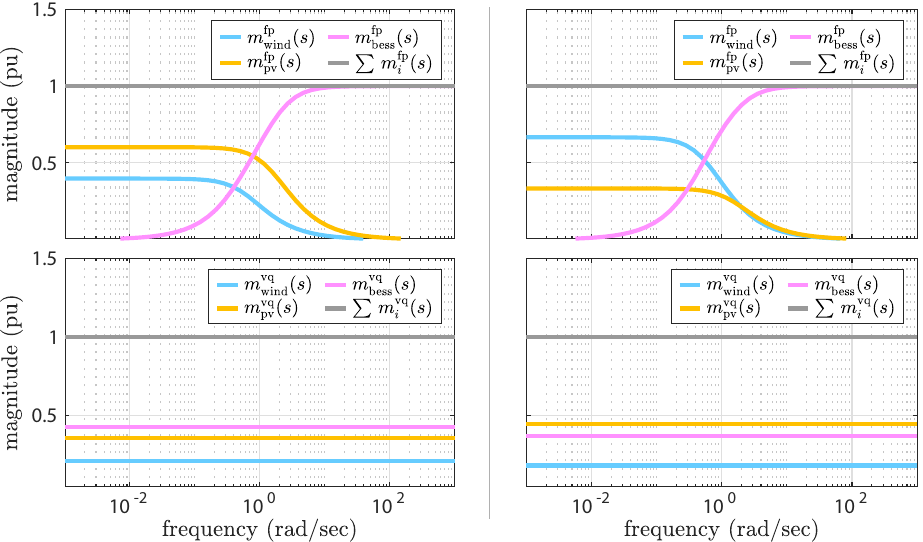}}
     \vspace{-0.25mm}
    \caption{Magnitude Bode plot snapshots of the ADPFs during nominal conditions (left) and after the PV generation capacity decrease (right).}
    \label{fig:ADPFs_cs1}
    \vspace{-2.5mm}
\end{figure}

The magnitude Bode plots of the selected ADPFs for the wind, PV and BESS during nominal power capacity conditions are shown on the left of \cref{fig:ADPFs_cs1}. In particular, following the method in \cref{sec:adaptive_divide_and_conquer}, the active power participation factors of the wind and PV are equipped with a low-pass filter behavior, where the dominant time constant for the roll-off is selected according to their resource dynamics modelled via $\tau_{\mathrm{dc}}$ in the associated converter-model in \cref{fig:converter_model_gform}. Moreover, the low-pass filter dc gains are chosen proportionately to the nominal active power capacity limit of the wind and PV, respectively. In order to satisfy the participation condition in \cref{eq:participation_conditions}, the active power ADPF of the BESS is specified to follow a high-pass filter behavior, intended to provide regulation on shorter time scales. Recall that the dc-voltage dynamics of each DER converter are controlled via the primary energy source, e.g., the wind, PV or BESS. In this regard, a careful selection of the active power ADPFs is very important, in order to ensure that the active power ac-side control of each power converter is designed such that a secure operation of the dc-voltage control within the response time and capacity limitations of the  primary energy source can be guaranteed. Namely, hitting the latter can cause a violation of the dc-voltage limits and result in a tripping of the converter.

Since the reactive power capability of the converters is independent of the dc-source technology, we select static ADPFs for the reactive power injection of all three DVPP units. In particular, the ADPF gains are selected proportionately to the reactive power capacity limit of each unit, which, in turn, is related to the active power capacity limit of the respective dc source via the PQ capability curve of the converter (see\cite{haberle2021control}).

Finally, for the local reference models of the $\mathrm{q}$-$\mathrm{v}$ control in \cref{eq:matching_conditions}, we augment the inverse transfer function ${T_\mathrm{des}^\mathrm{qv}}(s)^{-1}$ with a low-pass filter to improve the closed-loop performance by reducing the the sensitivity to high-frequency noise.

\begin{figure}[t!]
    \centering
    \scalebox{0.53}{\includegraphics[]{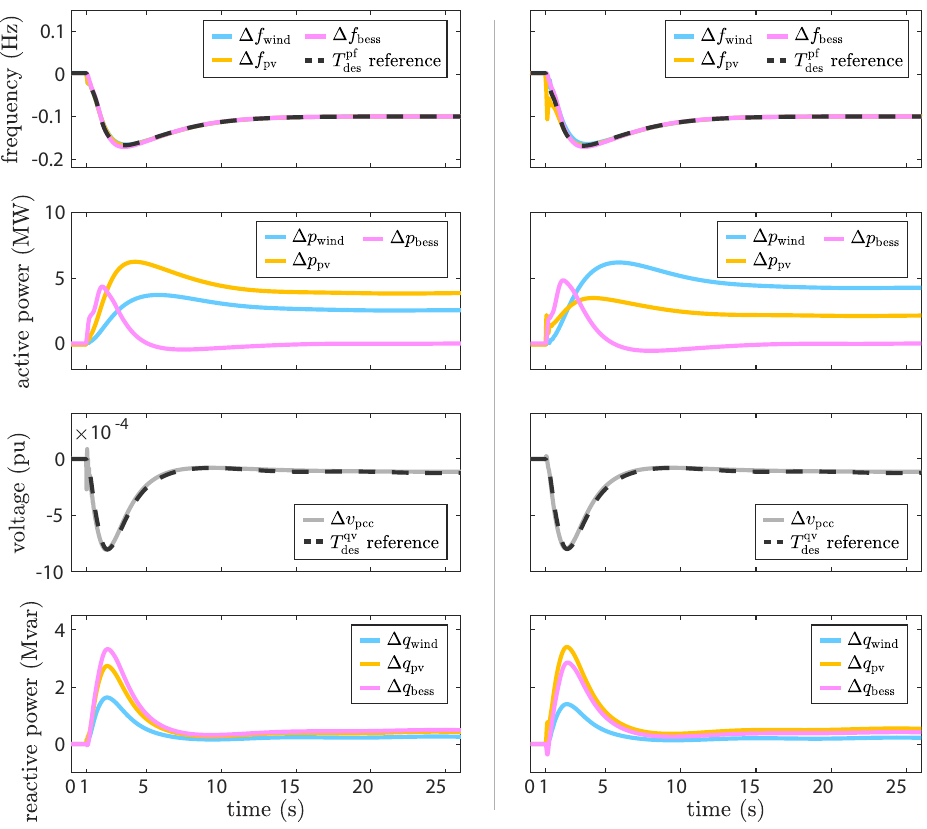}}
    \vspace{-0.75mm}
    \caption{System response of the grid-forming DVPP in case study I during a load increase (left) and a PV generation capacity decrease (right) at bus 2.}
    \vspace{-3mm}
    \label{fig:system_response_csI}
\end{figure}

We first simulate a 28 MW load increase at bus 2 and investigate the PCC's frequency and voltage magnitude response during nominal power capacity conditions. The wind and PV are operated under deloaded conditions with respect to their maximum power point, allowing them to put an active power reserve aside for participating in frequency regulation\cite{dreidy2017inertia}. Moreover, since we focus on time scales of seconds, the impact of the state of charge of the BESS is neglected. From the left of \cref{fig:system_response_csI}, it is apparent how the aggregate DVPP exhibits an accurate matching of the desired \textit{synchronized} frequency and voltage magnitude at the PCC (dashed lines), {while the individual DVPP units contribute according to their selected nominal ADPFs on the left of \cref{fig:ADPFs_cs1}}. Notice also that the coherency assumption \cref{eq:coherent_dynamics_PCC} is nearly perfectly satisfied. 

To investigate the online adaptability of the ADPFs during capacity fluctuations of weather-driven DERs, we simulate a decrease of the PV active power capacity limit {(e.g., caused by a cloud)}. This changes the PV active power set point and thus induces an equivalent active power generation deficiency of 28 MW as during the previous load increase at bus 2. By comparing the aggregate DVPP response during the PV generation decrease (\cref{fig:system_response_csI}, right) and the load increase (\cref{fig:system_response_csI}, left), we can see how the overall DVPP response remains nearly unaffected. In particular, the wind and BESS ADPFs are adapted online to compensate for the reduced active power DVPP control provided by the PV power plant (cf. ADPFs on the top left and right of \cref{fig:ADPFs_cs1}). On the other hand, while the reduced active power capacity limit decreases the PV contribution in the $\mathrm{p}$-$\mathrm{f}$ control of the DVPP, its reactive power contribution in the $\mathrm{q}$-$\mathrm{v}$ control of the DVPP is increased (cf. ADPFs on the bottom left and right in \cref{fig:ADPFs_cs1}). The reason is the increased reactive power capacity limit according to the P-Q capability curve of the PV converter\cite{haberle2021control}.
\begin{figure}[t!]
    \centering
    \vspace{-1mm}
    \usetikzlibrary{arrows}
\resizebox {0.96\columnwidth} {!} {
\begin{tikzpicture}[scale=0.39, every node/.style={scale=0.62}]

	\draw(-16.45,30.3) -- (-8.45,30.3);
	\draw [ultra thick](-12.4,31) -- (-12.4,29.5);
	\draw [ultra thick](-18.5,31) -- (-18.5,29.5);
	
	\draw [ultra thick](-15.4,31) -- (-15.4,29.5);
	
	\draw [ultra thick](-6.3,31) -- (-6.3,29.5);
	
	\draw [ultra thick](-9.4,31) -- (-9.4,29.5);
	\draw[ultra thick] (-14.7,28.8) -- (-13.1,28.8);
	
	\draw [ultra thick](-11.7,28.8) -- (-10.1,28.8);
	\draw [ultra thick](-13.2,27) -- (-11.6,27);

	\fill[black] (-18.5,30.3)circle (0.7 mm); 
	\fill[black]  (-15.4,30.3)circle (0.7 mm); 
	\fill[black] (-12.4,30.3)circle (0.7 mm); 
	\fill[black]  (-9.4,30.3)circle (0.7 mm); 
	\fill[black] (-6.3,30.3) circle (0.7 mm);

	\fill[black] (-10.4,28.8)circle (0.7 mm); 
	\fill[black] (-11.4,28.8)circle (0.7 mm); 
	\fill[black] (-13.4,28.8)circle (0.7 mm); 
	\fill[black] (-14.4,28.8)circle (0.7 mm); 
	\fill[black] (-12.9,27)circle (0.7 mm); 
	\fill[black] (-11.9,27)circle (0.7 mm); 
	\fill[black] (-12.4,27)circle (0.7 mm);

	\node at (-18.5,31.5) {2};
	\node at (-15.4,31.5) {7};
	\node at (-12.4,31.5) {8};
	\node at (-9.4,31.5) {9};
	\node at (-6.3,31.5) {3};
	\node at (-15.2,28.8) {5};
	\node at (-9.6,28.8) {6};
	\node at (-11.1,27) {4};
	\node at (-15,26.6) {1};

	\node (v2) at (-15.4,29.8) {};
	\node at (-6.3,29.8) {};
	\draw (-15.4,29.8) -- (-14.4,29.8) -- (-14.4,28.8);
	\fill[black] (-15.4,29.8) circle (0.7 mm); 
	\draw (-13.4,28.8) -- (-13.4,28.1) -- (-12.9,27.7) -- (-12.9,27);
	\draw (-11.4,28.8) -- (-11.4,28.1) -- (-11.9,27.7) -- (-11.9,27);
	
	\draw (-12.4,27) -- (-12.4,25.4) node (v1) {}-- (-13.35,25.4);
	\draw (-9.4,29.8) -- (-10.4,29.8) -- (-10.4,28.8);
	\fill[black] (-9.4,29.8) circle (0.7 mm); 
	\draw [-latex](-12.4,29.8) -- (-11.9,29.8) -- (-11.9,29);
	\fill[black] (-12.4,29.8) circle (0.7 mm); 
	\draw [-latex](-13.9,28.8) -- (-13.9,27.6);
	\fill[black](-13.9,28.8)circle (0.7 mm); 
	\draw [-latex](-10.9,28.8) -- (-10.9,27.6);
	\fill[black](-10.9,28.8)circle (0.7 mm);

	%\draw [-latex] (-4.5,4) rectangle (-2.5,2.5);

	\draw(-17.2,30.3)  circle (3.5 mm); 
	\draw(-16.8,30.3)  circle (3.5 mm);

	%\node at (-3.5,3.25) {DVPP};
	\draw [fill=black!20](-5.1,30.3) node (v8) {} circle (7 mm); 
	\draw(-13.7,25.4)  circle (3.5 mm); 
	\draw(-14.1,25.4)  circle (3.5 mm); 
	\draw  plot[smooth, tension=.7] coordinates {(-5.69,30.3) (-5.4,30.8) (v8) (-4.8,29.8) (-4.53,30.3)};
	\draw(-7.7,30.3)  circle (3.5 mm); 
	\draw(-8.1,30.3)  circle (3.5 mm); 
	\draw (-19.3,30.3) -- (-17.55,30.3);
	\draw (-5.8,30.3) -- (-7.35,30.3);
	\draw (-14.45,25.4) -- (-15.7,25.4);

	\draw[fill=black!20](-16.4,25.5) node {} circle (7 mm); 
	%(-12.7,16.5)
	\draw  plot[smooth, tension=.7] coordinates {(-16.9,25.5) (-16.687,26)(-16.4,25.5)(-16.14,25) (-15.9,25.5)};

	\node at (-5.2,29.2) {SG 3};

	\node at (-16.4,24.3) {SG 1};

	\draw [-latex](-9.4,30.7) -- (-10,30.7) -- (-10,31.5);

\draw[ultra thick] (-15,26.2) -- (-15,24.7);

\draw  [rounded corners =3,dashed,black!50,fill=black!5](-23.7,31.8) rectangle (-19.5,24.3);

\draw [fill=PV_foll] (-23.4,28) rectangle (-21.6,27);
\draw [fill=battery_form!60] (-23.4,26.8) rectangle (-21.6,25.8);
\draw [fill=battery_foll!60] (-23.4,25.6) rectangle (-21.6,24.6);
\node  [scale=0.9] at (-22.5,27.75) {PV};
\node [scale=0.75] at (-22.5,27.25) {(foll)};
\node  [scale=0.9] at (-22.5,26.55) {BESS};
\node  [scale=0.75] at (-22.5,26.05) {(form)};
\node  [scale=0.9] at (-22.5,25.35) {BESS};
\node  [scale=0.75] at (-22.5,24.85) {(foll)};
\draw (-19.8,26.3) node (v7) {} -- (-19.8,27.5) node (v6) {} -- (-20.2,27.5);
\draw (-19.8,26.3) -- (-20.2,26.3) --(-19.8,26.3) node (v4) {}  -- (-19.8,25) -- (-20.2,25);
\node [color=black!50] at (-21.5,23.8) {hybrid DVPP};
\draw (-20.9,25)  circle (3 mm); 
\draw (-20.5,25)  circle (3 mm); 
\draw (-20.9,27.5)  circle (3 mm);
\draw (-20.5,27.5)  circle (3 mm);
\draw (-20.9,26.3)  circle (3 mm); 
\draw (-20.5,26.3)  circle (3 mm); 

\draw (-21.2,27.5) -- (-21.6,27.5);

\draw (-21.2,26.3) -- (-21.6,26.3);
\draw[-latex] (-18.4,29.9) -- (-18,29.9) -- (-18,28.8);
\draw (-21.2,25) -- (-21.6,25);

\draw [fill=wind_form!60] (-23.4,31.6) rectangle (-21.6,30.6);
\draw [fill=wind_foll!60] (-23.4,30.4) rectangle (-21.6,29.4);
\draw [fill=PV_form] (-23.4,29.2) rectangle (-21.6,28.2);
\node at (-22.5,31.35) {wind};
\node  [scale=0.75] at (-22.5,30.85) {(form)};
\node at (-22.5,30.15) {wind};
\node [scale=0.75] at (-22.5,29.65) {(foll)};
\node  [scale=0.9] at (-22.5,28.95) {PV};
\node  [scale=0.75] at (-22.5,28.45) {(form)};
\draw (-19.8,29.9) node (v7) {} -- (-19.8,31.1) -- (-20.2,31.1);
\draw (-19.8,29.9) -- (-20.2,29.9) --(-19.8,29.9) node (v4) {}  -- (-19.8,28.7) node (v3) {} -- (-20.2,28.7);
\draw (-20.9,28.7)  circle (3 mm); 
\draw (-20.5,28.7)  circle (3 mm); 
\draw (-20.9,31.1)  circle (3 mm);
\draw (-20.5,31.1)  circle (3 mm);
\draw (-20.9,29.9)  circle (3 mm); 
\draw (-20.5,29.9)  circle (3 mm); 

\draw (-21.2,31.1) -- (-21.6,31.1);

\draw (-21.2,29.9) -- (-21.6,29.9);

\draw (-21.2,28.7) -- (-21.6,28.7);

\draw (-19.8,30.3) -- (-19.2,30.3) node (v5) {};

\draw (-19.8,27.5) --  (-19.8,28.7);
\end{tikzpicture}

}
       \vspace{-10.2mm}
    \caption{\textit{Case study II:} IEEE nine-bus system with a hybrid DVPP at bus 2.}
    \label{fig:system_model_case_study_II}
    \vspace{-3mm}
\end{figure}
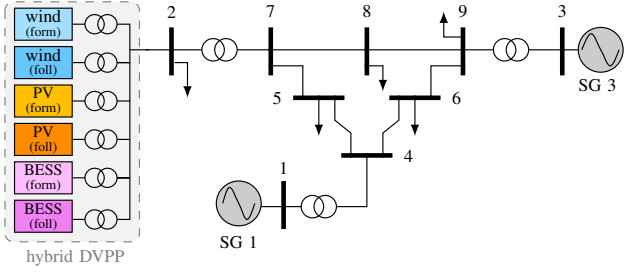
\begin{figure}[b!]
\vspace{-3mm}
    \centering
    \scalebox{0.56}{\includegraphics[]{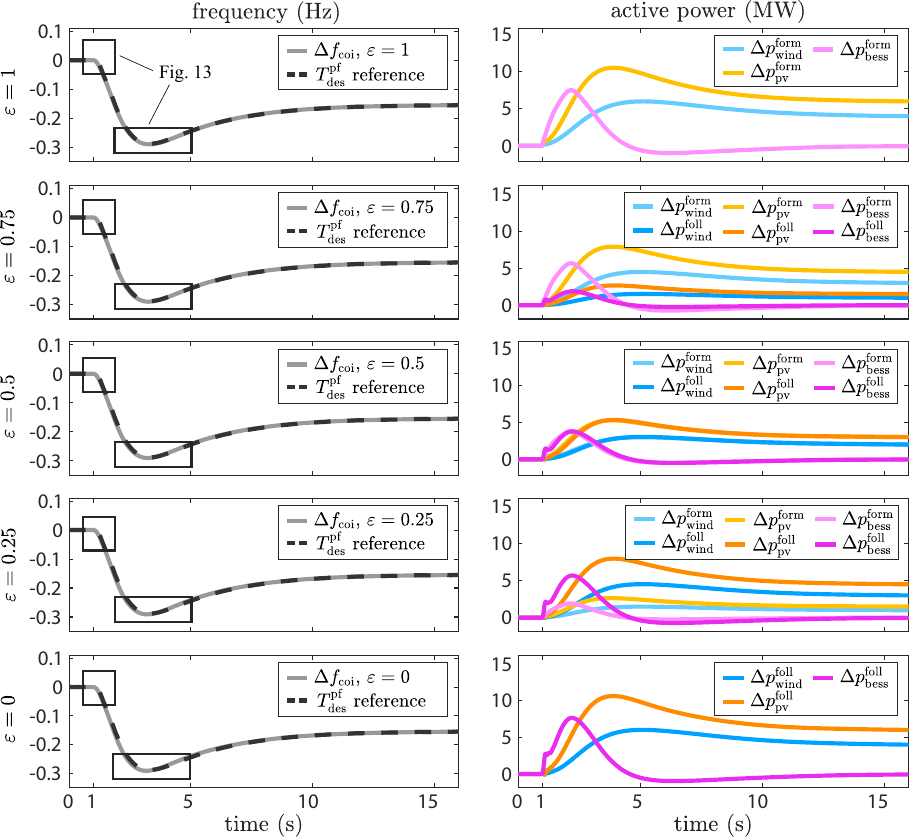}}
    \vspace{-1.5mm}
    \caption{Hybrid DVPP response in case study II for different $\varepsilon$ during a load increase at bus 2.}
    \label{fig:hybrid_DVPP_load}
    \vspace{-1mm}
\end{figure} 

\subsection{Case Study II: Investigation of Hybrid DVPP Performance}
We now extend the grid-forming base DVPP of case study~I to a hybrid DER aggregation, by splitting each DER type (wind, PV, BESS) into one grid-forming and one grid-following unit. As illustrated in \cref{fig:system_model_case_study_II}, this results in a hybrid DVPP composed of six converter-interfaced units. For the same $\mathrm{p}$-$\mathrm{f}$ and $\mathrm{v}$-$\mathrm{q}$ control specification as in \cref{eq:T_des_casestudy1}, we aim to investigate the hybrid DVPP performance during both a load increase as well as an outage of SG 1 and SG 3 in the grid, while varying the share of grid-forming and -following units. 

To do so, we start by considering the total power rating of each DER type, given by $S_\mathrm{wind,r}, S_\mathrm{pv,r}$ and $S_\mathrm{bess,r}$, respectively (\cref{tab:DVPP_parameters}), {which we split up} among the grid-forming and -following units using convex combinations as
\begin{align}
\begin{split}
    S_\mathrm{wind,r}^\mathrm{foll} = (1-\varepsilon)\cdot S_\mathrm{wind,r} &\text{ and }
    S_\mathrm{wind,r}^\mathrm{form} = \varepsilon \cdot S_\mathrm{wind,r},\\
    S_\mathrm{pv,r}^\mathrm{foll} = (1-\varepsilon)\cdot S_\mathrm{pv,r} &\text{ and }
    S_\mathrm{pv,r}^\mathrm{form} = \varepsilon \cdot S_\mathrm{pv,r},\\
    S_\mathrm{bess,r}^\mathrm{foll} = (1-\varepsilon)\cdot S_\mathrm{bess,r} &\text{ and }
    S_\mathrm{bess,r}^\mathrm{form} = \varepsilon \cdot S_\mathrm{bess,r},
\end{split}
\end{align}
where $\varepsilon\in\{0, 0.25, 0.5, 0.75, 1\}$ specifies the share of grid-forming units. In this regard, we also split the participation factors (\cref{fig:ADPFs_cs1}, left) of each DER type into two parts, i.e., 
\begin{align}
    m_i^k(s) = \underset{=:m_i^{k,\mathrm{form}}(s)}{\underbrace{\varepsilon \cdot m_i^k(s)}} + \underset{=:m_i^{k,\mathrm{foll}}(s)}{\underbrace{(1-\varepsilon) \cdot m_i^k(s)}},
\end{align}
where $i\hspace{-0.6mm}\in\hspace{-0.6mm}\{\mathrm{wind,pv,bess}\}$ and $k\hspace{-0.6mm}\in\hspace{-0.6mm}\{\mathrm{fp,vq}\}$. The local reference models specified for the grid-forming and -following units are obtained as described in \cref{sec:hybrid_DVPP}. Note that we consider nominal operating conditions, i.e., there is no temporal variation of any weather-dependent DER. Further, we report our simulation results only for the frequency response of the hybrid DVPP. However, similar observations can be made for the voltage magnitude response, accordingly. 
\begin{figure}[t!]
    \centering
    \vspace{-1mm}
    \scalebox{0.535}{\includegraphics[]{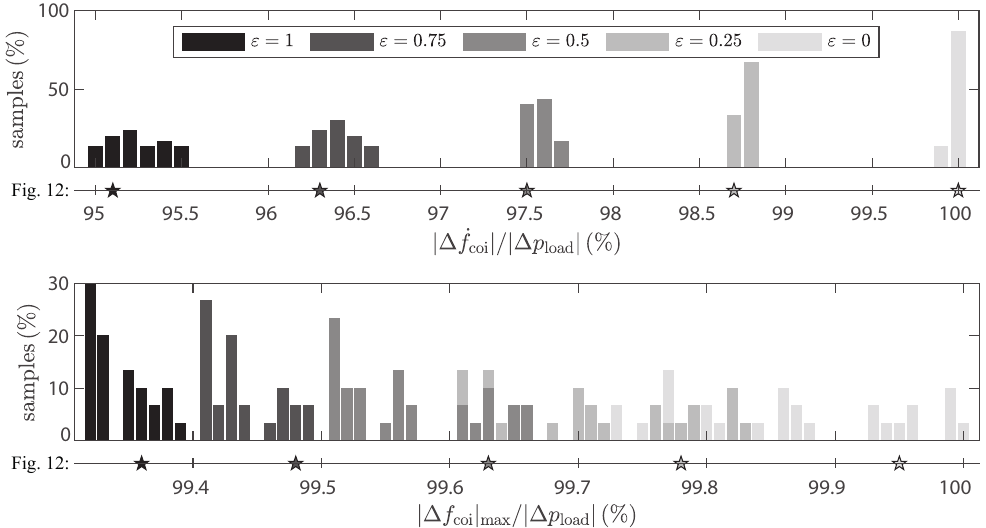}}
    \vspace{-5mm}
    \caption{Normalized distributions of the RoCoF and the nadir of the COI frequency for load disturbances $\Delta p_\mathrm{load}$ ranging from 10 MW to 50 MW at bus 2. The RoCoF $|\Delta \dot{f}_\mathrm{coi}|/|\Delta p_\mathrm{load}|$ and nadir $|\Delta {f}_\mathrm{coi}|_\mathrm{max}/|\Delta p_\mathrm{load}|$ are normalized by the maximum obtained value, respectively. The star symbols indicate the metrics of the time-domain simulations shown in \cref{fig:hybrid_DVPP_load}.}
    \vspace{-2mm}
    \label{fig:nadir_rocof_case_study_II}
\end{figure}
\begin{figure}[b!]
    \centering
    \vspace{-3mm}
    \scalebox{0.56}{\includegraphics[]{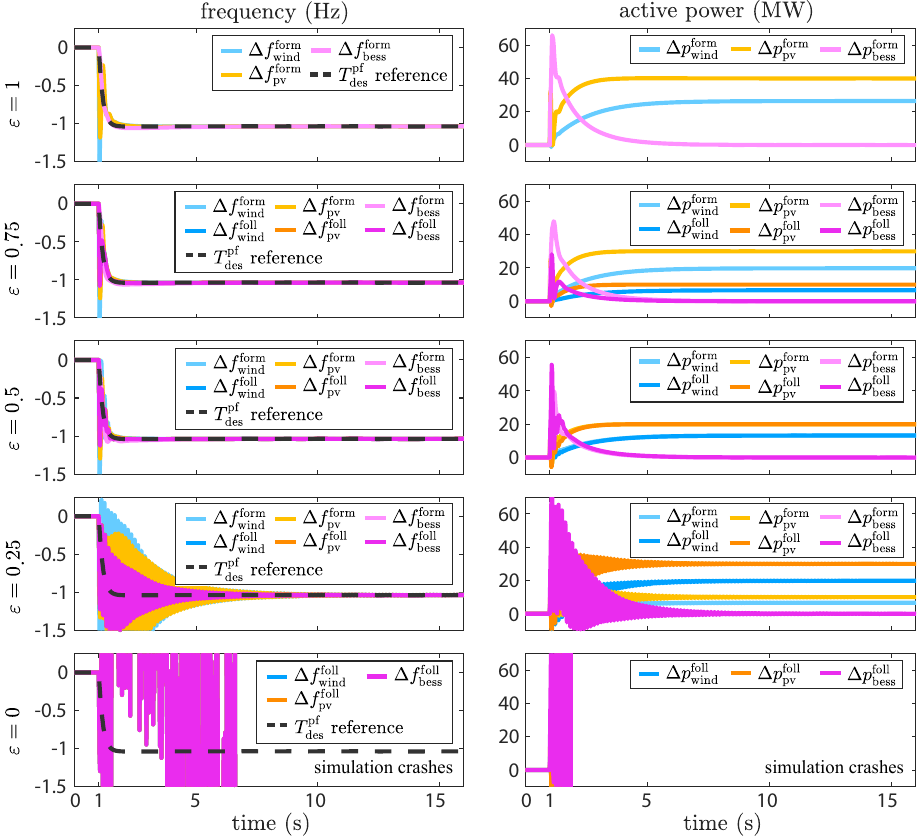}}
    \vspace{-1.5mm}
    \caption{Hybrid DVPP response in case study II for different $\varepsilon$ during the outage of all synchronous generators in the grid. For this simulation, we have slightly increased the DER ratings to stay within the local device limits during such a more severe disturbance.}
    \label{fig:hybrid_DVPP_SG}
    \vspace{-1mm}
\end{figure} 

\subsubsection*{Load increase} First, we demonstrate the conceptual success of the hybrid DVPP control for different values of $\varepsilon$. We impose a load increase of 42 MW at bus 2. The SGs connected to the system contribute approximately 75\% of the grid-rating, while the hybrid DVPP contributes approximately 25\% (cf. \cref{tab:system_parameters}). From the simulation results in \cref{fig:hybrid_DVPP_load}, it becomes apparent how the center-of-inertia (COI) frequency exhibits a very similar response behavior for the different values of $\varepsilon$ of the hybrid DVPP. In particular, the matching accuracy of the desired frequency response (dashed lines) proves to be nearly identical for all shares $\varepsilon$ of grid-forming units. However, when investigating the nadir and the rate-of-change-of-frequency (RoCoF) of the COI frequency, minor differences can be observed. As illustrated by the Monte Carlo simulations for different load disturbances in \cref{fig:nadir_rocof_case_study_II}, larger values of $\varepsilon$ generally result in a slightly lower nadir and RoCoF. Such observations are usually attributed to the fast inherent voltage source response behavior of the \textit{individual} grid-forming units \cite{osmoseh2020}. Nevertheless, the differences are marginal and of minor impact to the overall matching accuracy of the DVPP.

\subsubsection*{Outage of Synchronous Generators} For the same DVPP specification and participation factors as before, we investigate the hybrid DVPP performance during a more severe disturbance, namely an outage of the two remaining SGs in the nine-bus system. The simulation results are shown in \cref{fig:hybrid_DVPP_SG}. For decreasing values of $\varepsilon$, the associated loss in stiffness of the system frequency deteriorates the stability of the grid-following DER controls, and causes oscillatory interactions with the grid-forming units. In particular, after the SG outage for $\varepsilon=0$, there is no voltage source connected to the system anymore, such that the grid-following units lose synchronism. 
 
\begin{figure}[t!]
    \centering
    \vspace{-1mm}
    \usetikzlibrary{arrows}
\resizebox {0.96\columnwidth} {!} {
\begin{tikzpicture}[scale=0.39, every node/.style={scale=0.62}]
	
	\draw(-16.45,30.3) -- (-8.45,30.3);
	\draw [ultra thick](-12.4,31) -- (-12.4,29.5);
	\draw [ultra thick](-18.5,31) -- (-18.5,29.5);
	
	\draw [ultra thick](-15.4,31) -- (-15.4,29.5);
	
	\draw [ultra thick](-6.3,31) -- (-6.3,29.5);
	
	\draw [ultra thick](-9.4,31) -- (-9.4,29.5);
	\draw[ultra thick] (-14.7,28.8) -- (-13.1,28.8);
	
	\draw [ultra thick](-11.7,28.8) -- (-10.1,28.8);
	\draw [ultra thick](-13.2,27) -- (-11.6,27);

	\fill[black] (-18.5,30.3)circle (0.7 mm); 
	\fill[black]  (-15.4,30.3)circle (0.7 mm); 
	\fill[black] (-12.4,30.3)circle (0.7 mm); 
	\fill[black]  (-9.4,30.3)circle (0.7 mm); 
	\fill[black] (-6.3,30.3) circle (0.7 mm);

	\fill[black] (-10.4,28.8)circle (0.7 mm); 
	\fill[black] (-11.4,28.8)circle (0.7 mm); 
	\fill[black] (-13.4,28.8)circle (0.7 mm); 
	\fill[black] (-14.4,28.8)circle (0.7 mm); 
	\fill[black] (-12.9,27)circle (0.7 mm); 
	\fill[black] (-11.9,27)circle (0.7 mm); 
	\fill[black] (-12.9,27)circle (0.7 mm);

	\node at (-18.5,31.5) {2};
	\node at (-15.4,31.5) {7};
	\node at (-12.4,31.5) {8};
	\node at (-9.4,31.5) {9};
	\node at (-6.3,31.5) {3};
	\node at (-15.2,28.8) {5};
	\node at (-9.6,28.8) {6};
	\node at (-11.1,27) {4};
	\node at (-15.45,27.1) {1};

	\node (v2) at (-15.4,29.8) {};
	\node at (-6.3,29.8) {};
	\draw (-15.4,29.8) -- (-14.4,29.8) -- (-14.4,28.8);
	\fill[black] (-15.4,29.8) circle (0.7 mm); 
	\draw (-13.4,28.8) -- (-13.4,28.1) -- (-12.9,27.7) -- (-12.9,27);
	\draw (-11.4,28.8) -- (-11.4,28.1) -- (-11.9,27.7) -- (-11.9,27) node (v9) {};
	
	\draw (-12.9,27) -- (-12.9,25.9) node (v1) {}-- (-13.8,25.9);
	\draw (-9.4,29.8) -- (-10.4,29.8) -- (-10.4,28.8) node (v6) {};
	\fill[black] (-9.4,29.8) circle (0.7 mm); 
	\draw [-latex](-12.4,29.8) -- (-11.9,29.8) -- (-11.9,29);
	\fill[black] (-12.4,29.8) circle (0.7 mm); 
	\draw [-latex](-13.9,28.8) -- (-13.9,27.6);
	\fill[black](-13.9,28.8)circle (0.7 mm); 
	\draw [-latex](-12.4,27) -- (-12.4,28.1);
	\fill[black](-12.4,27)circle (0.7 mm);

	%\draw [-latex] (-4.5,4) rectangle (-2.5,2.5);
	\draw[fill=black!20](-20,30.3) node (v3) {} circle (7 mm); 
	\draw(-17.2,30.3)  circle (3.5 mm); 
	\draw(-16.8,30.3)  circle (3.5 mm); 
	
	\draw  plot[smooth, tension=.7] coordinates {(-20.5,30.3) (-20.287,30.8) (v3) (-19.76,29.8) (-19.5,30.3)};
	
	%\node at (-3.5,3.25) {DVPP};
	\draw [fill=black!20](-5.1,30.3) node (v8) {} circle (7 mm); 
	\draw(-14.15,25.9)  circle (3.5 mm); 
	\draw(-7.1,25.9)  circle (3.5 mm); 
	\draw  plot[smooth, tension=.7] coordinates {(-5.69,30.3) (-5.4,30.8) (v8) (-4.8,29.8) (-4.53,30.3)};
	\draw(-7.7,30.3)  circle (3.5 mm); 
	\draw(-8.1,30.3)  circle (3.5 mm); 
	\draw (-19.3,30.3) -- (-17.55,30.3);
	\draw (-5.8,30.3) -- (-7.35,30.3);
	\draw (-14.9,25.9) -- (-16.15,25.9);
		\draw(-14.55,25.9)  circle (3.5 mm);

	\draw[fill=black!20](-16.85,26) node {} circle (7 mm); 
	%(-12.7,16.5)
	\draw  plot[smooth, tension=.7] coordinates {(-17.35,26) (-17.137,26.5)(-16.85,26)(-16.59,25.5) (-16.35,26)};
	\node at (-20,29.2) {SG 2};

	\node at (-5.2,29.2) {SG 3};

	\node at (-18.4,26) {SG 1};

	\draw [-latex](-9.4,30.7) -- (-10,30.7) -- (-10,31.5);

\draw  [rounded corners =3,dashed,black!50,fill=black!5](-16.6,24.9) rectangle (-2.3,19);
\draw [fill=wind_form!60] (-16.2,22.2) rectangle (-14.4,21.3);
\draw [fill=PV_form] (-4.5,20.3) rectangle (-2.7,19.4);
\draw [fill=battery_form!60] (-13.4,20.2) rectangle (-11.6,19.3);
\node at (-15.3,21.75) {wind};
\node  [scale=0.9] at (-3.6,19.85) {PV};
\node  [scale=0.9] at (-12.5,19.75) {BESS};

\node [color=black!50] at (-17.6,19.9) {DVPP};
\node [color=black!50] at (-17.4,19.3) {area};

\draw (-10.9,19.7)  circle (3 mm); 
\draw (-10.5,19.7)  circle (3 mm); 
\draw (-13.7,21.8)  circle (3 mm);
\draw (-13.3,21.8)  circle (3 mm);
\draw (-5.5,19.8)  circle (3 mm); 
\draw (-5.1,19.8)  circle (3 mm); 

\draw (-14,21.8) -- (-14.4,21.8);

\draw (-4.8,19.8) -- (-4.5,19.8);

\draw (-11.2,19.7) -- (-11.6,19.7);

\draw[ultra thick] (-15.45,26.7) -- (-15.45,25.2);

\draw (-10.4,28.8) -- (-10.4,27) -- (-7.1,27) -- (-7.1,26.25);
\draw (-11.9,27)-- (-11.9,26.25);
	\draw(-7.1,25.5)  circle (3.5 mm); 
		\draw(-11.9,25.5)  circle (3.5 mm); 
				\draw(-11.9,25.9)  circle (3.5 mm); 
\draw (-11.9,25.15) -- (-11.9,24.5) node (v10) {}; 
\draw (-7.1,25.15) -- (-7.1,24.5);

\draw [ultra thick](-12.7,24.5) -- (-11.1,24.5);
\draw [ultra thick] (-9.1,24.5) -- (-5.1,24.5);
\draw  [ultra thick] (-12.7,22.7) -- (-11.1,22.7);
\draw  [ultra thick] (-13.1,20.8) -- (-11.5,20.8);
\draw  [ultra thick] (-9.8,22.7) -- (-8.2,22.7);
\draw  [ultra thick] (-6.4,22.7) -- (-4.7,22.7);
\draw[ultra thick]  (-10.2,20.8) -- (-8.6,20.8);
\draw [ultra thick] (-7.4,20.8) -- (-5.8,20.8);
\draw[ultra thick]  (-5.1,20.8) -- (-3.5,20.8);
\draw (-11.9,24.5) -- (-11.9,22.7);

\draw (-12.3,22.7) node (v4) {} -- (-12.3,20.8);
\draw (-11.5,22.7) node (v11) {} -- (-11.5,21.9) -- (-9.8,21.9) -- (-9.8,20.8) node (v5) {} -- (-9.4,20.8) node (v12) {} -- (-9.4,22.7);
\draw (-8.6,22.7) node (v13) {} -- (-8.6,24.5) -- (-5.6,24.5) -- (-5.6,22.7);
\draw (-6,22.7) -- (-6,21.7) -- (-6.6,21.7) -- (-6.6,20.8) node (v14) {}; 
\draw (-5.1,22.7) node (v7) {} -- (-5.1,21.7) -- (-4.3,21.7) -- (-4.3,20.8) node (v15) {};
\node at (-10.5,25.4) {POC 2};
\node at (-5.7,25.4) {POC 1};
\node at (-10.5,24.5) {d1b};
\node at (-10.6,22.7) {d7};
\node at (-13.6,20.8) {d8};
\node at (-4.5,24.5) {d1a};
\node at (-7.7,22.7) {d5};
\node at (-4.2,22.7) {d2};
\node at (-7.8,20.8) {d3};
\node at (-10.7,20.8) {d6};
\node at (-3,20.8) {d4};
\draw (-13,21.8) -- (-12.7,21.8) --(-12.7,20.8);
\draw (-10.2,19.7) -- (-9.8,19.7) --(-9.8,20.8);
\draw(-6.2,20.8)-- (-6.2,19.8) -- (-5.8,19.8); 
\draw[-latex] (-11.5,22.7) -- (-11.5,23.3) -- (-10.4,23.3);
\draw [-latex](-11.9,20.8) -- (-11.9,21.4) -- (-10.9,21.4); 
\draw [-latex](-9,20.8) node (v16) {} -- (-9,19.6); 
\draw [-latex](-9.4,22.7) -- (-9.4,23.9); 
\draw [-latex](-7,20.8) -- (-7,19.6);
\draw [-latex](-3.9,20.8) -- (-3.9,21.3)--(-2.8,21.3);
\draw (-9,20.8) -- (-9,21.3) -- (-7,21.3) -- (-7,20.8);
\end{tikzpicture}

}
       \vspace{-8.5mm}
    \caption{\textit{Case study III:} Modified IEEE nine-bus system with MV distribution grid connected at buses 4 and 6, similar to the setup in \cite{zhong2021coordinated}. To solely study the DVPP dynamics, the MV loads are modelled as constant power loads.}
    \vspace{-3mm}
    \label{fig:system_model_case_study_III}
\end{figure}
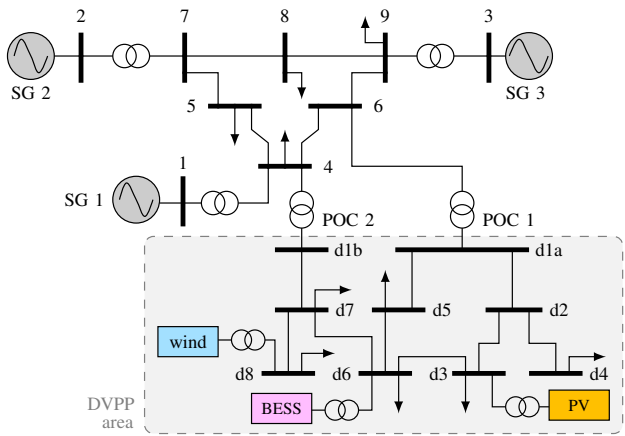

\subsection{Case Study III: Spatially Distributed DVPP}\label{sec:case_study_III}
In this case study, we investigate the performance of a spatially distributed arrangement of the base DVPP from case study~I. We restrict ourselves to a grid-forming DER aggregation, which however, could be easily generalized to a hybrid DVPP setup (cf. \cref{sec:spatially_distributed}). We consider the modified IEEE nine-bus system with a MV distribution grid connected at buses 4 and 6 (similar to the setup in \cite{zhong2021coordinated}), into which we have incorporated the DVPP units at different buses (\cref{fig:system_model_case_study_III}). As a starting point, we establish an idealized homogeneous $R/X=1$ ratio for all MV lines in the DVPP area. Moreover, we consider nominal operating conditions, i.e., there is no temporal variation of any weather-dependent DER.

\begin{figure}[t!]            
    \centering
    \scalebox{0.51}{\includegraphics[]{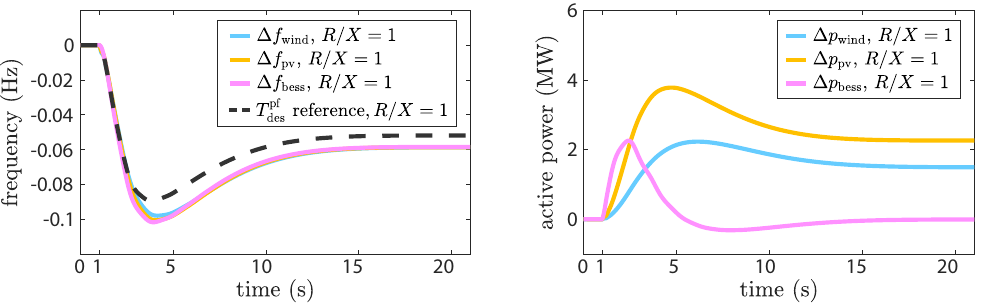}}
    \vspace{-1.5mm}
    \caption{System response of the spatially distributed DVPP in case study III for a \textit{classical $\mathrm{p}$-$\mathrm{f}$ control} during a load increase at bus 4 in the transmission system. The dashed lines indicate the reference frequency at the POCs.}
    \label{fig:classical_control_response}
\vspace{-2mm}
\end{figure}

We first investigate the DVPP behavior during a classical $\mathrm{p}$-$\mathrm{f}$ control implementation (as conventionally employed in a dominantly inductive grid). Namely, we naively employ the specification in \cref{eq:T_des_casestudy1} and the participation factors in \cref{fig:ADPFs_cs1} (left) of case study I to the spatially distributed DVPP units, without taking line impedances between the units into account. The voltage controls of the DVPP units are defined individually by means of local classical $\mathrm{q}$-$\mathrm{v}$ controls. To study the DVPP's frequency response behavior at the POCs, we employ a 20 MW load increase at bus~4 in the transmission system. As can be observed from the simulation results in \cref{fig:classical_control_response}, dis- regarding the line impedance characteristics during the DVPP control design results in a mismatch between the desired reference frequency at the POCs (dashed lines) and the actually obtained frequency response behavior. In particular, due to the active power losses of the non-inductive line impedances, the aggregated active power injections of the DVPP units do not coincide with the active power deviations obtained at the POCs. Of course, the size of the frequency mismatch strongly depends on the network topology, the size of the DVPP area, and the DER locations, and hence, might be acceptable in some cases, especially for rather small $R/X$ ratios. However, for general setups with possibly larger DVPP areas and larger $R/X$ ratios (e.g., in LV grids), the mismatch can be quite significant.

\begin{figure}[t!]
\centering
\begin{subfigure}{0.51\textwidth}
    \centering
    \vspace{-1mm}
    \scalebox{0.5}{\includegraphics[]{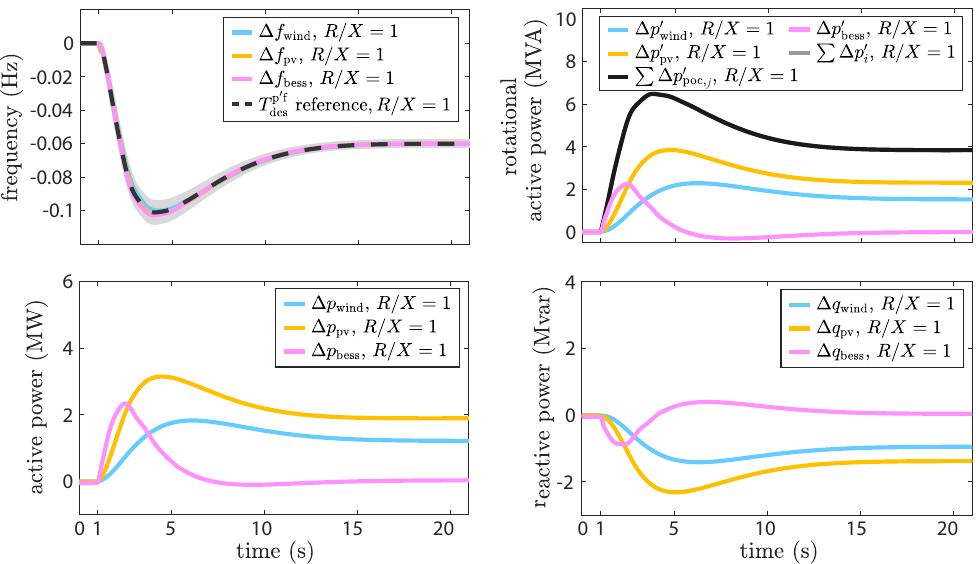}}
    \vspace{-1.75mm}
    \caption{Load increase at bus 4 in the transmission system (external disturbance).}
    \label{fig:external_disturbance}
\end{subfigure}
\begin{subfigure}{0.51\textwidth}
\vspace{2mm}
    \centering
    \scalebox{0.5}{\includegraphics[]{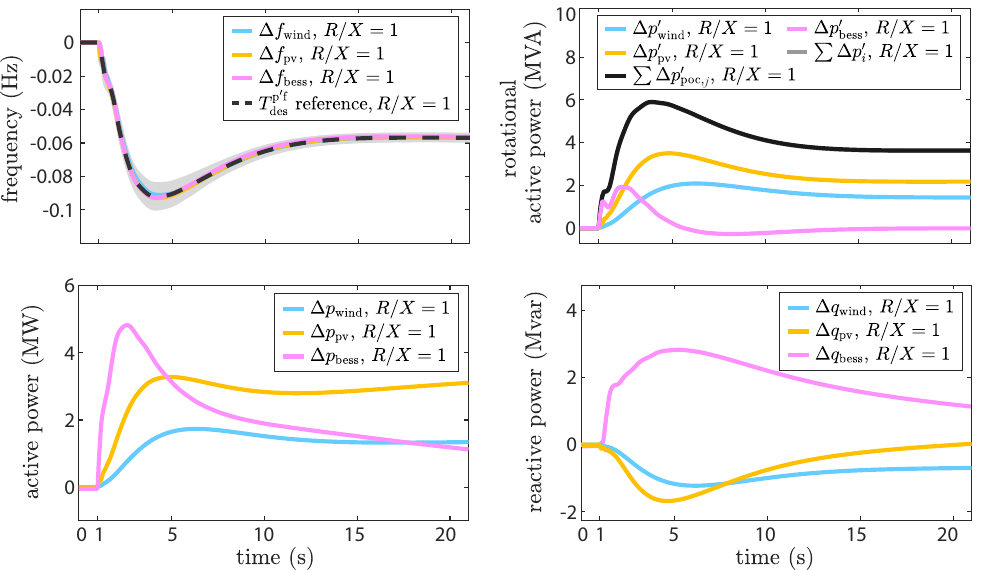}}
    \vspace{-1.75mm}
    \caption{Load increase at bus d5 in the distribution system (internal disturbance).}
    \label{fig:internal_disturbance}
\end{subfigure} 
 \caption{System responses of the spatially distributed DVPP in case study III for \textit{the proposed $\mathrm{p}'$-$\mathrm{f}$ rotational power control} during different disturbances for a homogeneous $R/X=1$ ratio of the MV lines, and random variations of each MV line between $R/X=0.4$ and $R/X=2$ (gray lines) \cite{cigre2014benchmark}.}
    \label{fig:system_response_csIII}
      \vspace{-3.3mm}
\end{figure}

In contrast, if we implement the proposed $\mathrm{p}'$-$\mathrm{f}$ rotational power control of the spatially distributed DVPP in \cref{sec:spatially_distributed}, such a mismatch can be completely eliminated by a more ef- fective coordination of the DVPP units, independent of the network topology, the size of the DVPP area, the DER locations, and the line impedances. To get an insightful comparison, for the $\mathrm{p}'$-$\mathrm{f}$ rotational power control, we employ the same specification in \cref{eq:T_des_casestudy1} and the participation factors in \cref{fig:ADPFs_cs1} (left) as for the prior decoupled $\mathrm{p}$-$\mathrm{f}$ control, allowing to reveal the analogy between ${p}'$ and ${p}$ in the frequency control setup in \cref{fig:DVPP_freq_ctrl_setup}. The local $\mathrm{q}'$-$\mathrm{v}$ controls of the DERs are chosen sufficiently conservative to satisfy the active and reactive power limitations in response time and capacity. As illustrated by the simulation results in \cref{fig:external_disturbance}, for the same 20 MW load increase at bus~4 in the transmission system, the rotational power control of the spatially distributed DVPP exhibits an accurate matching of the desired reference frequency response at the POCs (dashed lines). Moreover, note that the rotational power control of the DVPP reveals to be slightly more effective in terms of the active power injection effort of the individual DVPP units (\cref{fig:external_disturbance}, bottom left). Namely, while the frequency deviation obtained by the classical $\mathrm{p}$-$\mathrm{f}$ control of the DVPP (\cref{fig:classical_control_response}, left) is not overly dissimilar to the frequency deviation obtained by the $\mathrm{p'}$-$\mathrm{f}$ rotational DVPP control (\cref{fig:external_disturbance}, top left), the overall active power injection of the DVPP units is slightly larger in case of the $\mathrm{p}$-$\mathrm{f}$ DVPP control (\cref{fig:classical_control_response}, right). Note that these results are expected from the power electronics literature\cite{de2007voltage,yao2010design,bevrani2013intelligent,rocabert2012control}, where it is known that powers have to be rotated to make up for lossy line characteristics.

Next, after having investigated the rotational power control of the spatially distributed DVPP during an external disturbance outside of the distribution system, we now investigate an internal disturbance scenario by imposing a 20 MW load increase inside the DVPP area at bus d5 in the distribution grid. The simulation results are illustrated in \cref{fig:internal_disturbance}. Similar to the external disturbance scenario in \cref{fig:external_disturbance}, the spatially distributed DVPP again exhibits an accurate matching of the desired reference frequency response at the POCs (dashed lines). Moreover, it can be seen that although the external and internal disturbances are of equal amount of power, the DVPP responses with the individual DER behaviors are different, caused by the different locations of the disturbances.

Another important observation from \cref{fig:system_response_csIII} is the fact that the participation factors $m_i^\mathrm{fp'}(s)$ as in \cref{fig:ADPFs_cs1} (left) are imposed to limit the DERs' rotational powers $\Delta p_i'$, clearly recognizable by the obtained LPF and HPF response behaviors with different steady-state contributions. The actual power injections $\Delta p_i$ and $\Delta q_i$, however, are additionally influenced by the local $\mathrm{q}'$-$\mathrm{v}$ controls of the units, and hence deviate from the imposed behavior of the participation factors $m_i^\mathrm{fp'}(s)$ in different ways. As discussed in \cref{sec:spatially_distributed}, this phenomenon requires a careful choice of the participation factors $m_i^\mathrm{fp'}(s)$ and the local $\mathrm{q}'$-$\mathrm{v}$ controls in order to address the actual active and reactive power limitations of the DERs. 

Finally, we perform a Monte Carlo simulation to study the spatially distributed DVPP control for non-homogeneous $R/X$ ratios of the MV lines. Namely, we employ the previous two load increases (internal and external) for 24 different random combinations of $R/X$ ratios for the individual MV lines, varying in a range of $R/X=[0.4,2]$, which is considered typical for MV lines \cite{cigre2014benchmark}. The resulting frequency responses are illustrated with gray lines in \cref{fig:external_disturbance,fig:internal_disturbance}. We can see how the obtained responses are very similar to the initial response for a homogeneous $R/X=1$ ratio of all MV lines.

As a concluding remark on case studies I to III, it should be noted that the simulation results might differ for different choices of $T_\mathrm{des}(s)$, different grid topologies, or different operating points. While the presented results should be taken with caution from a quantitative perspective, they highlight the key features of our proposed DVPP control strategies. 

\subsection{Scalability Considerations}
Subsequent to the previous case studies, we would like to briefly discuss the scalability of the DVPP control setup to more complex test systems comprising multiple DVPPs or DVPPs with a larger number of DERs. In particular, since the local matching control of each DER within a DVPP is independent of the number of DVPP units, a larger number of DVPP units does not cause any scalability issues. Although the number of DERs within a DVPP does affect the selection of the ADPFs and their update strategy to ensure the participation conditions in \eqref{eq:participation_conditions} are satisfied, it needs to be done only once for each DER when setting up the DVPP ensemble (offline), and thus does not suffer from scalability issues during real-time operation. Further, the dc gains of the ADPFs are updated during real-time operation in a centralized or distributed way by means of communication, where only the communication infrastructure scales with the number of DVPP units.

A similar reasoning holds true when considering detailed aggregated converter modules within each DER unit (recall from \cref{sec:converter_model} that in our case studies, we have been using a converter model for each DER which represents an aggregation of multiple commercial converter modules\cite{purba2017reduced}). In this case, the ADPFs of each DER have to be split up proportionately (via a constant factor) to each single converter module according to the module's rating. The matching control is then employed independently for each converter module.

Beyond that, running multiple DVPPs (or DVPP areas) within the same grid is also feasible, since each DVPP is independent of the power grid to which it is connected to. In particular, while the DVPP input signal is different for different types of generators or other DVPPs in the network, the dynamic behavior of each aggregated DVPP is not affected. In other words, coexisting DVPPs do not influence the internal behavior of each other. An example case study including two independent DVPPs within the same power system is provided in our work in\cite{haberle2021control}.

\section{Conclusion}\label{sec:conclusion}

We have proposed novel grid-forming and spatially distributed DVPP control approaches with the objective to provide dynamic ancillary services across all temporal and spatial scales: from fast frequency and voltage support, and from HV transmission grids to LV distribution systems. The control methods study the nontrivial aggregation of grid-forming DERs and rely on our recently proposed adaptive divide-and-conquer strategy\cite{haberle2021control} that takes into account the local DER characteristics, and additionally handles temporal variability of weather-driven DERs. Our numerical case studies show the successful performance of our controls, and conceptually demonstrate how our DVPP control approaches can be used to facilitate the dynamic ancillary services provision by DERs.

Our currently ongoing research includes the design of multivariable and robust specifications for the desired DVPP dynamics to provide an optimal performance output in closed-loop with the power system, while simultaneously satisfying grid-code requirements and taking DVPP limitations into account. Moreover, future work should address the investigation of the DVPP performance for larger grid topologies comprising multiple DVPPs with a larger number of DERs, while being subject to different types of disturbances.

\appendix\label{sec:appendix_H_infinity}
In the following, we provide a brief comparison of the different matching control implementations for grid-forming converters as indicated in \cref{fig:converter_model_gform}, i.e., we compare the conventional controller based on cascaded PI-loops (option A) with the multivariable LPV $\mathcal{H}_\infty$ controller (option B). We restrict ourselves to grid-forming converter controls, however, a similar comparison can be also made for grid-following converters as in \cref{fig:converter_model_gfoll}.

The proposed LPV $\mathcal{H}_\infty$ state-feedback controller is designed based on a linearized representation of the converter model in \cref{fig:converter_model_gform}, and then applied to the full nonlinear model of the converter during simulations. An example on how to derive such a linearized converter model and the associated LPV $\mathcal{H}_\infty$ controller is given in \cite{haberle2021control}. Note that the multivariable $\mathcal{H}_\infty$ control design can be arbitrarily adapted from replacing only some of the inner control loops of the conventional converter control (see, e.g.,\cite{haberle2021control}) right through to capturing the entire converter control by one single controller. Further, the proposed $\mathcal{H}_\infty$ control can also be adapted to other types of converter topologies.

We compare the conventional controller with the multivariable LPV $\mathcal{H}_\infty$ controller in terms of the following aspects: 
\subsubsection{Tuning} The $\mathcal{H}_\infty$ controller is computed systematically by solving a constrained optimization problem (``single-shot'' design), which, however, needs to be tuned appropriately (cost function, weighting functions, etc.). The conventional controller is tuned loop by loop based on rich engineering experience and extensive parameter studies (e.g., through eigenvalue loci or Nyquist diagrams) to ensure a smooth interaction of the cascaded PI loops\cite{wen2015analysis,pogaku2007modeling,suul2009design}. Ill-tuned control gains can result in a poor tracking performance, high-frequency distortions, or even cause instabilities \cite{suul2016impedance,huang2019grid}.
\subsubsection{Design Objectives} Various design objectives (e.g., tracking performance or high-frequency disturbance rejection) can be incorporated into the $\mathcal{H}_\infty$ optimization problem via ellipsoidal constraints or weighting functions \cite{boyd1994linear,huang2020h,kammer2018convex,chen2021generalized}. The tuning of the conventional controller is based on iterative trial and simulation, and it does not allow for a systematic incorporation of several control objectives at once, since eigenvalue loci and Nyquist diagrams can hardly deal with multiple parameters. 
\subsubsection{Optimality} The $\mathcal{H}_\infty$ optimization problem computes an optimal controller out of a set of feasible (i.e., stabilizing) controllers. The conventional controller is within this set of feasible controllers, i.e., suboptimal, but (if well tuned) can achieve a performance close to the optimal $\mathcal{H}_\infty$ controller.
\subsubsection{Parameter Variations} The LPV $\mathcal{H}_\infty$ controller can be designed for parameter-variations (e.g., during changing power capacities), where the entire parameter space is taken into account during control design. On the other hand, the conventional controller suffers in performance during parameter variations, as the PI gains might need to be re-tuned.

\begin{figure}[t!]
    \centering
    \scalebox{0.49}{\includegraphics[]{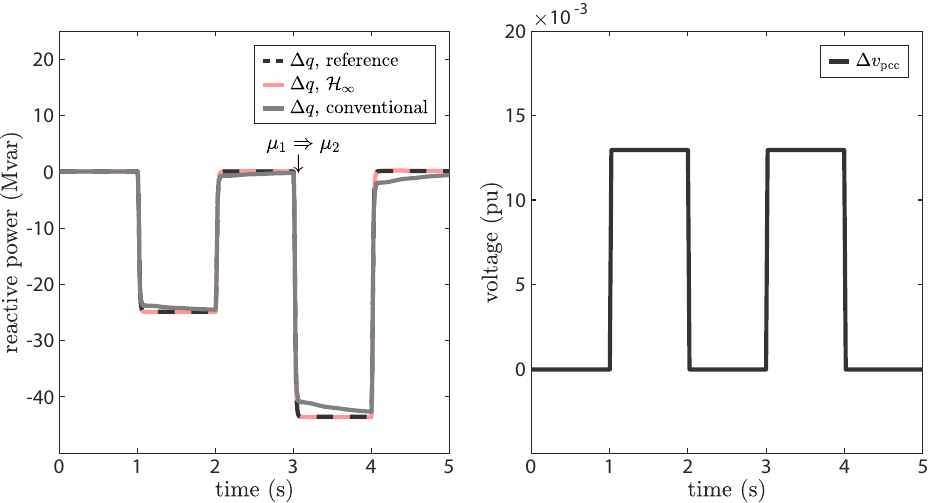}}
    \vspace{-1.25mm}
    \caption{Comparing the matching accuracy of the reference reactive power response for the conventional and the LPV $\mathcal{H}_\infty$ controller during a repeated step change of $\Delta v_\mathrm{pcc}$ and a change in $\mu(t)$ at $t=3$s.}
    \label{fig:performance_tracking}
    \vspace{-2.5mm}
\end{figure}

\begin{figure}[t!]
    \centering
    \scalebox{0.49}{\includegraphics[]{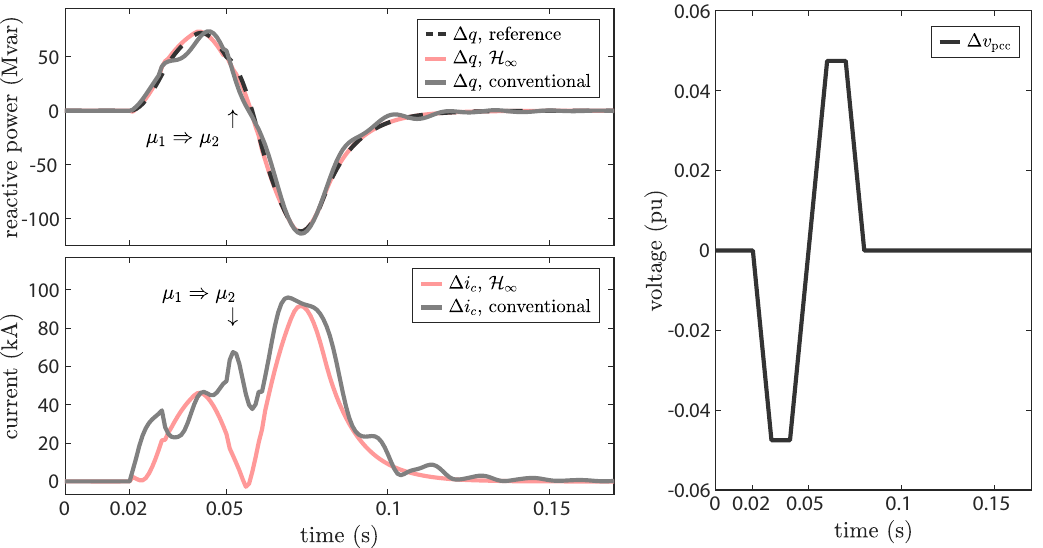}}
    \vspace{-1.25mm}
    \caption{Comparing the high frequency disturbance rejection of the converter current magnitude and the reactive power response for the conventional and the LPV $\mathcal{H}_\infty$ controller during a fast repeated, sign-inverted pulse-like change of $\Delta v_\mathrm{pcc}$ and a change in $\mu(t)$ at $t=0.05$s.}
    \label{fig:high_frequency_disturbance}
    \vspace{-3.5mm}
\end{figure}

Next, we compare the conventional matching controller with the LPV $\mathcal{H}_\infty$ matching controller in a simple numerical experiment. We consider a grid-forming converter-interfaced battery system connected to an infinite bus at the PCC. Since, by design choice, the $\mathcal{H}_\infty$ controller only captures the cascaded PI-loops of the $\mathrm{v}$-$\mathrm{q}$ control (cf. \cref{fig:converter_model_gform}), we only compare the performance of the converter's $\mathrm{v}$-$\mathrm{q}$ control during voltage magnitude variations at the PCC, and fix the frequency of the infinite bus to 50~Hz. We specify a reference model for the $\mathrm{v}$-$\mathrm{q}$ control to be matched as $\Delta q(s)=\tfrac{-100\mu(t)}{0.01s+1}\Delta v_\mathrm{pcc}(s)$, where $\mu(t)$ is a time-varying parameter to account for capacity variations. We study the performance of the two controllers with respect to their reactive power matching accuracy and their high frequency disturbance rejection capability during changing values of $\mu(t)$. The exemplary simulation results in \cref{fig:performance_tracking,fig:high_frequency_disturbance} show how the $\mathcal{H}_\infty$ controller generally exhibits a superior tracking performance and a better high frequency disturbance rejection. Notice that, depending on the tuning of the cascaded PI-loops, the conventional controller behavior can improve or worsen.

\fontdimen2\font=0.4ex
\renewcommand{\baselinestretch}{0.94}

\bibliographystyle{IEEEtran}
\bibliography{IEEEabrv,mybibliography}

\newpage
\begin{IEEEbiography}[{\includegraphics[width=1in,height=1.25in,clip,keepaspectratio]{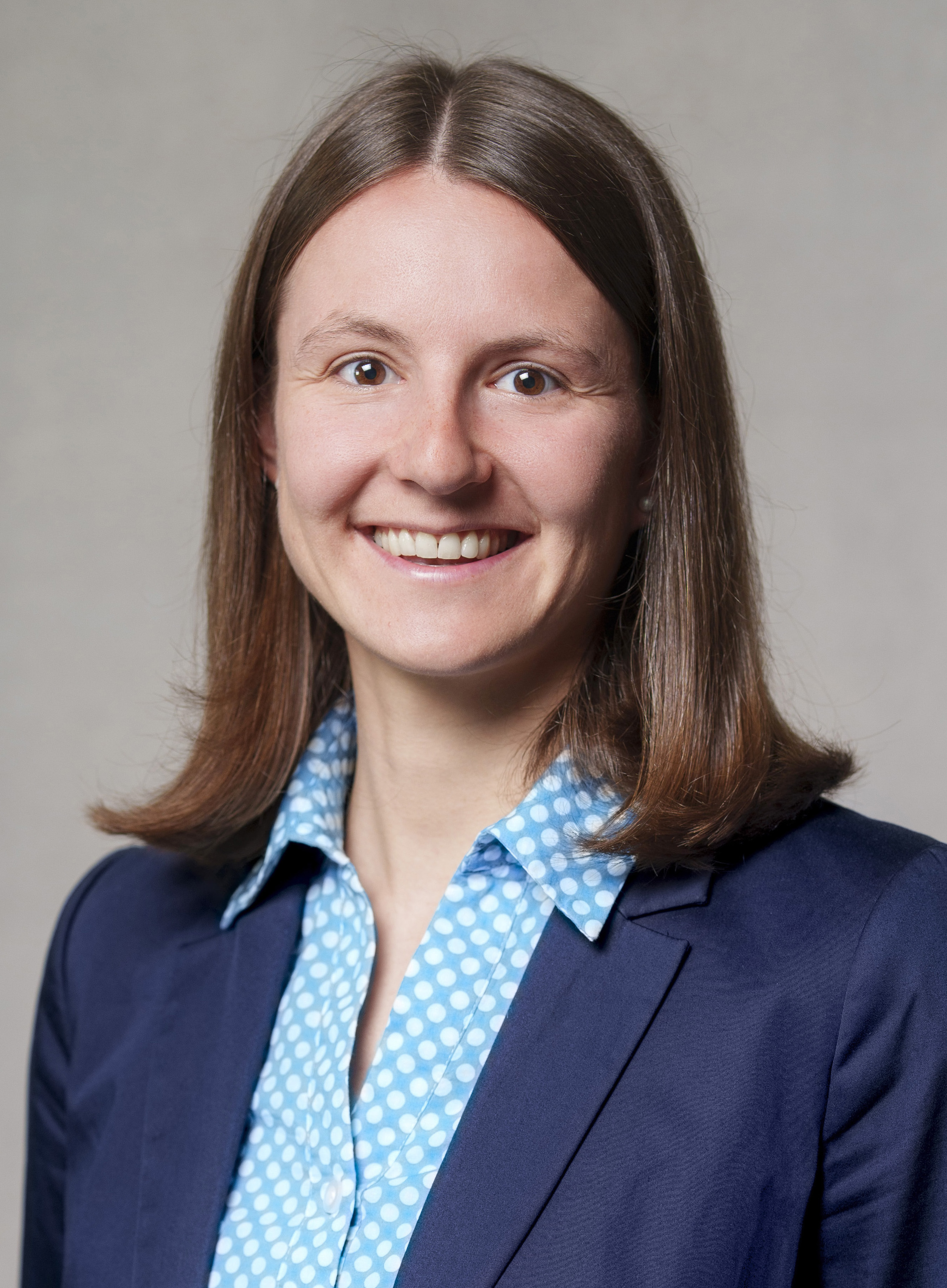}}]{Verena Häberle} is a Ph.D. student with the Automatic Control Laboratory at ETH Zurich, Switzerland, since June 2020. She received the B.Sc. and M.Sc. degree in electrical engineering and information technology from ETH Zurich, in 2018 and 2020, respectively. For her outstanding academic achievements during her Master's thesis at the Automatic Control Laboratory, ETH Zurich, under Professor Florian Dörfler, she was honored with the ETH Medal and the SGA Award from the Swiss Society of Automatic Control. Her research focuses on the control design of dynamic virtual power plants in future power systems.
\end{IEEEbiography}

\begin{IEEEbiography}[{\includegraphics[width=1in,height=1.25in,clip,keepaspectratio]{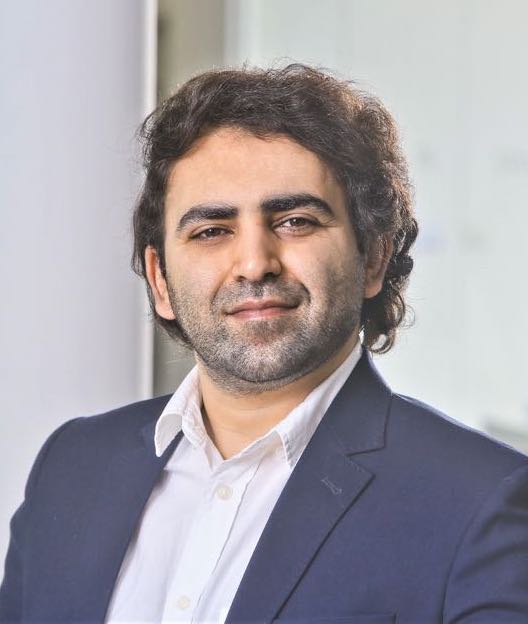}}]{Ali Tayyebi} received the B.Sc. degree in electrical engineering from the University of Tehran, Tehran, Iran, in 2012, the M.Sc. degree in engineering mathematics (joint MATHMODS program) from the University of L'Aquila, L'Aquila, Italy, and the University of Hamburg, Hamburg, Germany, in 2014, the second M.Sc. degree in sustainable transportation and electric power systems (joint STEPS program) from La Sapienza, Rome, Italy, University of Nottingham, Nottingham, U.K., and the University of Oviedo, Oviedo, Spain, in 2016.,In 2016, he joined Austrian Institute of Technology (AIT), Vienna, Austria, as a master thesis candidate and afterward continued there as a Research Assistant. In 2017, he started his joint Ph.D. project with AIT and Automatic Control Laboratory, Swiss Federal Institute of Technology (ETH) Zürich, Zürich, Switzerland. In 2022, he joined Hitachi Energy Corporate Research, Västerås, Sweden, as a Research Scientist., Mr. Tayyebi has won the IEEE PES General Meeting (PESGM) 2020 best paper award. From 2014 to 2016, he was the recipient of EU scholarship for master studies.
\end{IEEEbiography}

\begin{IEEEbiography}[{\includegraphics[width=1in,height=1.25in,clip,keepaspectratio]{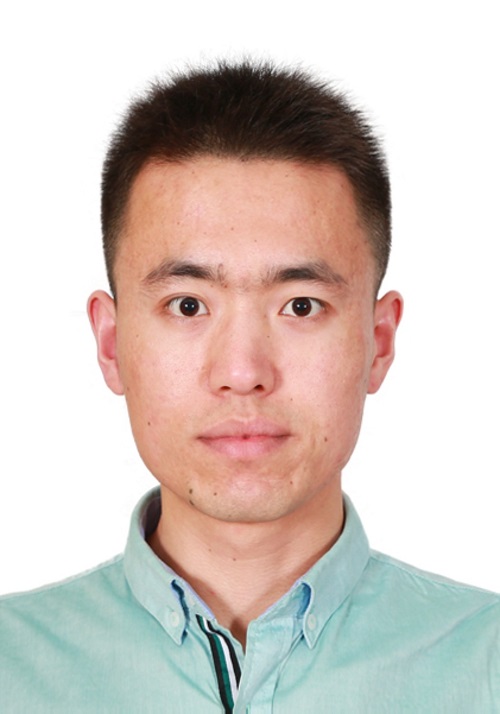}}]{Xiuqiang He} (Member, IEEE) received his B.S. degree and Ph.D. degree in control science and engineering from Tsinghua University, China, in 2016 and 2021, respectively. Since 2021, he has been a Postdoctoral Researcher with the Automatic Control Laboratory, ETH Zürich, Switzerland. His current research interests include stability issues in future power systems and grid ancillary services by renewables. Dr. He was the recipient of the Beijing Outstanding Graduates Award and the Outstanding Doctoral Dissertation Award of Tsinghua University.
\end{IEEEbiography}

\begin{IEEEbiography}[{\includegraphics[width=1in,height=1.25in,clip,keepaspectratio]{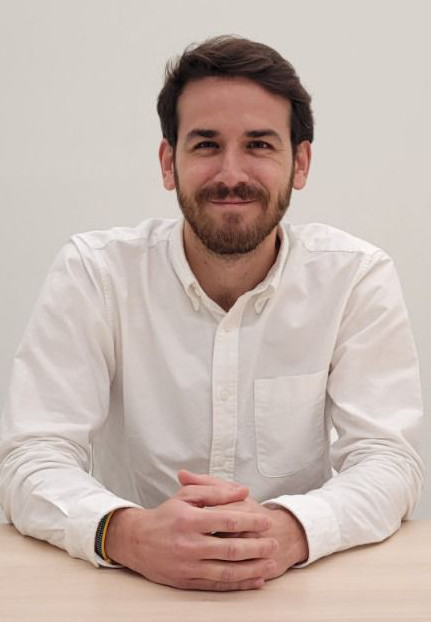}}]{Eduardo Prieto-Araujo} (S’12-M’16-SM’21) received the degree in industrial engineering from the School of Industrial Engineering of Barcelona (ETSEIB), Technical University of Catalonia (UPC), Barcelona, Spain, in 2011 and the Ph.D. degree in electrical engineering from the UPC in 2016. He joined CITCEA-UPC research group in 2010 and currently he is a Serra Hunter Associate Professor with the Electrical Engineering Department, UPC. During 2021, he was a visiting professor at the Automatic Control Laboratory, ETH Zurich. In 2022, he co-founded the start-up eRoots Analytics focused on the analysis of modern power systems. His main interests are renewable generation systems, control of power converters for HVDC applications, interaction analysis between converters and power electronics dominated power systems.
\end{IEEEbiography}

\begin{IEEEbiography}[{\includegraphics[width=1in,height=1.25in,clip,keepaspectratio]{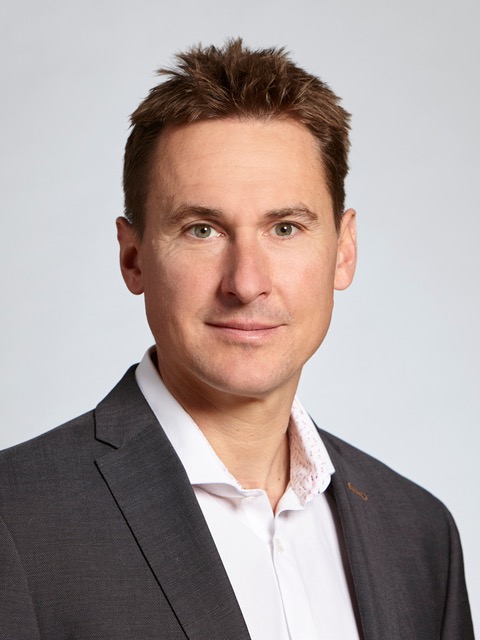}}]{Florian Dörfler} is an Associate Professor at the Automatic Control Laboratory at ETH Zürich. He received his Ph.D. degree in Mechanical Engineering from the University of California at Santa Barbara in 2013, and a Diplom degree in Engineering Cybernetics from the University of Stuttgart in 2008. From 2013 to 2014 he was an Assistant Professor at the University of California Los Angeles. He has been serving as the Associate Head of the ETH Zürich Department of Information Technology and Electrical Engineering from 2021 until 2022. His primary research interests are centered around control, optimization, and system theory with applications in network systems, in particular electric power grids. He is a recipient of the distinguished young research awards by IFAC (Manfred Thoma Medal 2020) and EUCA (European Control Award 2020). His students were winners or finalists for Best Student Paper awards at the European Control Conference (2013, 2019), the American Control Conference (2016), the Conference on Decision and Control (2020), the PES General Meeting (2020), the PES PowerTech Conference (2017), the International Conference on Intelligent Transportation Systems (2021), and the IEEE CSS Swiss Chapter Young Author Best Journal Paper Award (2022). He is furthermore a recipient of the 2010 ACC Student Best Paper Award, the 2011 O. Hugo Schuck Best Paper Award, the 2012-2014 Automatica Best Paper Award, the 2016 IEEE Circuits and Systems Guillemin-Cauer Best Paper Award, the 2022 IEEE Transactions on Power Electronics Prize Paper Award, and the 2015 UCSB ME Best PhD award.
\end{IEEEbiography}

\end{document}